\documentclass[a4paper,fleqn]{cas-sc}

\usepackage[numbers]{natbib}

\usepackage{amssymb}
\usepackage{amsmath}
\usepackage{graphicx}
\usepackage{booktabs}
\usepackage{multirow}
\usepackage{float}
\usepackage{array}
\usepackage{bm}
\usepackage{pdflscape}
\usepackage{caption}
\usepackage{calc}

\newcommand{\ohdr}[2]{\makebox[\widthof{#1}][c]{#2}}

\begin{document}

\shorttitle{Unified Benchmark Study Based on LSTM-PINN}
\shortauthors{C.Lin et~al.}

\title[mode=title]{A Unified Benchmark Study of Shock-Like Problems in Two-Dimensional Steady Electrohydrodynamic Flow Based on LSTM-PINN}
\author[1]{Chao Lin}[orcid=0009-0002-4669-3466]
\credit{Calculation, data analyzing and manuscript writing}\fnref{co-first}
\author[1]{Ze Tao}[orcid=0009-0004-0202-3641]
\credit{Calculation, data analyzing and manuscript writing}\fnref{co-first}
\affiliation[1]{organization={Nanophotonics and Biophotonics Key Laboratory of Jilin Province, School of Physics, Changchun University of Science and Technology},
                city={Changchun},
                postcode={130022},
                country={P.R. China}}
\author[1]{Fujun Liu}[orcid=0000-0002-8573-450X]
\credit{Review and Editing}
\cormark[1]
\fntext[co-first]{The authors contribute equally to this work.}
\ead{fjliu@cust.edu.cn}
\cortext[1]{Corresponding author}

\begin{abstract}
Accurately resolving steady electrohydrodynamic (EHD) flows presents a formidable computational challenge due to the strong nonlinear coupling between charged-particle density, velocity fields, and electric potential. These interactions frequently induce sharp transition layers, crossing fronts, and multiscale spatial structures, which notoriously degrade the predictive accuracy of standard mesh-free solvers like Physics-Informed Neural Networks (PINNs). To systematically address this bottleneck, we formulate a unified four-variable operator framework and develop a comprehensive benchmark suite for two-dimensional steady EHD shock-like problems. The benchmark comprises eight rigorously designed cases featuring diverse front geometries, such as oblique, curved, and intersecting layers, alongside complex multiscale patterns. Under strictly identical configurations, including governing equations, source terms, sampling strategies, and loss formulations, we evaluate a Standard MLP-based PINN, a Residual Attention PINN (ResAtt-PINN), and an LSTM-PINN that leverages pseudo-sequential spatial encoding. Extensive numerical experiments demonstrate that the LSTM-PINN consistently achieves the highest predictive accuracy across all eight cases. It successfully reconstructs sharp gradients and intricate multiscale structures where other architectures fail or over-smooth. Furthermore, the LSTM backbone efficiently captures long-range spatial correlations while maintaining an exceptionally low computational overhead and GPU memory footprint. These findings not only establish the LSTM-PINN as a robust and efficient solver for strongly coupled PDEs with shock-like features, but also provide the computational physics community with a standardized, reproducible benchmark for future algorithmic evaluations.
\end{abstract}

\begin{highlights}
    \item Unified benchmark proposed for strongly coupled steady EHD shock-like problems.
    \item A novel unified four-variable operator framework is mathematically formulated.
    \item Eight rigorous cases test sharp fronts, curved interfaces and multiscale patterns.
    \item LSTM-PINN rigorously outperforms Standard and ResAtt-PINNs across all test cases.
    \item The LSTM backbone captures sharp spatial gradients with low memory overhead.
\end{highlights}


\maketitle

\section{Introduction}

Steady electrohydrodynamic (EHD) transport is a key process in microfluidic chips, electroosmotic micropumps, and other electrically driven microsystems \cite{Woias2005Micropumps,Wang2009EOPReview,Zeng2001EOMicropump,Iranshahi2024EHD}, where it governs microscale pumping, transport regulation, and field manipulation. The primary computational difficulty in modeling these systems arises from the strong nonlinear coupling among charged-species distributions, flow fields, and electric potentials. This complex interaction frequently generates sharp transition layers, localized fronts, and multiscale spatial structures. Consequently, these inherent physical features make steady charged-fluid problems both practically important and highly demanding to resolve accurately using standard numerical techniques.

A large body of work has been devoted to the numerical modeling of EHD and related transport phenomena to address these computational demands. While classical mesh-based methods can achieve high accuracy in regular domains and for relatively smooth fields, their computational cost increases rapidly once thin layers, strong gradients, or geometrically intricate structures appear. To overcome these grid-based limitations, physics-informed neural networks (PINNs) have attracted considerable attention as a mesh-free alternative for solving partial differential equation constrained problems \cite{Raissi2019PINN}, and recent reviews have further highlighted their growing use in complex fluid dynamics \cite{Zhao2024PINNReview}. Despite this widespread adoption, standard MLP-based PINNs remain fragile in regimes characterized by sharp gradients, shock-like fronts, and multiscale dynamics. Under these conditions, network optimization becomes exceptionally difficult, and the recovered solutions tend to be overly smooth or locally inaccurate \cite{Abbasi2025ShockPINN,Tao2026LSTMPINN}. Recent developments have attempted to address complex physical interactions through interface-dominated multiphase dynamics and geometry-generalizable architectures, such as interval-constrained PINNs for tracking gas and liquid interfaces in underwater explosions and PI-KAN-PointNet for inverse incompressible flow over numerous irregular geometries \cite{Xing2025IntervalConstrainedPINN,Kashefi2025PIKANPointNet}. However, robustly capturing localized, shock-like structures within strongly coupled EHD fields remains a distinct and unresolved challenge for conventional network designs.

A promising step toward resolving the limitations of standard architectures in capturing sharp spatial gradients was recently made by Tao et al., who proposed the LSTM-PINN framework \cite{Tao2026LSTMPINN,tao2025lstm2}. By replacing the conventional MLP backbone with stacked long short-term memory (LSTM) layers and introducing a pseudo-sequential spatial encoding strategy, their method fundamentally improves the representation of long-range spatial correlations. This architectural shift achieves better convergence, stability, and predictive accuracy than conventional PINNs \cite{xing2025modeling,tao2025analytical,zhang2023physics}. The success of this approach suggests that recurrent memory mechanisms are especially well-suited for target fields containing highly structured fronts and strongly coupled spatial patterns, which are the exact characteristics of complex EHD flows.

Motivated by the potential of recurrent architectures to handle such localized features, the present work develops a unified benchmark framework specifically tailored for two-dimensional steady charged-fluid problems with shock-like structures. Utilizing this framework, we systematically compare a Standard PINN, a Residual Attention PINN (ResAtt-PINN), and the LSTM-PINN under strictly controlled and fully comparable settings. Starting from a modified steady governing system, we formulate a unified four-variable operator framework and construct eight representative benchmark cases that span a diverse array of multiscale spatial structures. Through rigorous qualitative and quantitative evaluations, this study extends the utility of the LSTM-PINN to a highly demanding charged-fluid setting. Ultimately, this work provides the computational physics community with a standardized benchmark for evaluating emerging network architectures on steady EHD problems, directly supporting the broader demand for systematic and reproducible benchmark studies in physics-informed machine learning \cite{PINNacle2024Benchmark}.

\section{Mathematical Formulation}

\subsection{Unified Governing Equations}

Although the eight benchmark cases considered in this study differ in geometric pattern and field structure, they all belong to the same coupled system of two-dimensional steady electrohydrodynamic flow. The unified unknown vector is defined as
\[
\mathbf{q}=(n,u_x,u_y,\phi)^T,
\]
where \(n\) denotes the charged particle number density, \(u_x\) and \(u_y\) are the two velocity components, and \(\phi\) is the electric potential.

Unlike the original LSTM-PINN paper by Tao et al. \cite{Tao2026LSTMPINN}, which mainly focused on a steady EHD model for a two-dimensional velocity field \((u,v)\), the present work considers a four-variable benchmark setting under the same unified governing framework. The governing equations are written as
\begin{equation}
\frac{\partial (n u_x)}{\partial x}+\frac{\partial (n u_y)}{\partial y}=S_n(x,y),
\end{equation}

\begin{equation}
u_x\frac{\partial u_x}{\partial x}
+u_y\frac{\partial u_x}{\partial y}
+\frac{a^2}{n}\frac{\partial n}{\partial x}
+\frac{\partial \phi}{\partial x}
-\nu\left(\frac{\partial^2 u_x}{\partial x^2}+\frac{\partial^2 u_x}{\partial y^2}\right)
=S_{u_x}(x,y),
\end{equation}

\begin{equation}
u_x\frac{\partial u_y}{\partial x}
+u_y\frac{\partial u_y}{\partial y}
+\frac{a^2}{n}\frac{\partial n}{\partial y}
+\frac{\partial \phi}{\partial y}
-\nu\left(\frac{\partial^2 u_y}{\partial x^2}+\frac{\partial^2 u_y}{\partial y^2}\right)
=S_{u_y}(x,y),
\end{equation}

\begin{equation}
-\varepsilon^2\left(\frac{\partial^2 \phi}{\partial x^2}+\frac{\partial^2 \phi}{\partial y^2}\right)
-(n-n_{\mathrm{bg}})
=S_{\phi}(x,y).
\end{equation}

For compact notation, the above system can also be expressed in operator form as
\[
\mathcal{L}(\mathbf{q};\lambda)=0,
\]
where \(\lambda\) denotes the set of physical parameters and case-specific parameters.

The first equation represents conservation of charged-particle flux, the second and third equations represent momentum balance in the two coordinate directions, and the fourth equation is a Poisson-type electric-potential equation. Since the system simultaneously contains nonlinear convection, diffusion, electric-potential gradients, and density gradients, it can naturally produce localized fronts, crossing layers, radial transition zones, and multiscale pockets. For the eight benchmark settings considered in this work, the resulting problems remain analytically tractable, thereby providing a direct basis for quantitative error evaluation under the same unified physical shell.

\subsection{Unified Physical Interpretation of the Benchmark Cases}

To give the eight benchmark cases a clearer practical interpretation, all of them are uniformly viewed as abstract steady electrohydrodynamic flow patterns in microfluidic chips or electro-osmotic micropumps \cite{Iranshahi2024EHD,Woias2005Micropumps,Wang2009EOPReview}. Under this physical shell, \(n\) corresponds to the local charged-particle density, \((u_x,u_y)\) corresponds to the two-dimensional velocity field, and \(\phi\) corresponds to the electric-potential distribution. Meanwhile, the parameters \(a^2\), \(\nu\), \(\varepsilon^2\), and \(n_{\mathrm{bg}}\) correspond to the effective compressibility response, viscous diffusion strength, electrostatic coupling coefficient, and background particle density, respectively.

Under this interpretation, the eight cases are no longer isolated mathematical functions; instead, they represent different steady EHD field patterns within the same device-level physical shell. The benchmark cases considered in this work are summarized as follows:

\begin{table}[H]
\centering
\footnotesize
\setlength{\tabcolsep}{4pt}
\renewcommand{\arraystretch}{1.05}
\caption{Unified physical parameters and structural descriptions of the eight benchmark cases.}
\label{tab:benchmark_cases}
\resizebox{\textwidth}{!}{%
\begin{tabular}{c p{8.8cm} c c c c}
\toprule
Case & Description & $a^2$ & $\nu$ & $\varepsilon^2$ & $n_{\mathrm{bg}}$ \\
\midrule
Case 01 & Vertical internal shock layer with mild transverse waviness & 0.30 & 0.020 & 0.080 & 1.00 \\
Case 02 & Horizontal shock with cross-flow modulation & 0.28 & 0.018 & 0.090 & 1.00 \\
Case 03 & Single oblique shock, smooth background and electrostatic tilt & 0.32 & 0.020 & 0.085 & 1.00 \\
Case 04 & Two interacting oblique layers, one compressive and one expansive & 0.35 & 0.022 & 0.080 & 1.00 \\
Case 05 & Curved radial front with weak rotating background flow & 0.27 & 0.016 & 0.075 & 1.00 \\
Case 06 & Electro-shear layer with localized density pocket & 0.25 & 0.020 & 0.095 & 1.00 \\
Case 07 & Double front in $x$ with secondary transverse layer and Gaussian forcing pocket & 0.30 & 0.024 & 0.080 & 1.00 \\
Case 08 & Hybrid oblique-front + local pocket, hardest multiscale benchmark & 0.34 & 0.018 & 0.070 & 1.00 \\
\bottomrule
\end{tabular}%
}
\end{table}

\section{Methodology}

\subsection{LSTM-PINN Architecture}

The network design adopted in this study is directly motivated by the LSTM-PINN architecture proposed by Tao et al. \cite{Tao2026LSTMPINN}. In this work, the algorithmic framework and the basic design idea of the recurrent physics-informed backbone are therefore inherited from the LSTM-PINN method reported by Tao et al. \cite{Tao2026LSTMPINN}. Unlike traditional PINNs, which feed spatial coordinates directly into a multilayer perceptron, LSTM-PINN first transforms the spatial coordinates \((x,y)\) into sequence-compatible inputs through pseudo-sequential spatial encoding, then models long-range spatial correlations by stacked LSTM layers, and finally produces the predictions of physical field variables through a linear output head.

\begin{figure}[H]
    \centering
    \includegraphics[width=0.99\textwidth]{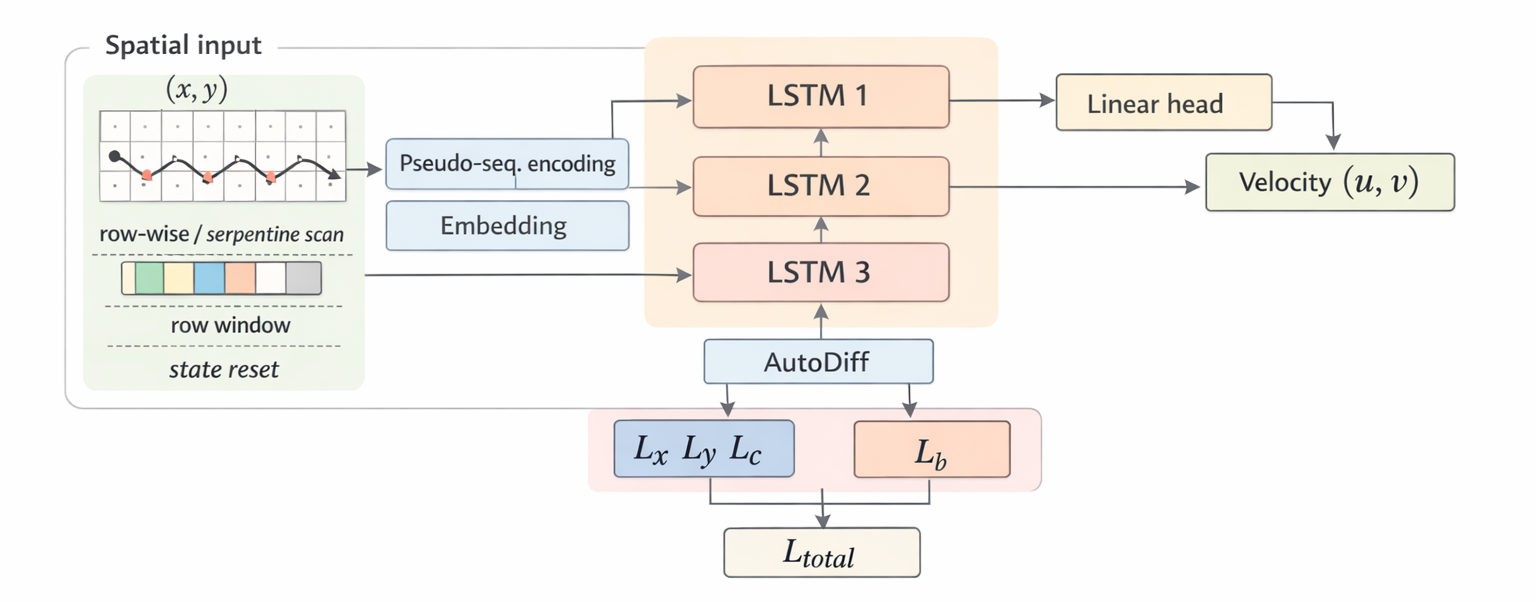}
    \caption{Conceptual architecture of the LSTM-PINN workflow adapted from Tao et al., showing the pseudo-sequential encoding, stacked recurrent backbone, output head, and physics-informed loss construction. In the present study, the same design idea is extended from the original two-output formulation to the four-variable field \((n,u_x,u_y,\phi)\).}
    \label{fig:lstm_pinn_architecture}
\end{figure}

In the original LSTM-PINN paper, the network output corresponds to a two-dimensional velocity field \((u,v)\). In the present study, however, the same architectural idea is extended to the four-variable steady electrohydrodynamic field \((n,u_x,u_y,\phi)\). For consistency at the task level, the comparative models considered in this work are all configured to predict the same four-variable field, so that the subsequent performance differences can be attributed primarily to the backbone architecture rather than to differences in output definition. Figure~\ref{fig:lstm_pinn_architecture} illustrates the LSTM-PINN architecture used in this work.

\begin{figure}[H]
    \centering
    \includegraphics[width=0.99\textwidth]{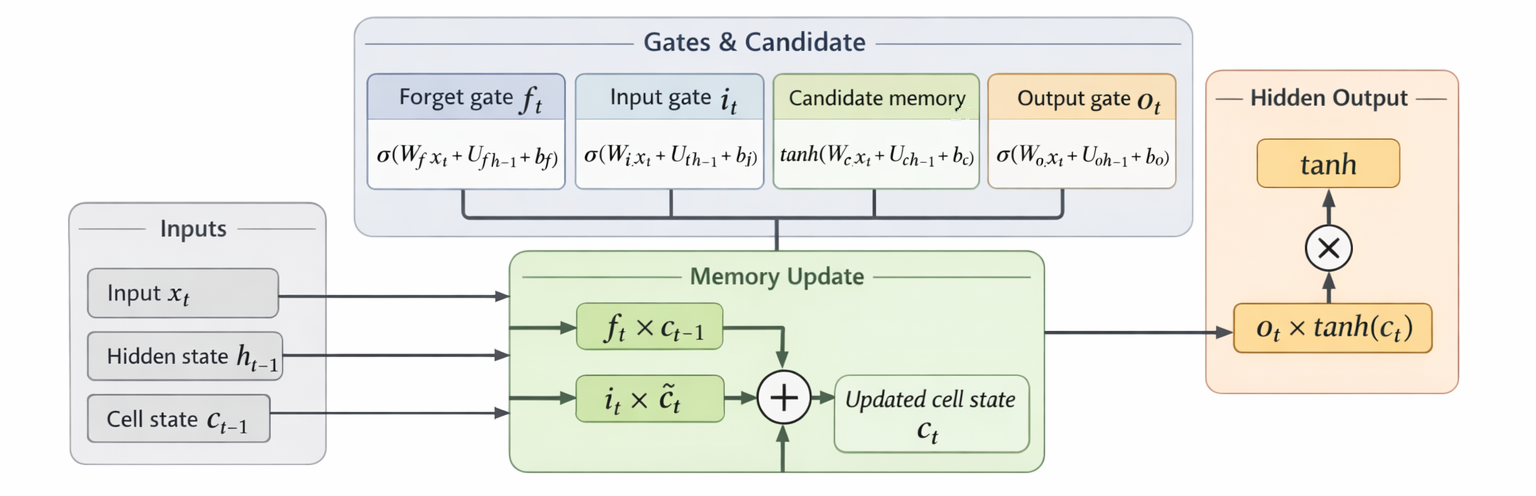}
    \caption{Gate-level structure of an LSTM unit, illustrating the coordinated roles of the input gate, forget gate, output gate, hidden state, and cell state in information updating and propagation.}
    \label{fig:lstm_gate_structure}
\end{figure}

To further clarify the internal information flow of the recurrent backbone, Figure~\ref{fig:lstm_gate_structure} shows the gate-level structure of an LSTM unit. The input gate, forget gate, and output gate jointly control how pseudo-spatial information is written into, retained in, and read out from the cell state. This gated memory mechanism provides the structural basis for modeling long-range spatial dependence in the pseudo-sequential encoding framework.

In the original formulation, the core update equations of an LSTM unit are \cite{Hochreiter1997LSTM}
\begin{equation}
c_t=f_t\odot c_{t-1}+i_t\odot \tilde{c}_t,
\end{equation}
\begin{equation}
h_t=o_t\odot \tanh(c_t),
\end{equation}
where the candidate memory state and the gate functions satisfy
\begin{equation}
\tilde{c}_t=\tanh(W_c x_t+U_c h_{t-1}+b_c),
\end{equation}
\begin{equation}
i_t=\sigma(W_i x_t+U_i h_{t-1}+b_i),\quad
f_t=\sigma(W_f x_t+U_f h_{t-1}+b_f),\quad
o_t=\sigma(W_o x_t+U_o h_{t-1}+b_o).
\end{equation}

These equations indicate that the LSTM selectively filters and updates historical information through the input, forget, and output gates, thereby enabling stable hidden-state propagation along pseudo-spatial sequences and improving the representation of complex spatial patterns. In the original PINN framework, the total loss was formulated as
\begin{equation}
\mathcal{L}_{\mathrm{total}}=\mathcal{L}_x+\mathcal{L}_y+\mathcal{L}_c+\mathcal{L}_b.
\end{equation}

Although the present study does not directly adopt the original two-velocity governing equations, it preserves the core idea of a stacked LSTM plus physics-informed loss framework \cite{Tao2026LSTMPINN}. At the same time, the adapted LSTM-PINN is placed in a unified four-variable benchmark setting together with Standard PINN and ResAtt-PINN under the same prediction target and physical constraints.

\subsection{Unified Residual Construction and Loss Function}

During the training of all three models, the network output is uniformly denoted by
\[
\hat{\mathbf{q}}(x,y)=\left(\hat{n},\hat{u}_x,\hat{u}_y,\hat{\phi}\right)^T.
\]
For each benchmark case, the physical residuals are defined by inserting the network prediction together with the corresponding case-specific source terms \(S_n\), \(S_{u_x}\), \(S_{u_y}\), and \(S_\phi\) into the unified governing equations:
\begin{equation}
R_n=\partial_x(\hat{n}\hat{u}_x)+\partial_y(\hat{n}\hat{u}_y)-S_n,
\end{equation}
\begin{equation}
R_{u_x}=\hat{u}_x\partial_x \hat{u}_x+\hat{u}_y\partial_y \hat{u}_x+\frac{a^2}{\hat{n}}\partial_x \hat{n}+\partial_x \hat{\phi}-\nu \Delta \hat{u}_x-S_{u_x},
\end{equation}
\begin{equation}
R_{u_y}=\hat{u}_x\partial_x \hat{u}_y+\hat{u}_y\partial_y \hat{u}_y+\frac{a^2}{\hat{n}}\partial_y \hat{n}+\partial_y \hat{\phi}-\nu \Delta \hat{u}_y-S_{u_y},
\end{equation}
\begin{equation}
R_{\phi}=-\varepsilon^2 \Delta \hat{\phi}-(\hat{n}-n_{\mathrm{bg}})-S_{\phi}.
\end{equation}

Accordingly, the physics loss at the interior collocation points is defined as
\begin{equation}
\mathcal{L}_{\mathrm{PDE}}
=
\frac{1}{N_i}\sum_{j=1}^{N_i}
\left(
R_n^2+R_{u_x}^2+R_{u_y}^2+R_\phi^2
\right),
\end{equation}
while the supervised boundary loss is defined as
\begin{equation}
\mathcal{L}_{\mathrm{BC}}
=
\frac{1}{N_b}\sum_{j=1}^{N_b}
\left[
(\hat{n}-n)^2+(\hat{u}_x-u_x)^2+(\hat{u}_y-u_y)^2+(\hat{\phi}-\phi)^2
\right].
\end{equation}

The final total loss is therefore
\begin{equation}
\mathcal{L}_{\mathrm{total}}
=
\lambda_{\mathrm{PDE}}\mathcal{L}_{\mathrm{PDE}}
+
\lambda_{\mathrm{BC}}\mathcal{L}_{\mathrm{BC}}.
\end{equation}

Since Standard PINN, ResAtt-PINN, and LSTM-PINN use exactly the same data split, boundary supervision, source-term construction, and loss definition, together with the same governing equations and prediction target, the performance differences discussed later can be attributed directly to the backbone architecture itself rather than to changes in the physical constraints or supervision setting. The present formulation preserves the standard physics-informed residual-learning principle introduced in PINNs while adapting it to the current four-variable benchmark setting \cite{Raissi2019PINN,Tao2026LSTMPINN}.

\section{Results and Analysis}

\subsection{Case 1: Vertical Internal Shock Layer}

Case 01 represents a vertical internal shock layer with mild transverse waviness. Under the unified physical shell, it serves as the simplest front-dominated benchmark and mainly tests whether a model can accurately recover the position, thickness, and continuity of a relatively simple transition layer.

\begin{figure}[H]
    \centering
    \includegraphics[width=0.95\textwidth]{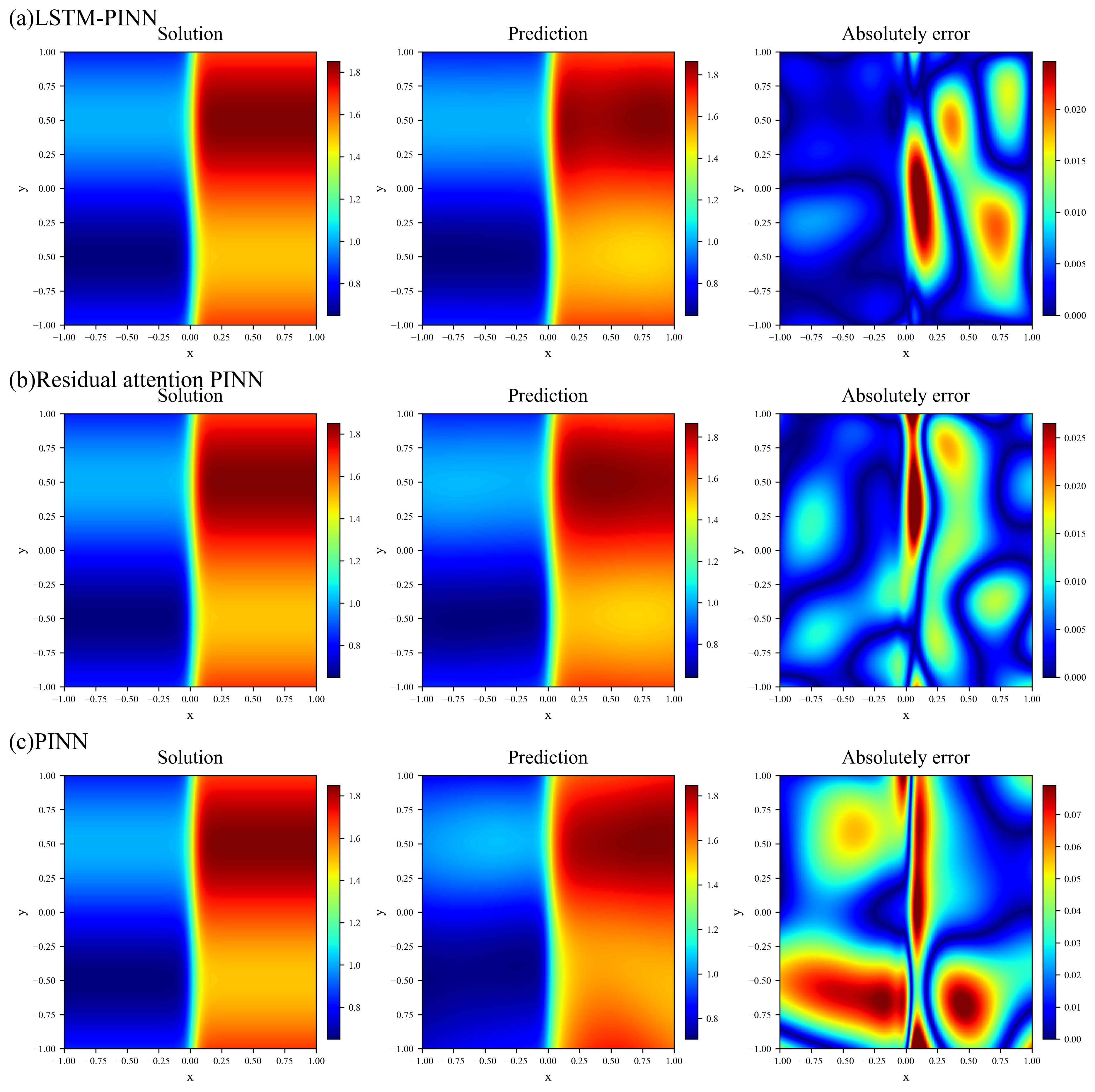}
    \caption{3$\times$3 reconstruction comparison of the \(n\)-field for Case 01. The first row shows the reference solution, the second row shows the predictions of the three models, and the third row shows the corresponding absolute errors.}
    \label{fig:case01_n_compare}
\end{figure}

\begin{figure}[H]
    \centering
    \includegraphics[width=0.95\textwidth]{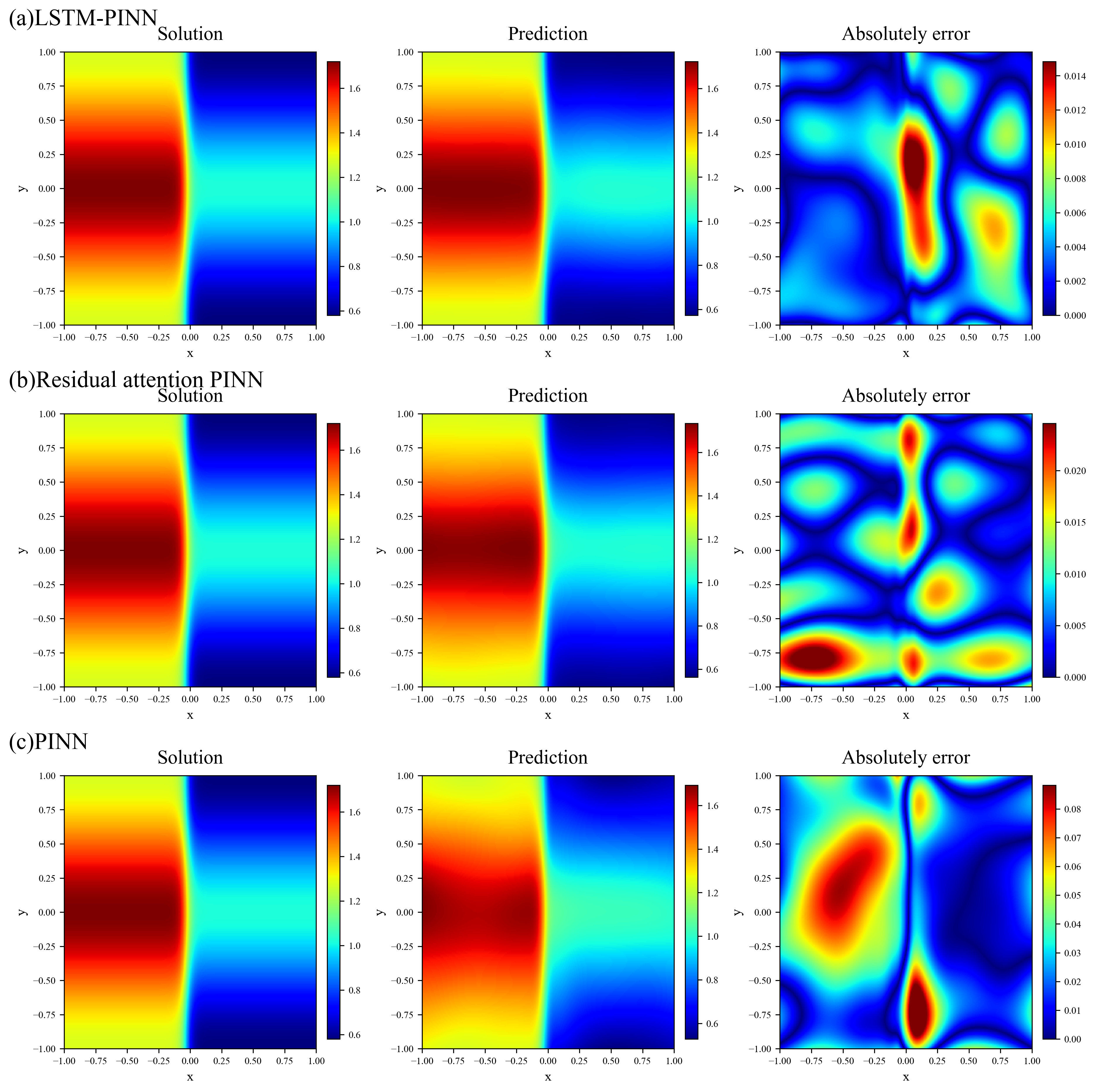}
    \caption{3$\times$3 reconstruction comparison of the \(u_x\)-field for Case 01.}
    \label{fig:case01_ux_panel_3x3}
\end{figure}

\begin{figure}[H]
    \centering
    \includegraphics[width=0.95\textwidth]{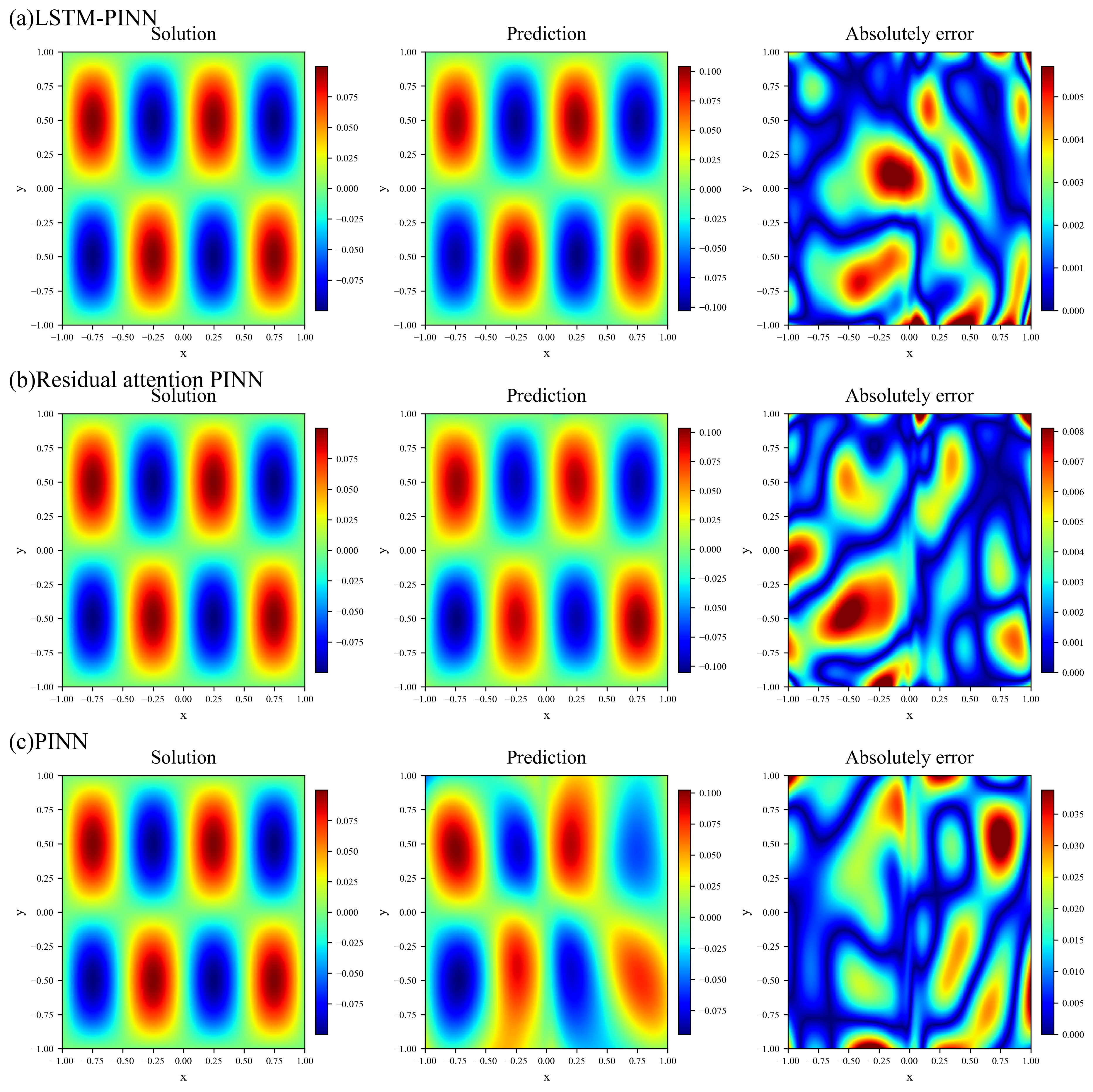}
    \caption{3$\times$3 reconstruction comparison of the \(u_y\)-field for Case 01.}
    \label{fig:case01_uy_panel_3x3}
\end{figure}

\begin{figure}[H]
    \centering
    \includegraphics[width=0.95\textwidth]{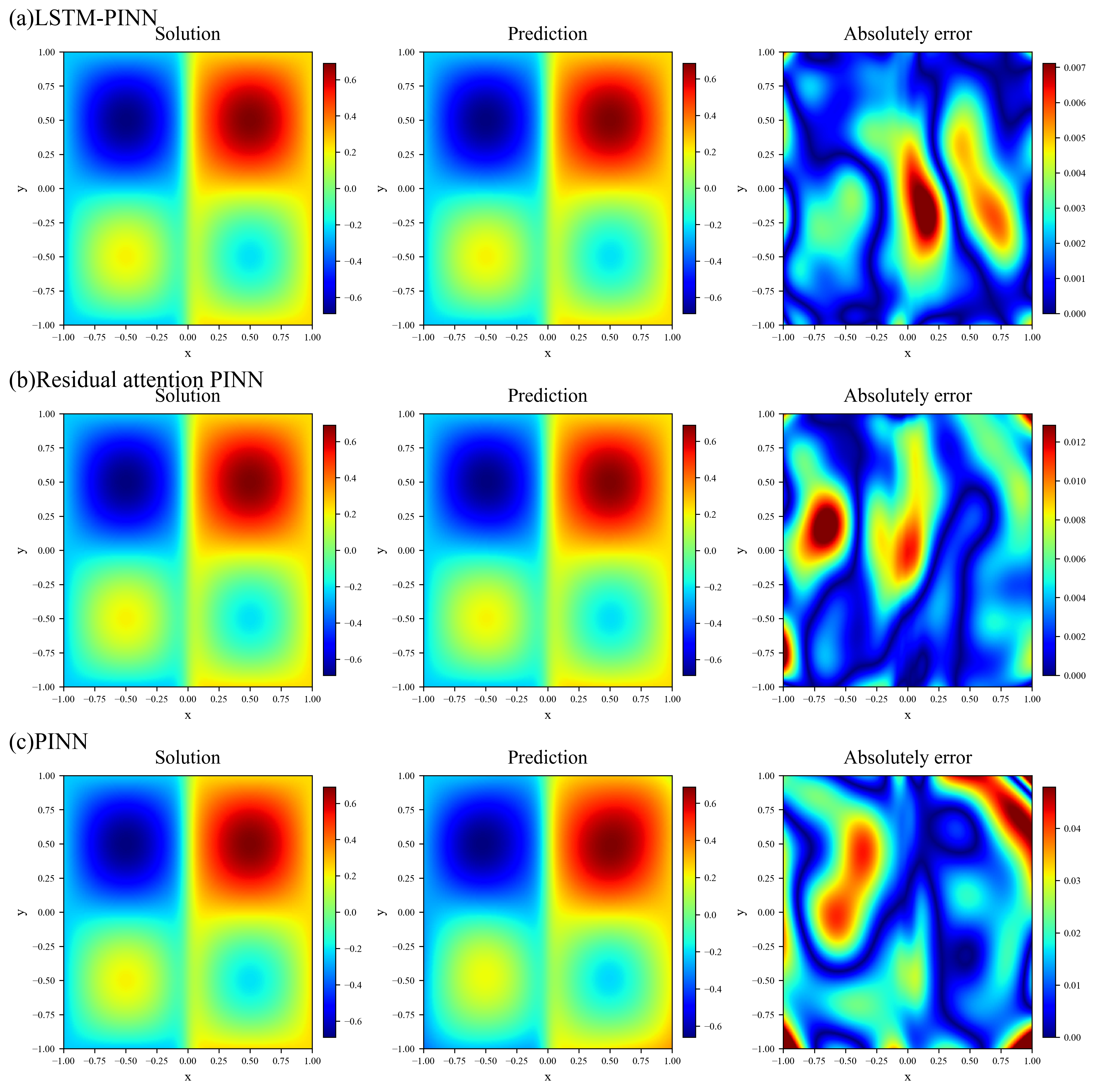}
    \caption{3$\times$3 reconstruction comparison of the \(\phi\)-field for Case 01.}
    \label{fig:case01_phi_panel_3x3}
\end{figure}

As shown in Fig.~\ref{fig:case01_n_compare}, all three models identify the approximate location of the vertical front, but LSTM-PINN reconstructs the sharpest transition and produces the closest layer thickness to the reference solution. Its error band remains narrower along the front, indicating better preservation of layer continuity across the domain. ResAtt-PINN ranks second and captures the main front correctly, but its error remains more visible in the front-adjacent region. By contrast, Standard PINN exhibits stronger diffusion on both sides of the layer and a visibly broader smearing band. The same ranking is also supported by the associated \(u_x\), \(u_y\), and \(\phi\) fields in Figs.~\ref{fig:case01_ux_panel_3x3}, \ref{fig:case01_uy_panel_3x3}, and \ref{fig:case01_phi_panel_3x3}. In the \(u_x\) panel, LSTM-PINN keeps the vertical jump sharp and confines the main error to a thin strip near the interface, whereas Standard PINN generates much broader off-interface distortion in both the upper-left and lower-left regions. In the oscillatory \(u_y\) field, LSTM-PINN better preserves the phase-separated band structure across the front, while Standard PINN visibly deforms the right-side lobes and introduces stronger amplitude leakage into the background. The \(\phi\) panel shows the same tendency: LSTM-PINN retains the quadrant-like potential pattern with the most compact front-adjacent error, whereas the other two models, especially Standard PINN, exhibit a wider and less coherent distortion footprint across the coupled variables.

\begin{landscape}
\begin{figure}[p]
    \centering
    \includegraphics[width=1\linewidth]{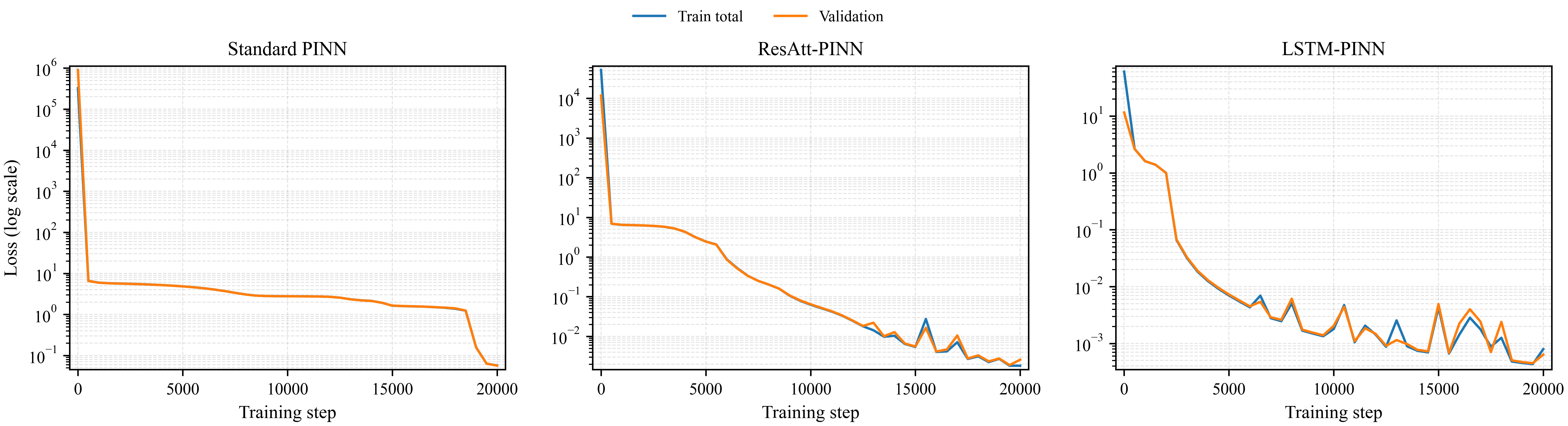}
    \caption{Training and validation loss curves for Case 01.}
    \label{fig:case01_loss_compare}

    \vspace{3em}

    \includegraphics[width=1\linewidth]{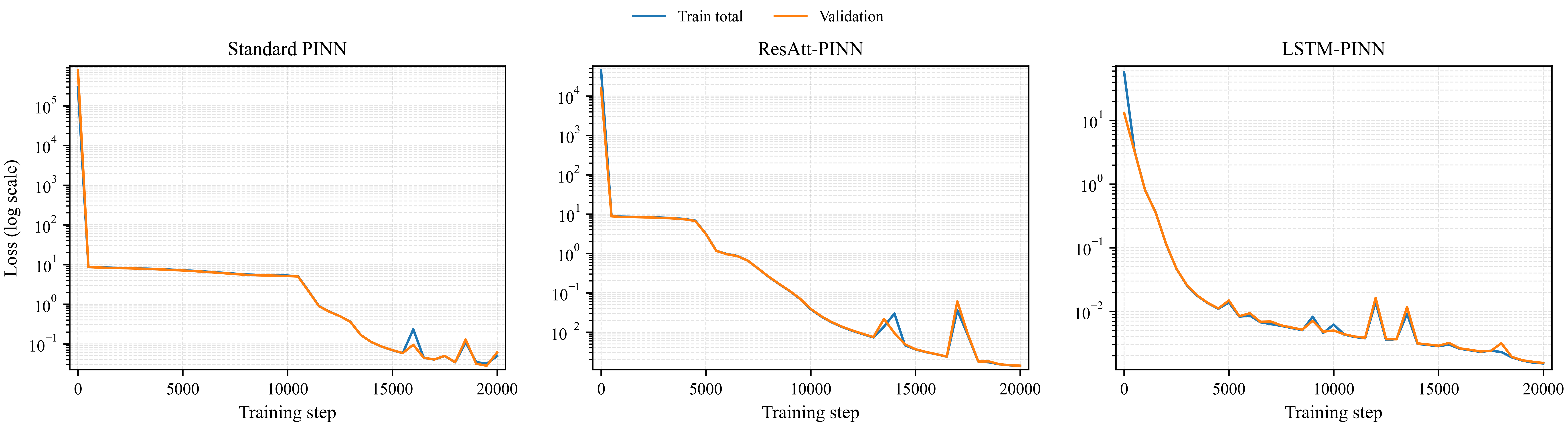}
    \caption{Training and validation loss curves for Case 02.}
    \label{fig:case02_loss_compare}
\end{figure}
\end{landscape}

Table~\ref{tab:case01_metrics} quantitatively confirms the same trend: LSTM-PINN achieves the smallest overall RMSE of 0.0046, compared with 0.0066 for ResAtt-PINN and 0.0296 for Standard PINN. The same ranking is also observed for overall MSE, MAE, and \(L_2\), which further confirms that LSTM-PINN not only reduces the average error level but also better suppresses relatively large local deviations near the shock layer. Even in this simplest shock-layer scenario, LSTM-PINN already shows the best front-resolution capability. The corresponding loss-evolution curves are shown in Fig.~\ref{fig:case01_loss_compare}. In this case, Standard PINN undergoes a sharp initial loss drop but then remains in a long plateau-like regime at approximately the \(\mathcal{O}(10^{0})\) level for most of training, indicating limited ability to further reduce the coupled residuals after capturing the coarse front structure. ResAtt-PINN converges much deeper, while LSTM-PINN enters the low-loss regime earlier and reaches the lowest terminal loss overall. The close agreement between the training and validation curves also suggests that this advantage is associated with more effective optimization rather than an evident train--validation separation.

\begin{center}
\captionsetup{hypcap=false}
\captionof{table}{Quantitative comparison for Case 01.}
\label{tab:case01_metrics}
\small
\setlength{\tabcolsep}{5pt}
\renewcommand{\arraystretch}{1.12}
\resizebox{\textwidth}{!}{%
\begin{tabular}{lcccccc}
\toprule
Model & \ohdr{overall RMSE}{RMSE} & \ohdr{overall MSE}{MSE} & \ohdr{overall MAE}{MAE} & \ohdr{overall \(L_2\)}{\(L_2\)} & Training time (s) & Peak GPU memory (GB) \\
\midrule
Standard PINN & 0.0296 & 8.76e-04 & 0.0226 & 0.0331 & \textbf{715.3} & 3.455 \\
ResAtt-PINN & 0.0066 & 4.37e-05 & 0.0048 & 0.0074 & 5304.8 & 5.108 \\
LSTM-PINN & \textbf{0.0046} & \textbf{2.09e-05} & \textbf{0.0030} & \textbf{0.0051} & 3366.8 & \textbf{1.699} \\
\bottomrule
\end{tabular}}
\end{center}

\subsection{Case 2: Horizontal Shock Layer}

Case 02 corresponds to a horizontal shock layer with cross-flow modulation. Compared with Case~01, the dominant variation direction is rotated from the \(x\)-direction to the \(y\)-direction, which helps verify whether the superiority of the model depends on front orientation.

\begin{figure}[H]
    \centering
    \includegraphics[width=0.95\textwidth]{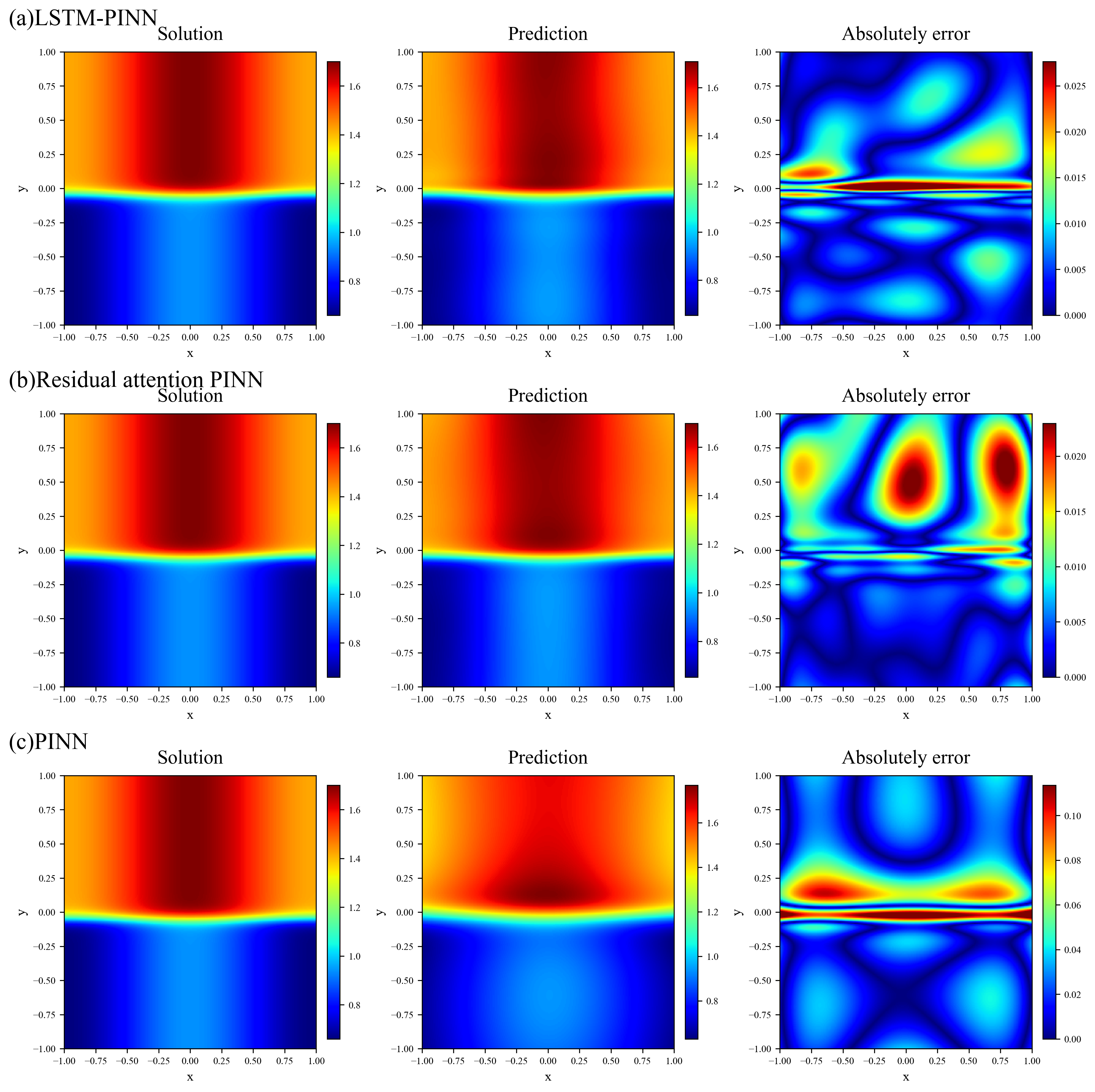}
    \caption{3$\times$3 reconstruction comparison of the \(n\)-field for Case 02, highlighting the recovery of the horizontal shock layer and the cross-flow modulation.}
    \label{fig:case02_n_compare}
\end{figure}

\begin{figure}[H]
    \centering
    \includegraphics[width=0.95\textwidth]{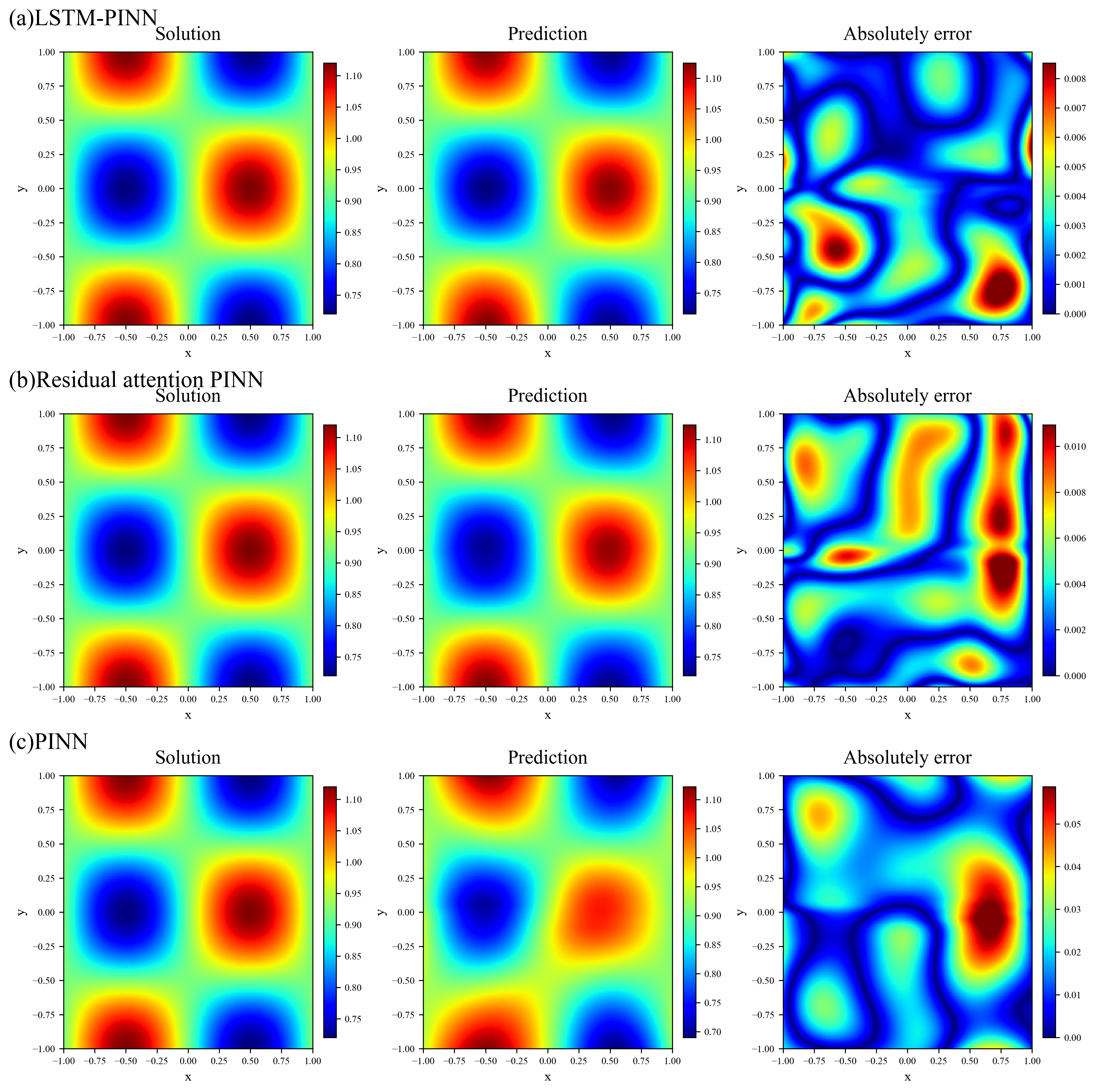}
    \caption{3$\times$3 reconstruction comparison of the \(u_x\)-field for Case 02.}
    \label{fig:case02_ux_panel_3x3}
\end{figure}

\begin{figure}[H]
    \centering
    \includegraphics[width=0.95\textwidth]{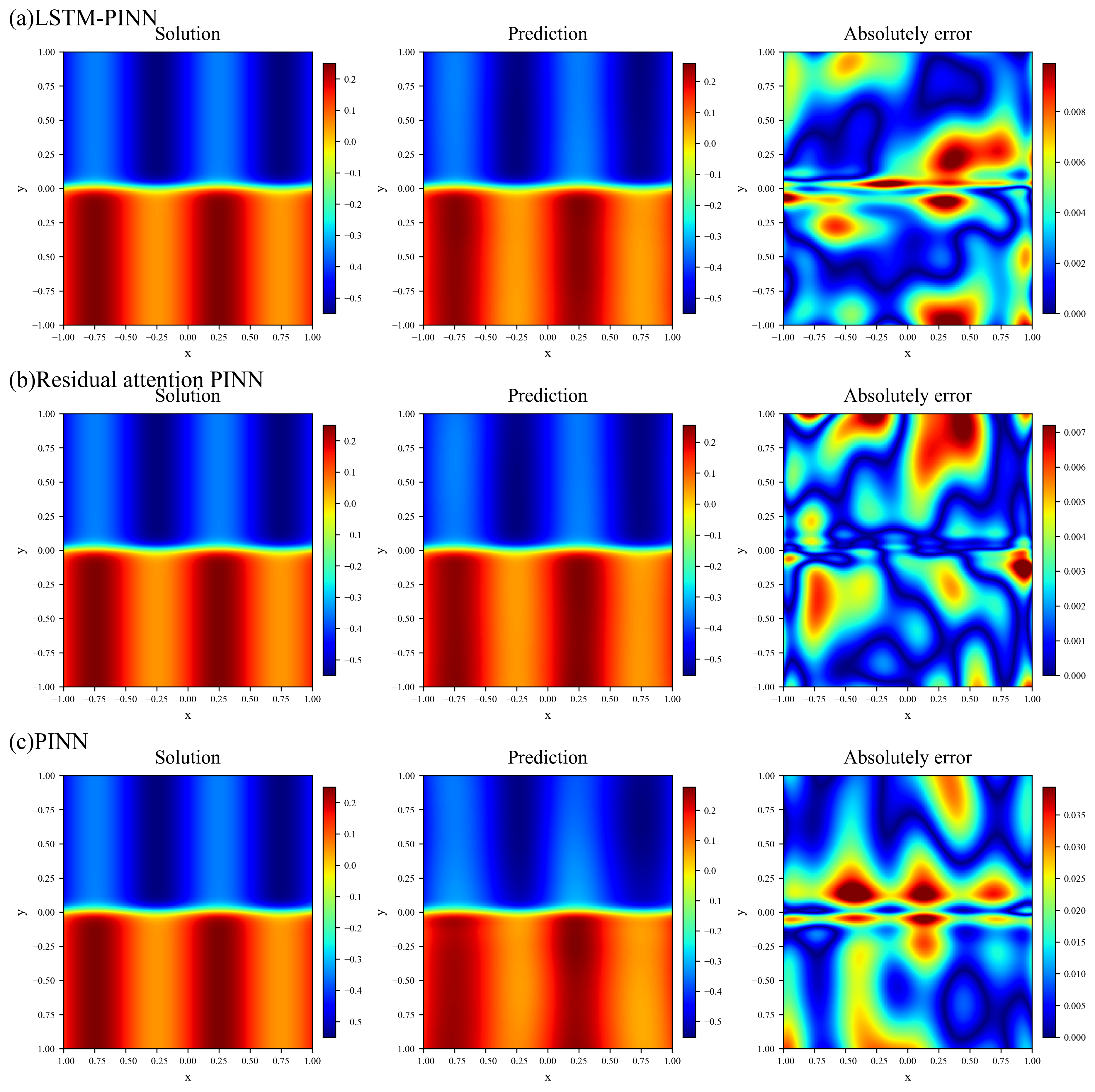}
    \caption{3$\times$3 reconstruction comparison of the \(u_y\)-field for Case 02.}
    \label{fig:case02_uy_panel_3x3}
\end{figure}

\begin{figure}[H]
    \centering
    \includegraphics[width=0.95\textwidth]{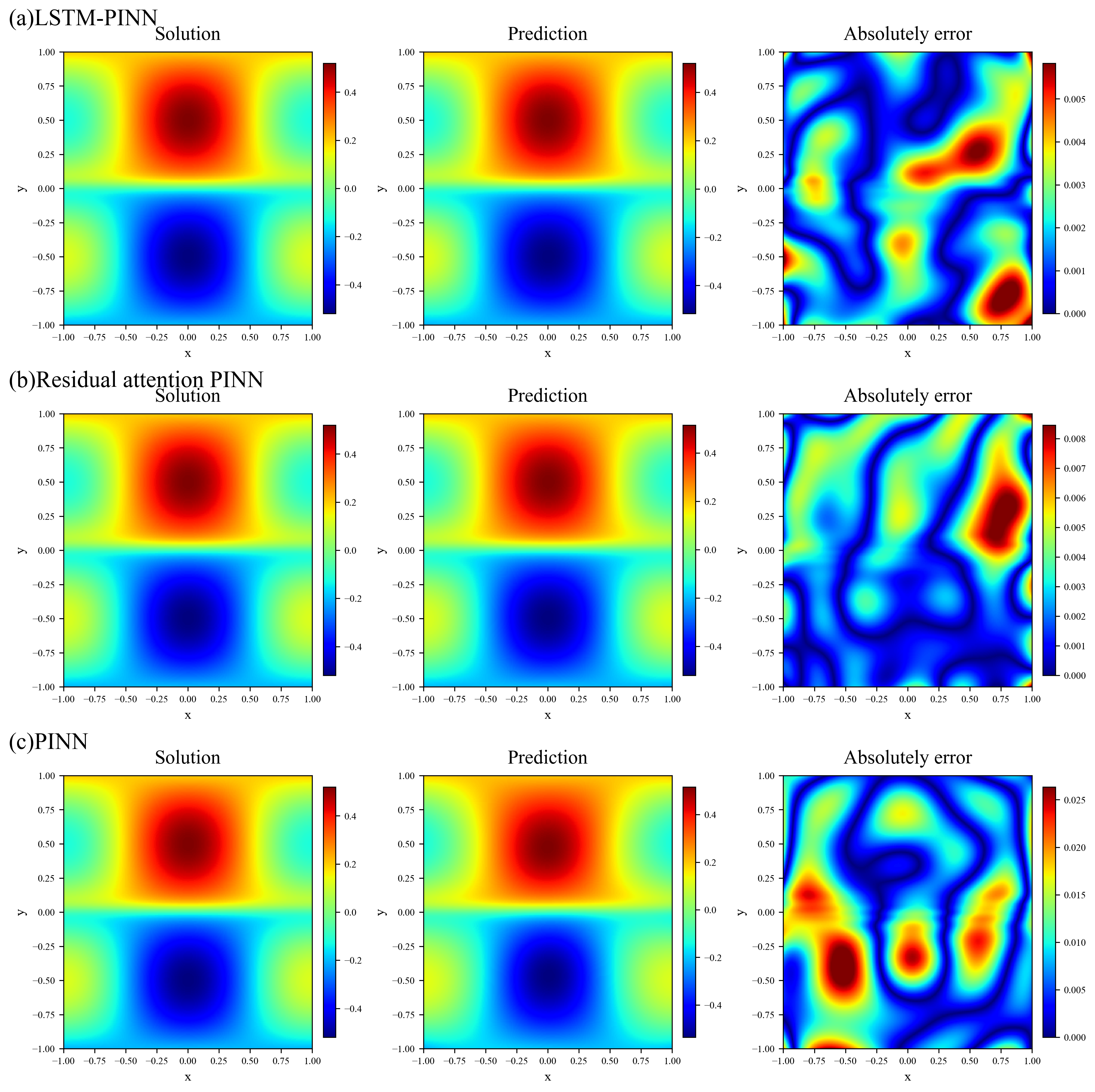}
    \caption{3$\times$3 reconstruction comparison of the \(\phi\)-field for Case 02.}
    \label{fig:case02_phi_panel_3x3}
\end{figure}

The reconstruction results shown in Fig.~\ref{fig:case02_n_compare} show that LSTM-PINN still maintains the smallest error in the horizontal layer position, the transverse oscillation amplitude, and the transition in the smooth region. This suggests that the advantage of the LSTM backbone is not limited to vertically aligned fronts. Standard PINN exhibits a broader error band around the layer, while ResAtt-PINN again lies between the two. This cross-variable consistency is also visible in the corresponding \(u_x\), \(u_y\), and \(\phi\) reconstructions in Figs.~\ref{fig:case02_ux_panel_3x3}, \ref{fig:case02_uy_panel_3x3}, and \ref{fig:case02_phi_panel_3x3}. In the checkerboard-like \(u_x\) field, LSTM-PINN preserves both the horizontal interface location and the relative amplitude balance of the four lobes, whereas Standard PINN noticeably distorts the lower-right lobe and spreads the error farther away from the layer. In the \(u_y\) field, LSTM-PINN maintains a thinner transition band and better keeps the weak streamwise modulation aligned across the horizontal front, while Standard PINN again produces a thicker interfacial strip. The \(\phi\) panel leads to the same conclusion: the LSTM-PINN error stays concentrated near the layer, whereas the error of Standard PINN expands into a much broader footprint over both the upper and lower subregions.

Quantitatively, LSTM-PINN yields the smallest overall RMSE of 0.0047, slightly better than the 0.0050 achieved by ResAtt-PINN and much better than the 0.0231 of Standard PINN. The corresponding overall MSE, MAE, and \(L_2\) errors follow the same trend, indicating that the advantage of LSTM-PINN remains consistent across different overall error measures. The corresponding training and validation loss curves are shown in Fig.~\ref{fig:case02_loss_compare}, where LSTM-PINN maintains the lowest terminal loss among the three models.

\begin{center}
\captionsetup{hypcap=false}
\captionof{table}{Quantitative comparison for Case 02.}
\label{tab:case02_metrics}
\small
\setlength{\tabcolsep}{5pt}
\renewcommand{\arraystretch}{1.12}
\resizebox{\textwidth}{!}{%
\begin{tabular}{lcccccc}
\toprule
Model & \ohdr{overall RMSE}{RMSE} & \ohdr{overall MSE}{MSE} & \ohdr{overall MAE}{MAE} & \ohdr{overall \(L_2\)}{\(L_2\)} & Training time (s) & Peak GPU memory (GB) \\
\midrule
Standard PINN & 0.0231 & 5.36e-04 & 0.0165 & 0.0289 & \textbf{713.4} & 3.455 \\
ResAtt-PINN & 0.0050 & 2.53e-05 & 0.0036 & 0.0063 & 5028.5 & 10.190 \\
LSTM-PINN & \textbf{0.0047} & \textbf{2.18e-05} & \textbf{0.0032} & \textbf{0.0058} & 3355.6 & \textbf{1.699} \\
\bottomrule
\end{tabular}}
\end{center}

\subsection{Case 3: Single Oblique Shock}

Case 03 is a single oblique shock with a smooth background and electrostatic tilt. It can be interpreted as an abstract EHD pattern affected by a biased electric field or an inclined channel boundary. Compared with axis-aligned fronts, the oblique front simultaneously couples gradients in both coordinate directions and therefore imposes a higher requirement on spatial representation capability.

\begin{figure}[H]
    \centering
    \includegraphics[width=0.95\textwidth]{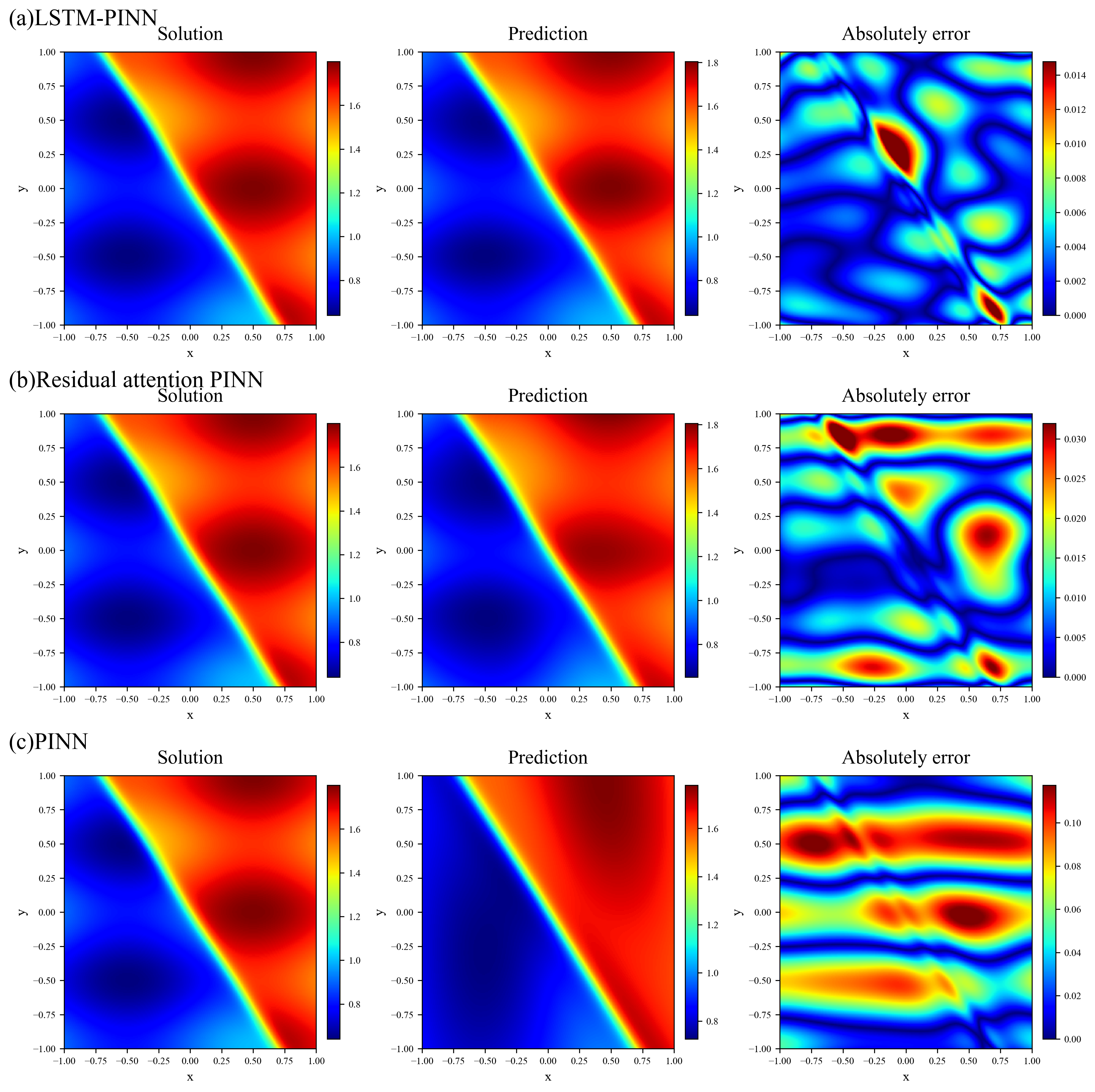}
    \caption{3$\times$3 reconstruction comparison of the \(n\)-field for Case 03, with emphasis on the recovery of the angle and thickness of the oblique shock layer.}
    \label{fig:case03_n_compare}
\end{figure}

\begin{figure}[H]
    \centering
    \includegraphics[width=0.95\textwidth]{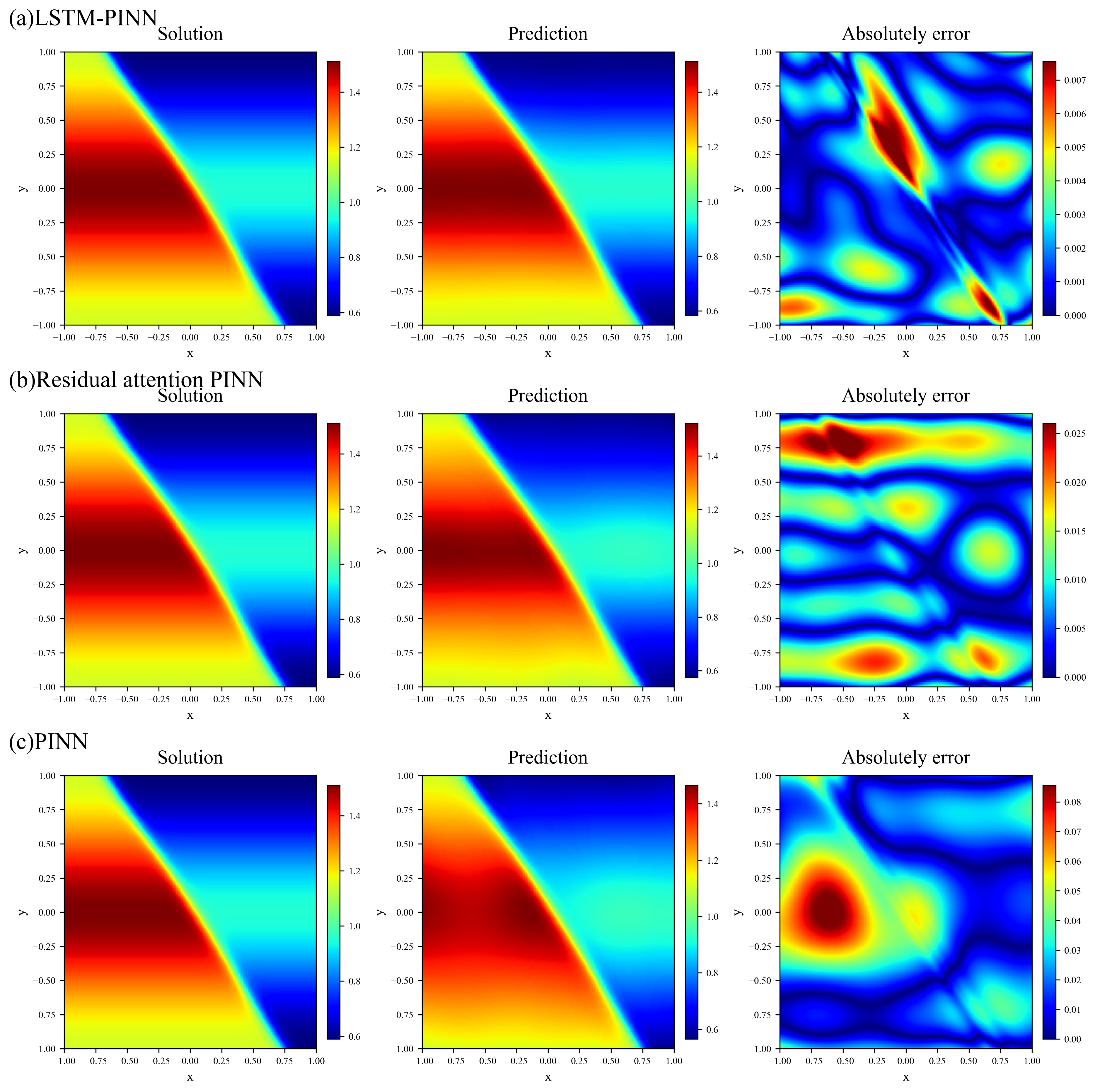}
    \caption{3$\times$3 reconstruction comparison of the \(u_x\)-field for Case 03.}
    \label{fig:case03_ux_panel_3x3}
\end{figure}

\begin{figure}[H]
    \centering
    \includegraphics[width=0.95\textwidth]{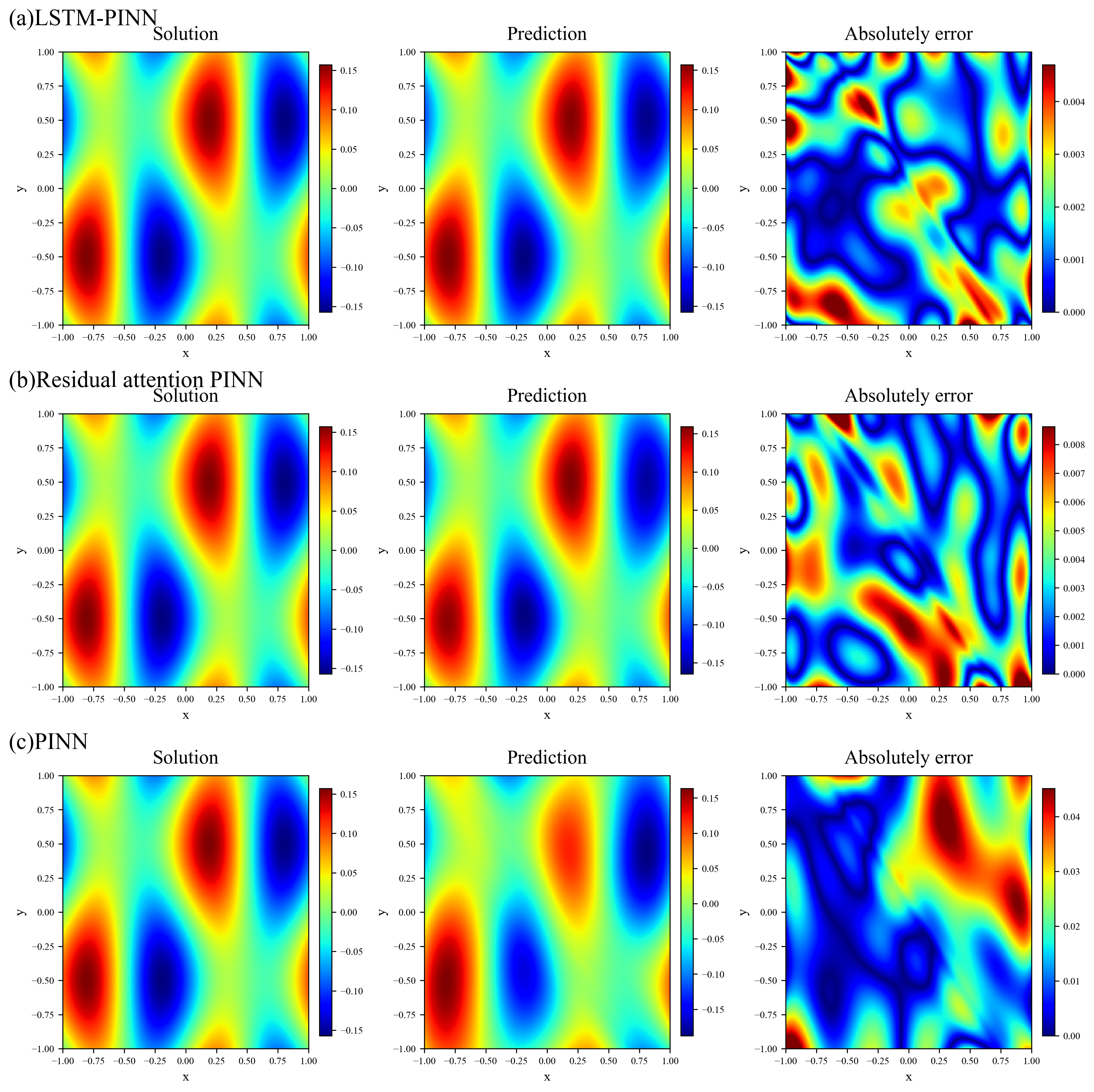}
    \caption{3$\times$3 reconstruction comparison of the \(u_y\)-field for Case 03.}
    \label{fig:case03_uy_panel_3x3}
\end{figure}

\begin{figure}[H]
    \centering
    \includegraphics[width=0.95\textwidth]{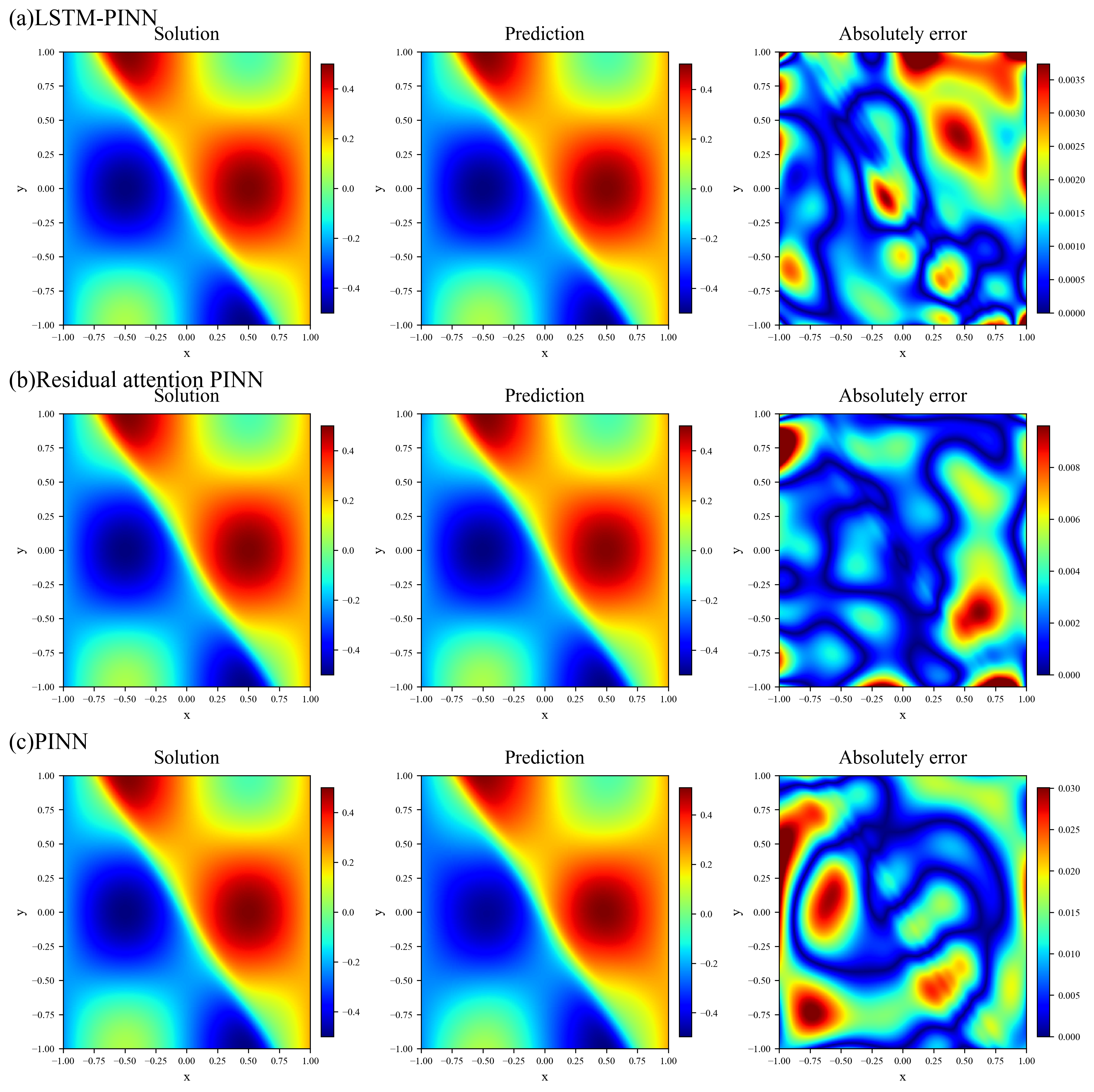}
    \caption{3$\times$3 reconstruction comparison of the \(\phi\)-field for Case 03.}
    \label{fig:case03_phi_panel_3x3}
\end{figure}

The qualitative comparison in Fig.~\ref{fig:case03_n_compare} shows that LSTM-PINN preserves both the angle and continuity of the oblique layer more faithfully, whereas Standard PINN is more prone to layer blurring and local displacement. This case more clearly reveals the advantage of sequential spatial encoding in handling non-axis-aligned structures. The same conclusion is supported by the accompanying \(u_x\), \(u_y\), and \(\phi\) fields in Figs.~\ref{fig:case03_ux_panel_3x3}, \ref{fig:case03_uy_panel_3x3}, and \ref{fig:case03_phi_panel_3x3}. In the \(u_x\) panel, LSTM-PINN keeps the oblique jump narrow and nearly perfectly aligned with the reference interface, whereas Standard PINN bends the transition zone and produces a large off-interface error lobe on the left side of the front. In the \(u_y\) and \(\phi\) panels, the same advantage appears as a noticeably cleaner preservation of the oblique separation line and a much lower level of front-parallel error spreading. ResAtt-PINN remains substantially better than Standard PINN, but LSTM-PINN still gives the most accurate interface angle and the most compact front-adjacent error field across the coupled variables.

\begin{landscape}
\begin{figure}[p]
    \centering
    \includegraphics[width=1\linewidth]{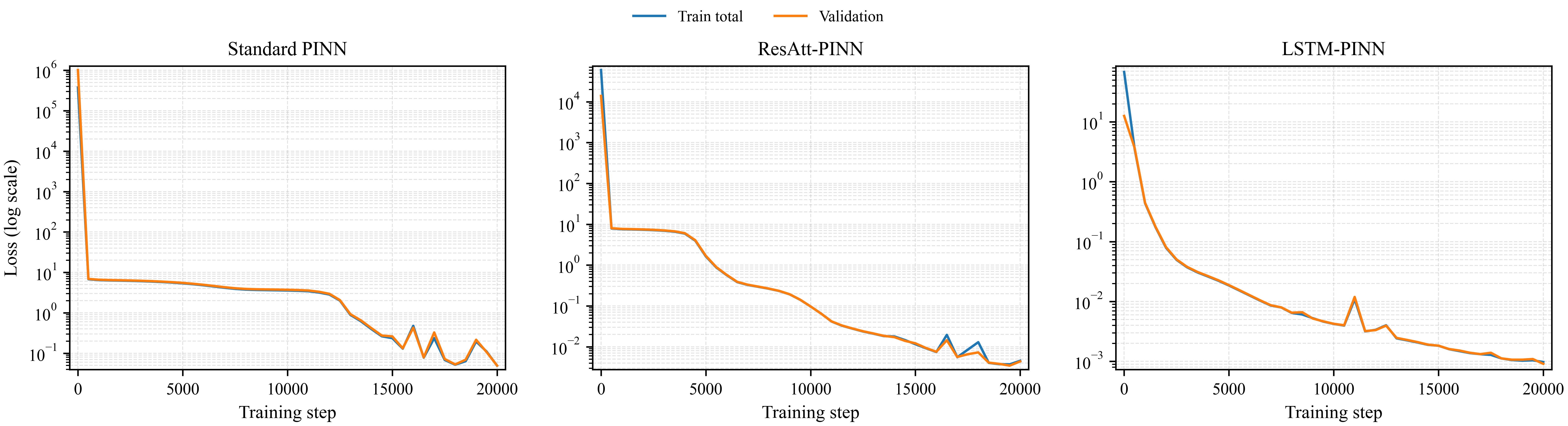}
    \caption{Training and validation loss curves for Case 03.}
    \label{fig:case03_loss_compare}

    \vspace{3em}

    \includegraphics[width=1\linewidth]{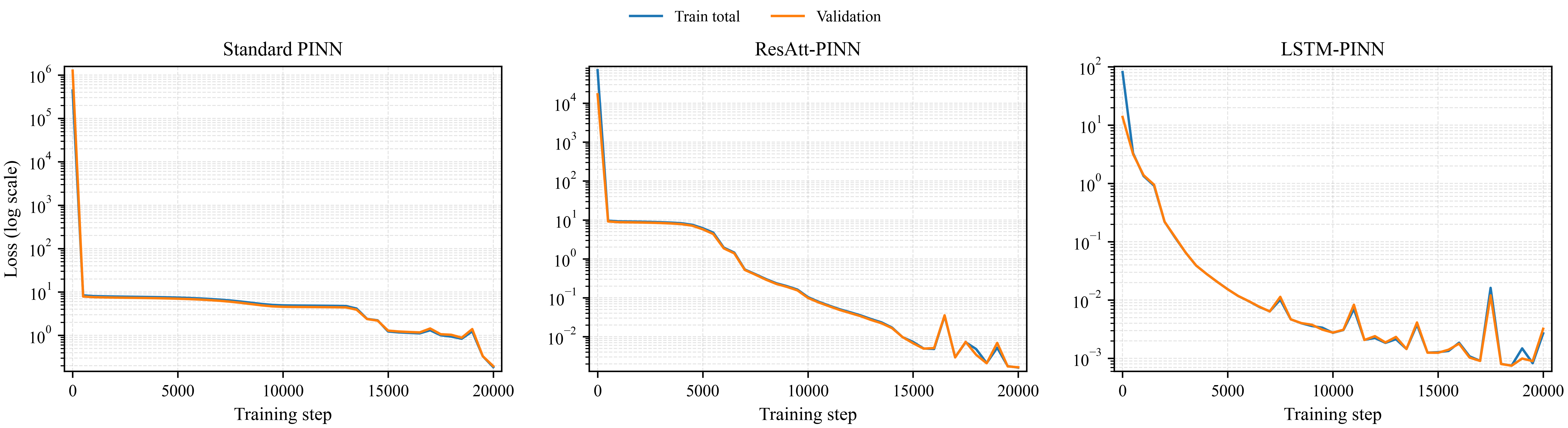}
    \caption{Training and validation loss curves for Case 04.}
    \label{fig:case04_loss_compare}
\end{figure}
\end{landscape}

The quantitative results are even more striking: LSTM-PINN obtains an overall RMSE of 0.0029, substantially lower than 0.0085 for ResAtt-PINN and 0.0361 for Standard PINN. The overall MSE, MAE, and \(L_2\) metrics show the same ranking, which further indicates that LSTM-PINN not only improves the average reconstruction quality but also better controls relatively large local errors along the oblique interface. The corresponding loss behavior is shown in Fig.~\ref{fig:case03_loss_compare}.

\begin{center}
\captionsetup{hypcap=false}
\captionof{table}{Quantitative comparison for Case 03.}
\label{tab:case03_metrics}
\small
\setlength{\tabcolsep}{5pt}
\renewcommand{\arraystretch}{1.12}
\resizebox{\textwidth}{!}{%
\begin{tabular}{lcccccc}
\toprule
Model & \ohdr{overall RMSE}{RMSE} & \ohdr{overall MSE}{MSE} & \ohdr{overall MAE}{MAE} & \ohdr{overall \(L_2\)}{\(L_2\)} & Training time (s) & Peak GPU memory (GB) \\
\midrule
Standard PINN & 0.0361 & 1.30e-03 & 0.0257 & 0.0421 & \textbf{711.7} & 3.455 \\
ResAtt-PINN & 0.0085 & 7.30e-05 & 0.0061 & 0.0100 & 4974.2 & 10.190 \\
LSTM-PINN & \textbf{0.0029} & \textbf{8.16e-06} & \textbf{0.0021} & \textbf{0.0033} & 2933.4 & \textbf{1.699} \\
\bottomrule
\end{tabular}}
\end{center}

\subsection{Case 4: Crossing Oblique Layers}

Case 04 contains two interacting oblique layers, one compressive and one expansive. Their interaction creates several local high-gradient regions near the crossing center, making this benchmark more challenging than the single-front cases. Under the unified EHD interpretation, this case represents the interaction of multiple transition structures within the same device-scale flow pattern.

\begin{figure}[H]
    \centering
    \includegraphics[width=0.95\textwidth]{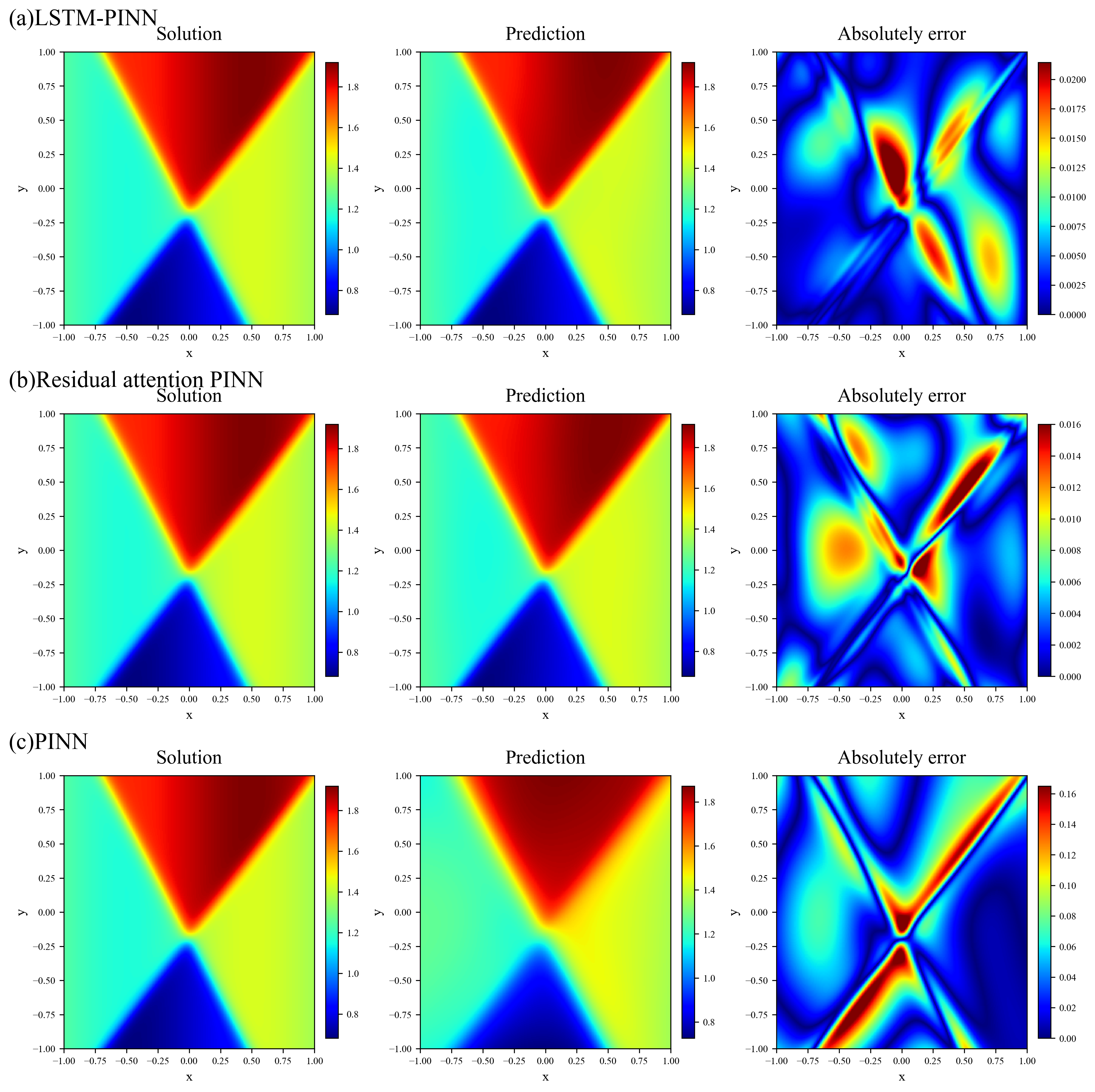}
    \caption{3$\times$3 reconstruction comparison of the \(n\)-field for Case 04, focusing on the interaction region of the crossing oblique layers.}
    \label{fig:case04_n_compare}
\end{figure}

\begin{figure}[H]
    \centering
    \includegraphics[width=0.95\textwidth]{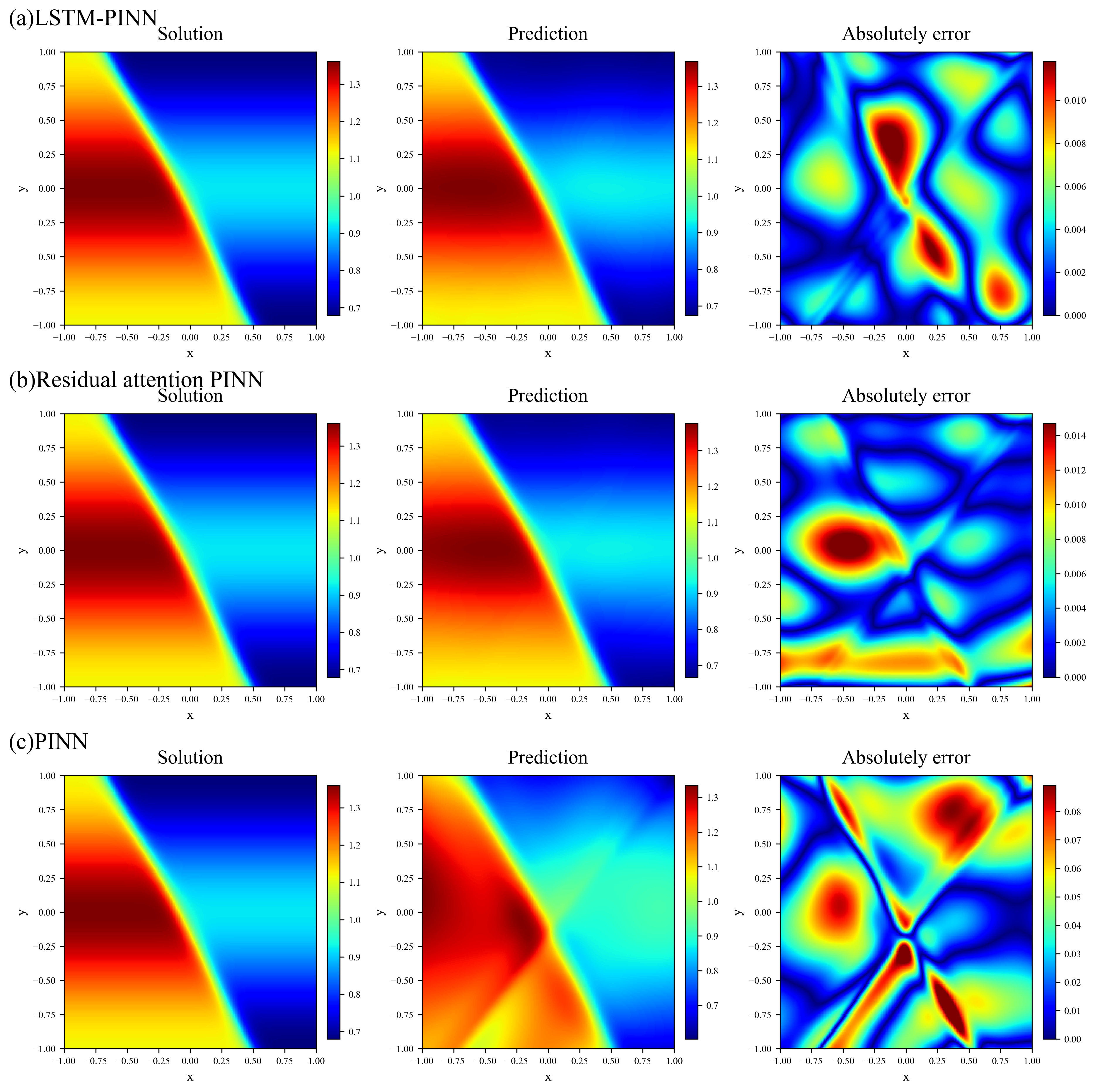}
    \caption{3$\times$3 reconstruction comparison of the \(u_x\)-field for Case 04.}
    \label{fig:case04_ux_panel_3x3}
\end{figure}

\begin{figure}[H]
    \centering
    \includegraphics[width=0.95\textwidth]{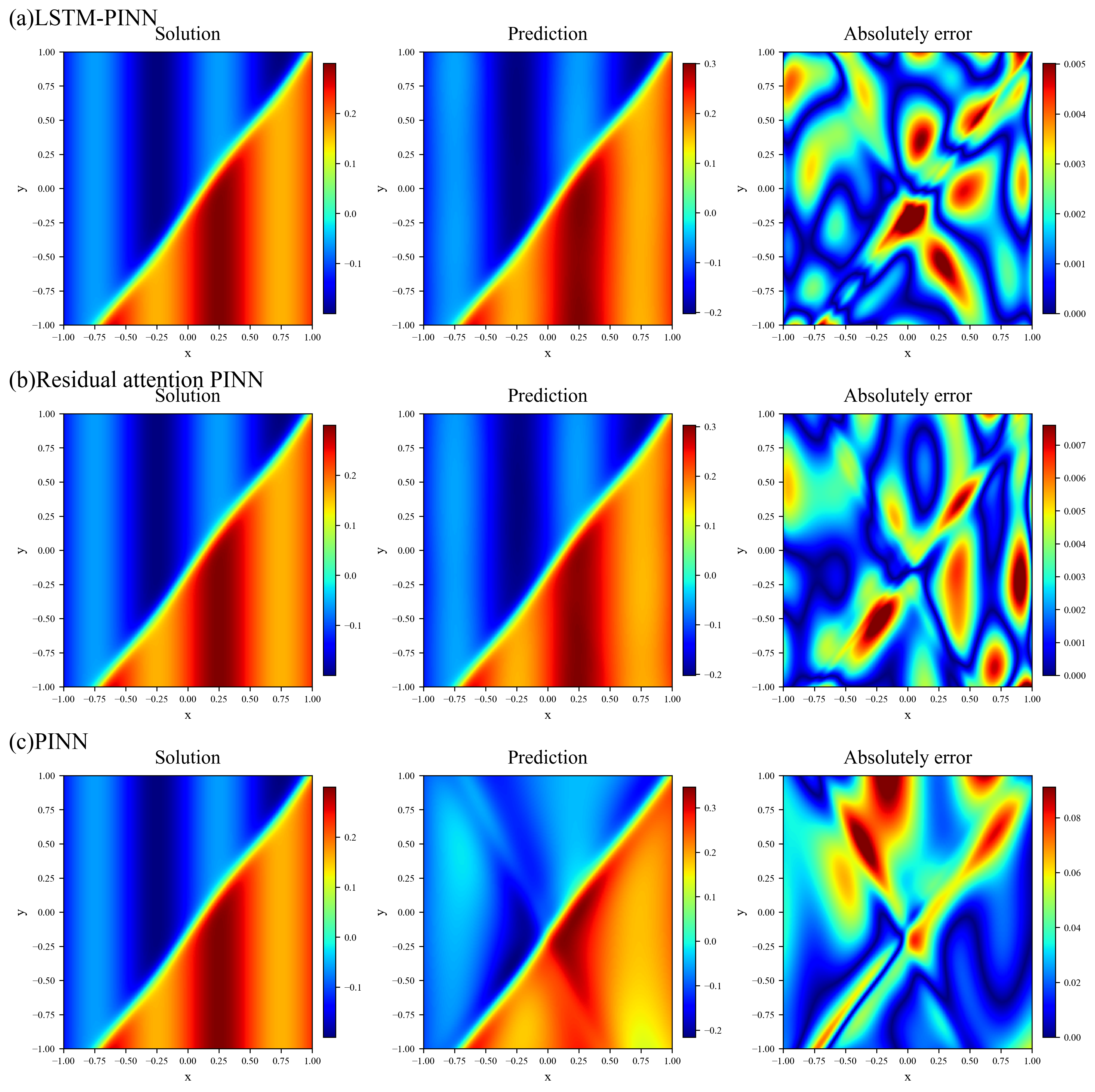}
    \caption{3$\times$3 reconstruction comparison of the \(u_y\)-field for Case 04.}
    \label{fig:case04_uy_panel_3x3}
\end{figure}

\begin{figure}[H]
    \centering
    \includegraphics[width=0.95\textwidth]{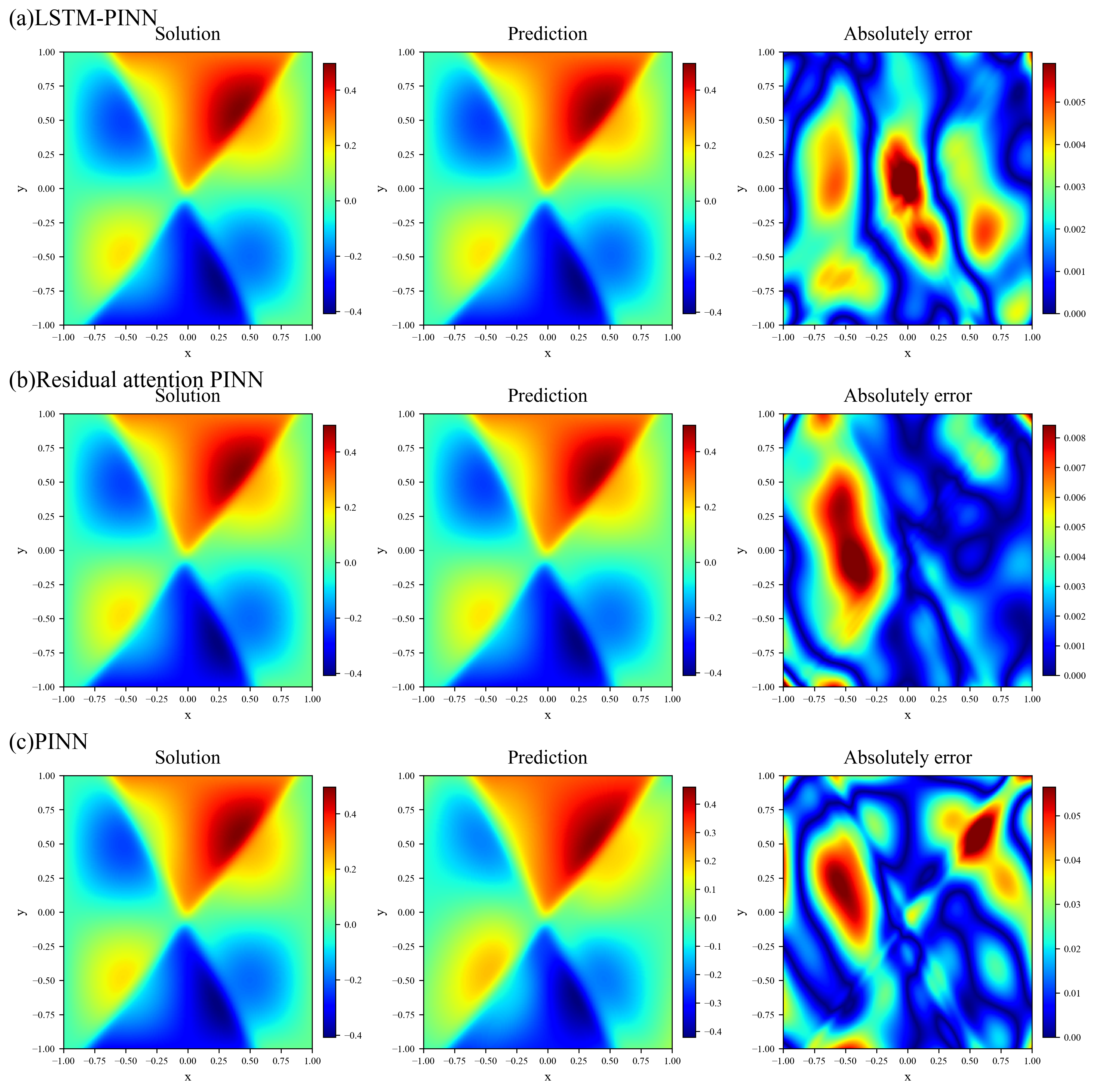}
    \caption{3$\times$3 reconstruction comparison of the \(\phi\)-field for Case 04.}
    \label{fig:case04_phi_panel_3x3}
\end{figure}

As shown in Fig.~\ref{fig:case04_n_compare}, both LSTM-PINN and ResAtt-PINN can preserve the main structures, but the error of LSTM-PINN is more concentrated and lower around the intersection center. By contrast, Standard PINN suffers from a much larger distortion in the crossing region. This case further demonstrates that LSTM-PINN is better at maintaining global structural consistency when multiple fronts interact. The added \(u_x\), \(u_y\), and \(\phi\) comparisons in Figs.~\ref{fig:case04_ux_panel_3x3}, \ref{fig:case04_uy_panel_3x3}, and \ref{fig:case04_phi_panel_3x3} reinforce this observation. In the \(u_x\) and \(u_y\) panels, LSTM-PINN resolves the intersecting branches with the clearest separation at the crossing center and avoids the spurious broadening that is visible in Standard PINN, especially along the outgoing diagonal arms. The \(\phi\) panel is particularly informative: the LSTM-PINN prediction keeps the central four-sector structure and its narrow junction region far more cleanly than the other two models, whereas Standard PINN produces the strongest diffuse error halo and the most obvious loss of branch sharpness near the interaction core. ResAtt-PINN stays closer to LSTM-PINN, but its error support is still slightly wider around the crossing center.

The quantitative comparison confirms this observation. LSTM-PINN achieves the smallest overall RMSE of 0.0041, slightly lower than 0.0044 for ResAtt-PINN and far lower than 0.0413 for Standard PINN. The same ordering is also reflected by the overall MSE, MAE, and \(L_2\) values, showing that its advantage in the crossing region is not limited to a single error metric. The corresponding optimization curves are shown in Fig.~\ref{fig:case04_loss_compare}.

\begin{center}
\captionsetup{hypcap=false}
\captionof{table}{Quantitative comparison for Case 04.}
\label{tab:case04_metrics}
\small
\setlength{\tabcolsep}{5pt}
\renewcommand{\arraystretch}{1.12}
\resizebox{\textwidth}{!}{%
\begin{tabular}{lcccccc}
\toprule
Model & \ohdr{overall RMSE}{RMSE} & \ohdr{overall MSE}{MSE} & \ohdr{overall MAE}{MAE} & \ohdr{overall \(L_2\)}{\(L_2\)} & Training time (s) & Peak GPU memory (GB) \\
\midrule
Standard PINN & 0.0413 & 1.71e-03 & 0.0317 & 0.0472 & \textbf{740.1} & 3.455 \\
ResAtt-PINN & 0.0044 & 1.93e-05 & 0.0033 & 0.0050 & 4996.7 & 10.190 \\
LSTM-PINN & \textbf{0.0041} & \textbf{1.70e-05} & \textbf{0.0028} & \textbf{0.0047} & 2917.4 & \textbf{1.699} \\
\bottomrule
\end{tabular}}
\end{center}

\subsection{Case 5: Curved Radial Front}

Case 05 features a curved radial front with a weak rotating background flow. It can be interpreted as a ring-like transition layer generated near a local electrode or field-enhancement region. Compared with straight fronts, a curved interface is more difficult to reconstruct because errors may accumulate along the arc-shaped boundary.

\begin{figure}[H]
    \centering
    \includegraphics[width=0.95\textwidth]{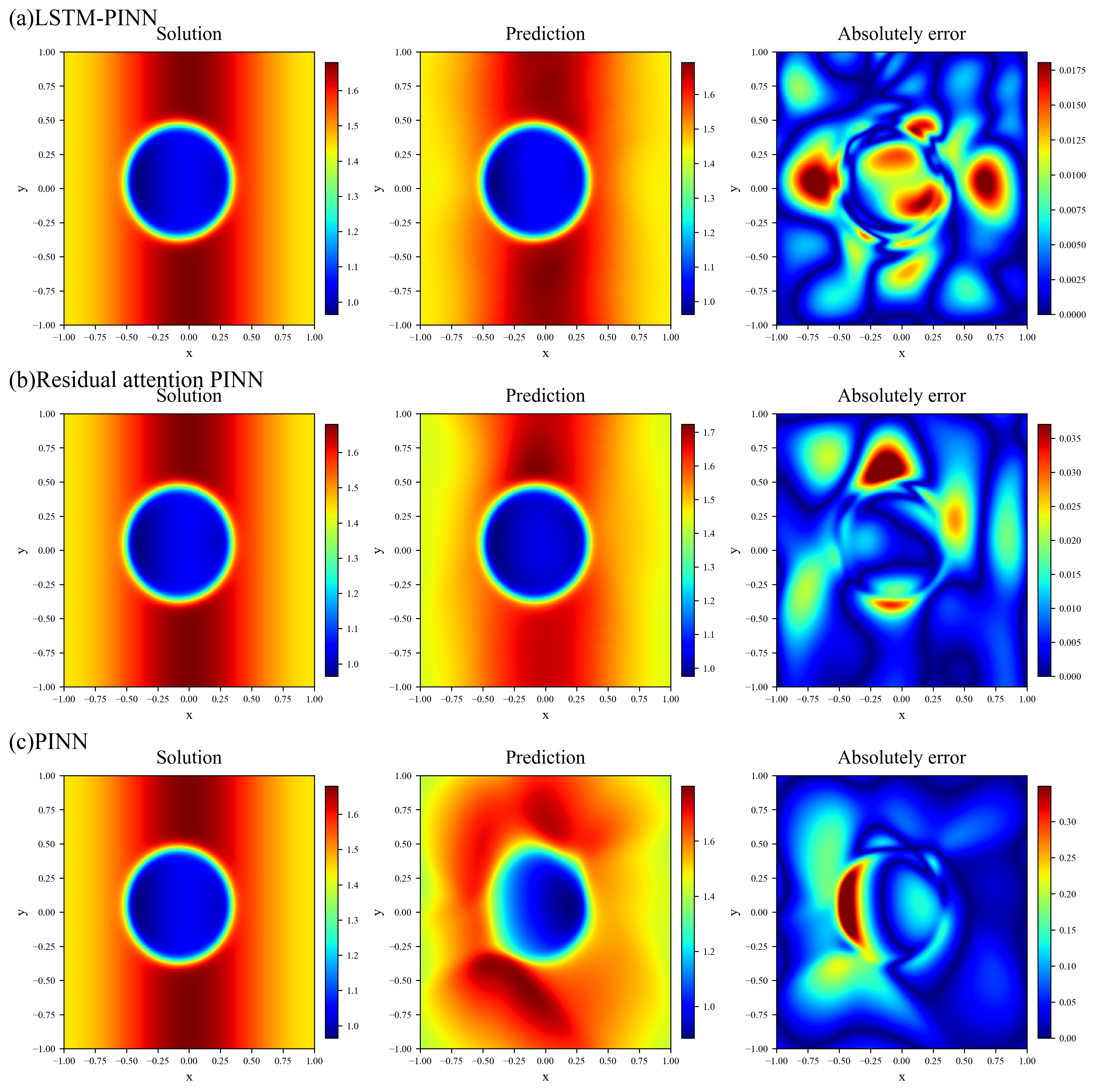}
    \caption{3$\times$3 reconstruction comparison of the \(n\)-field for Case 05, highlighting the reconstruction of the curved radial front.}
    \label{fig:case05_n_compare}
\end{figure}

\begin{figure}[H]
    \centering
    \includegraphics[width=0.95\textwidth]{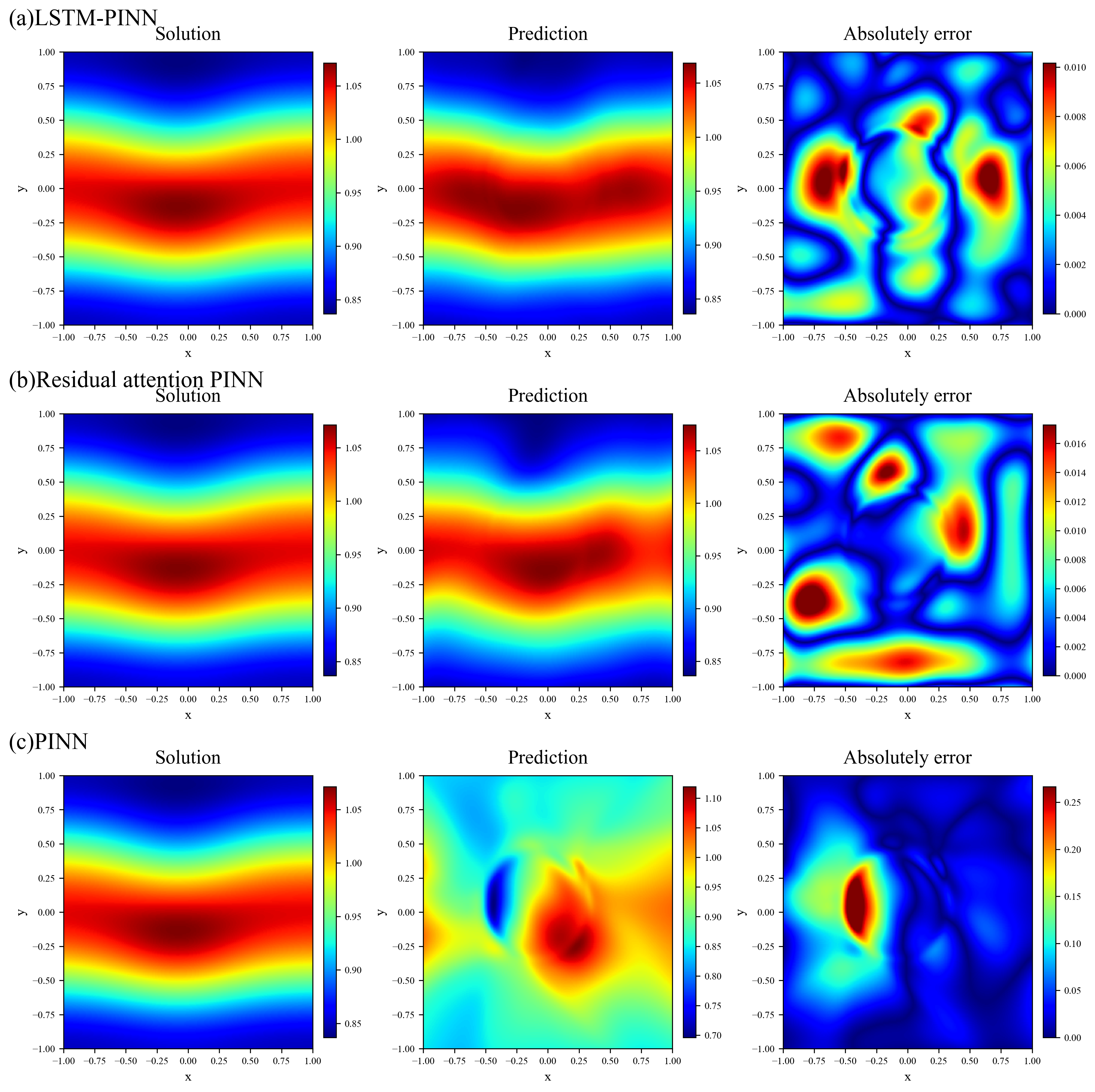}
    \caption{3$\times$3 reconstruction comparison of the \(u_x\)-field for Case 05.}
    \label{fig:case05_ux_panel_3x3}
\end{figure}

\begin{figure}[H]
    \centering
    \includegraphics[width=0.95\textwidth]{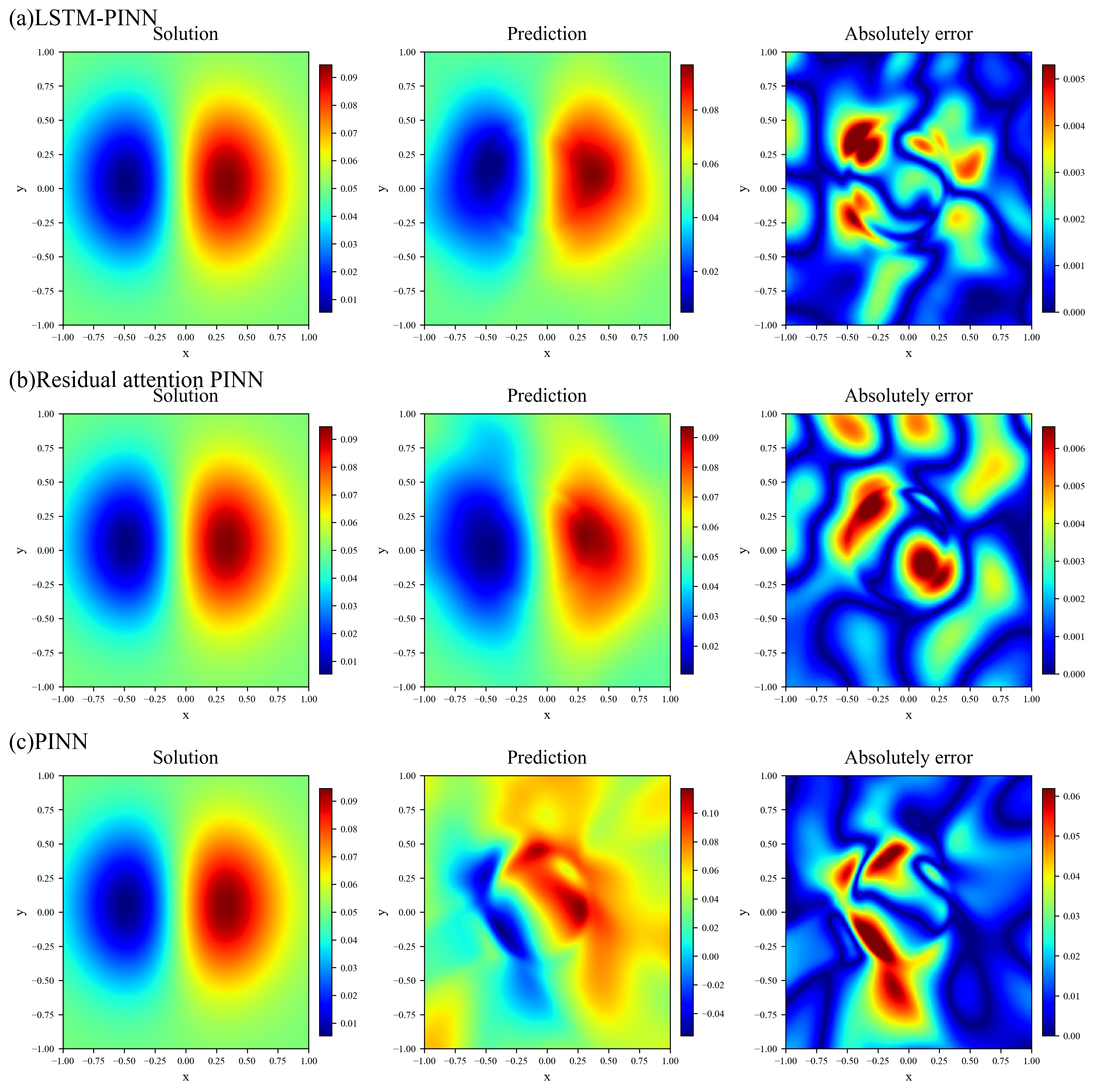}
    \caption{3$\times$3 reconstruction comparison of the \(u_y\)-field for Case 05.}
    \label{fig:case05_uy_panel_3x3}
\end{figure}

\begin{figure}[H]
    \centering
    \includegraphics[width=0.95\textwidth]{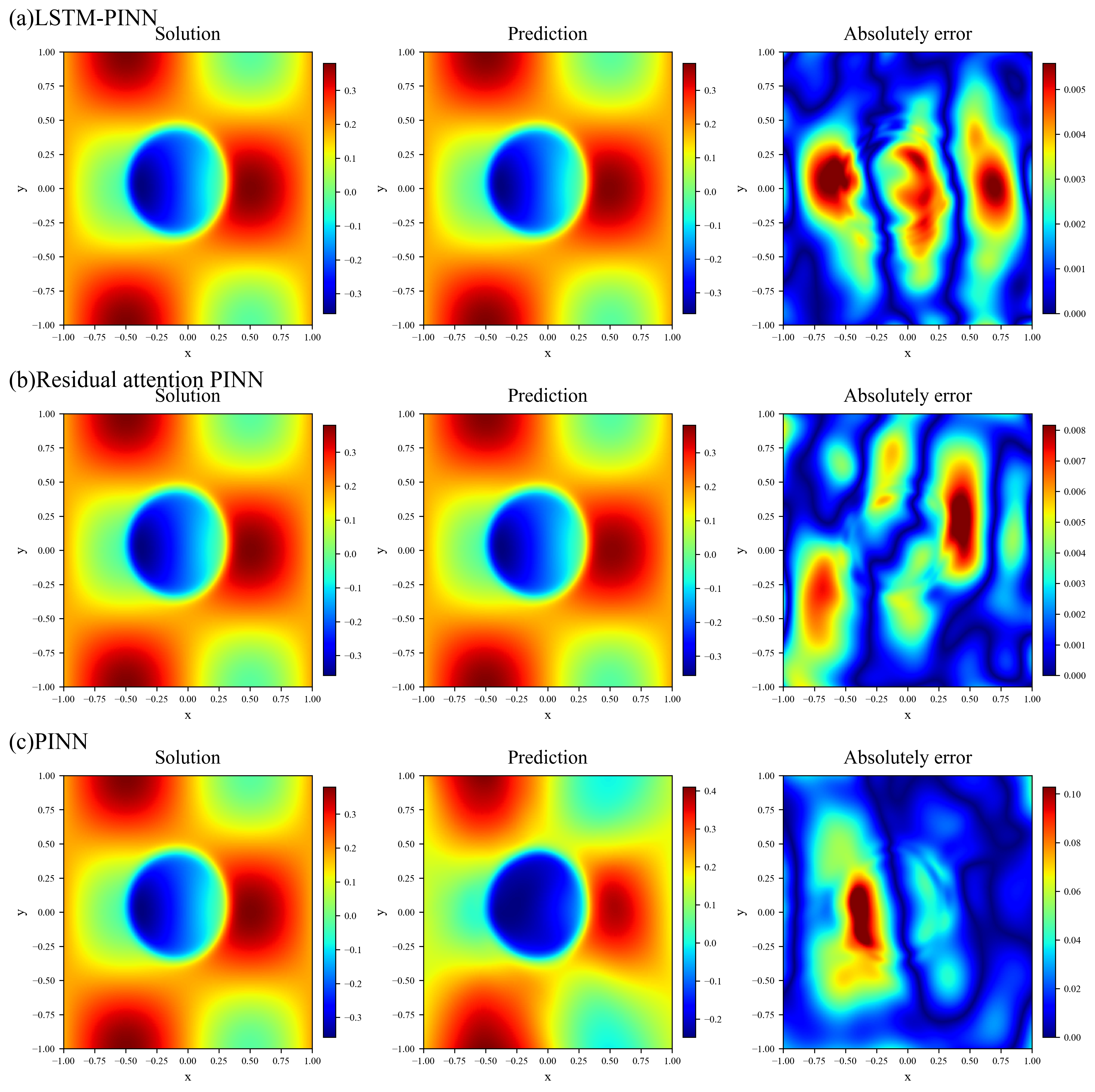}
    \caption{3$\times$3 reconstruction comparison of the \(\phi\)-field for Case 05.}
    \label{fig:case05_phi_panel_3x3}
\end{figure}

The visual comparison in Fig.~\ref{fig:case05_n_compare} shows that LSTM-PINN preserves the curved contour of the radial front more accurately, whereas ResAtt-PINN exhibits a slight amplitude bias and Standard PINN tends to flatten or overly smooth the curved interface. This indicates that LSTM-PINN can better retain geometric fidelity for nonplanar structures. The same geometric advantage is consistently reflected in the associated \(u_x\), \(u_y\), and \(\phi\) fields shown in Figs.~\ref{fig:case05_ux_panel_3x3}, \ref{fig:case05_uy_panel_3x3}, and \ref{fig:case05_phi_panel_3x3}. In the \(u_x\) panel, LSTM-PINN maintains the horizontal-band background while keeping the circular disturbance sharply closed, whereas Standard PINN breaks this structure into a deformed central blob with substantial leakage into the surrounding field. In the dipole-like \(u_y\) field, LSTM-PINN better preserves the left-right symmetry around the circular interface, while Standard PINN introduces a strong asymmetric distortion and shifts the high-gradient region away from the true ring. The \(\phi\) panel leads to the same conclusion: LSTM-PINN follows the circular boundary with the most uniform ring thickness and the most compact error support, whereas the other two models, especially Standard PINN, show more obvious radial flattening and off-boundary contamination.

\begin{landscape}
\begin{figure}[p]
    \centering
    \includegraphics[width=1\linewidth]{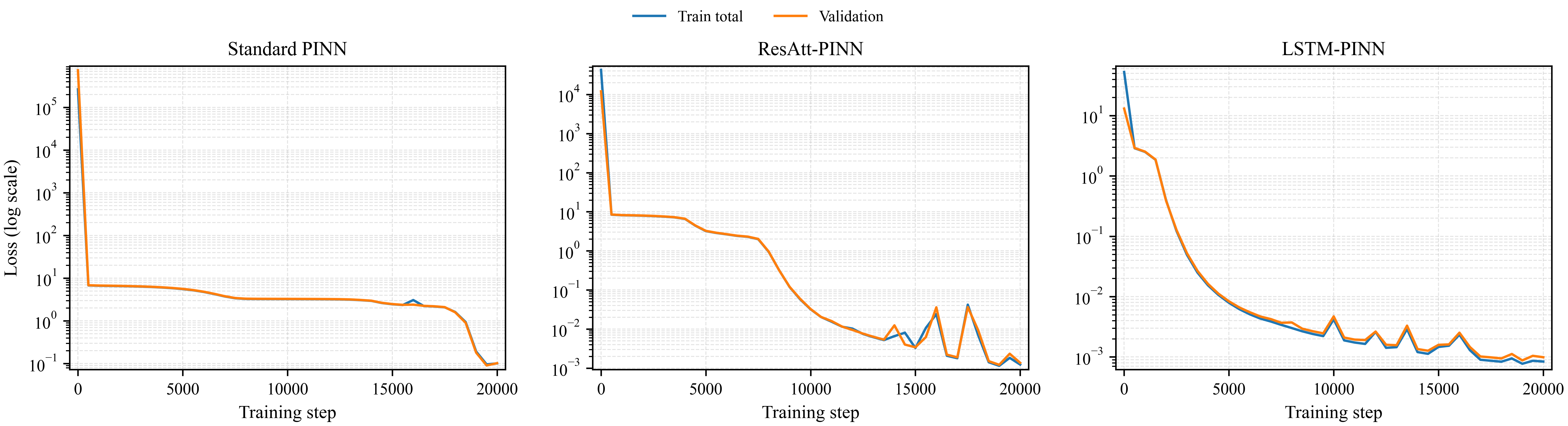}
    \caption{Training and validation loss curves for Case 05.}
    \label{fig:case05_loss_compare}

    \vspace{3em}

    \includegraphics[width=1\linewidth]{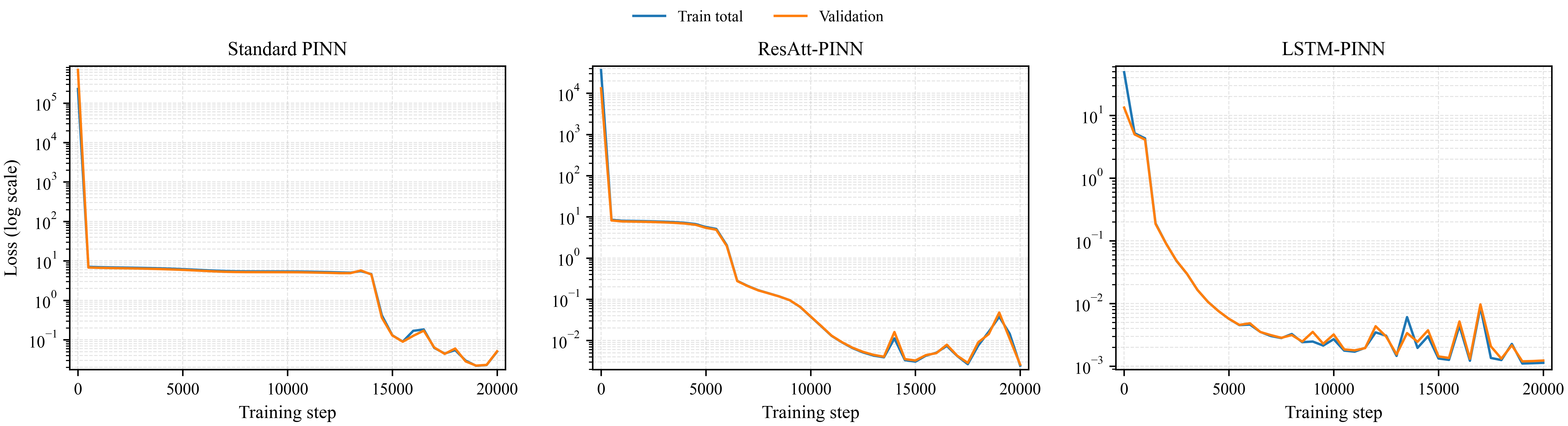}
    \caption{Training and validation loss curves for Case 06.}
    \label{fig:case06_loss_compare}
\end{figure}
\end{landscape}

The numerical metrics again support this conclusion. LSTM-PINN reaches an overall RMSE of 0.0037, outperforming ResAtt-PINN at 0.0062 and Standard PINN at 0.0534. The corresponding overall MSE, MAE, and \(L_2\) metrics are also the smallest for LSTM-PINN, which is consistent with its better preservation of the curved front geometry. The corresponding loss curves are shown in Fig.~\ref{fig:case05_loss_compare}.

\begin{center}
\captionsetup{hypcap=false}
\captionof{table}{Quantitative comparison for Case 05.}
\label{tab:case05_metrics}
\small
\setlength{\tabcolsep}{5pt}
\renewcommand{\arraystretch}{1.12}
\resizebox{\textwidth}{!}{%
\begin{tabular}{lcccccc}
\toprule
Model & \ohdr{overall RMSE}{RMSE} & \ohdr{overall MSE}{MSE} & \ohdr{overall MAE}{MAE} & \ohdr{overall \(L_2\)}{\(L_2\)} & Training time (s) & Peak GPU memory (GB) \\
\midrule
Standard PINN & 0.0534 & 2.85e-03 & 0.0316 & 0.0603 & \textbf{662.8} & 3.455 \\
ResAtt-PINN & 0.0062 & 3.81e-05 & 0.0040 & 0.0070 & 6082.3 & 10.190 \\
LSTM-PINN & \textbf{0.0037} & \textbf{1.40e-05} & \textbf{0.0024} & \textbf{0.0042} & 3042.0 & \textbf{1.699} \\
\bottomrule
\end{tabular}}
\end{center}

\subsection{Case 6: Electro-Shear Layer with Localized Density Pocket}

Case 06 corresponds to an electro-shear layer near the wall with an additional localized density pocket. This benchmark can be used to describe locally enhanced transport in the near-wall region of a micropump. Its characteristic difficulty lies in the superposition of a dominant shear layer and a local anomalous pocket.

\begin{figure}[H]
    \centering
    \includegraphics[width=0.95\textwidth]{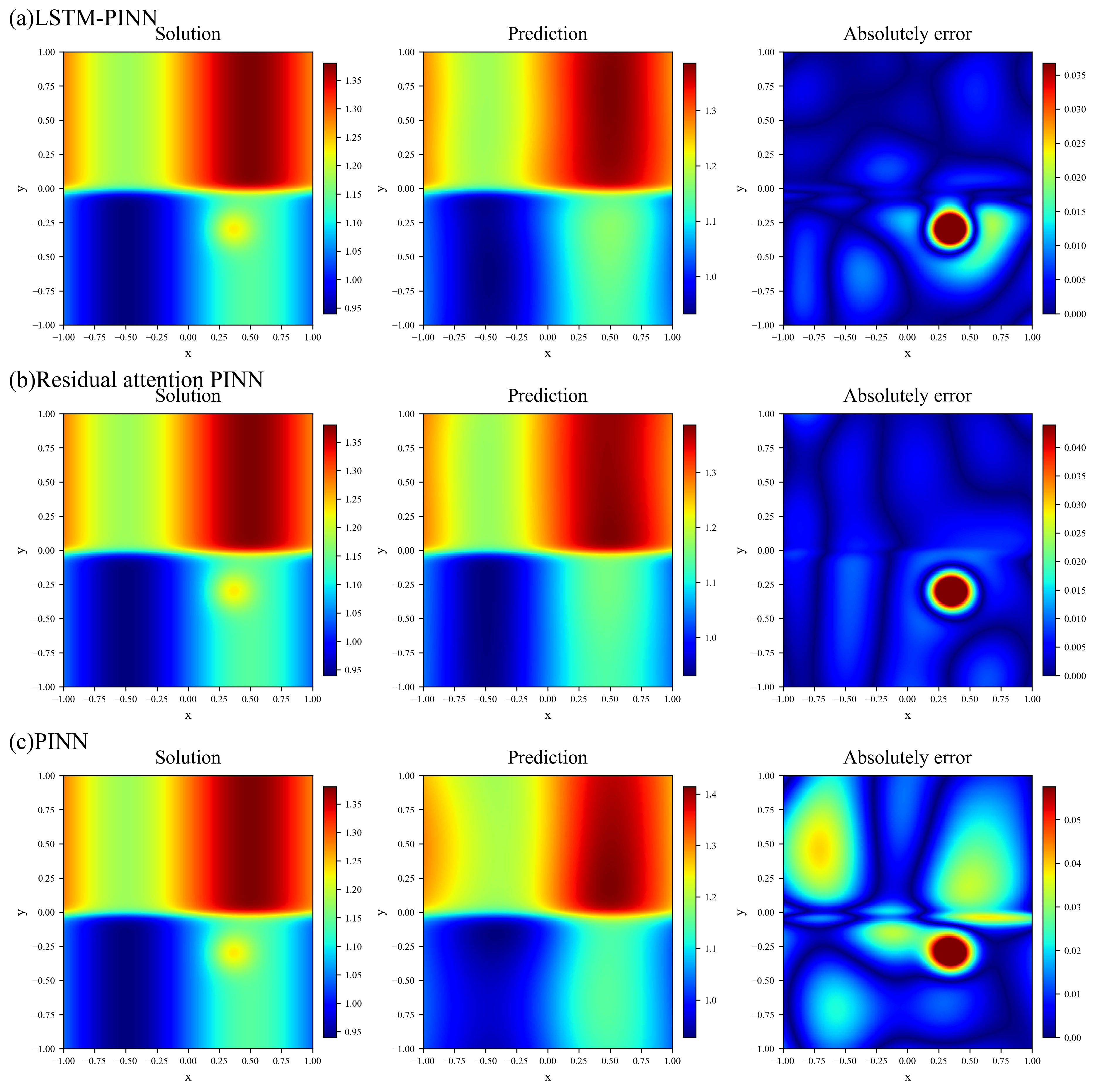}
    \caption{3$\times$3 reconstruction comparison of the \(n\)-field for Case 06.}
    \label{fig:case06_n_panel_3x3}
\end{figure}

\begin{figure}[H]
    \centering
    \includegraphics[width=0.95\textwidth]{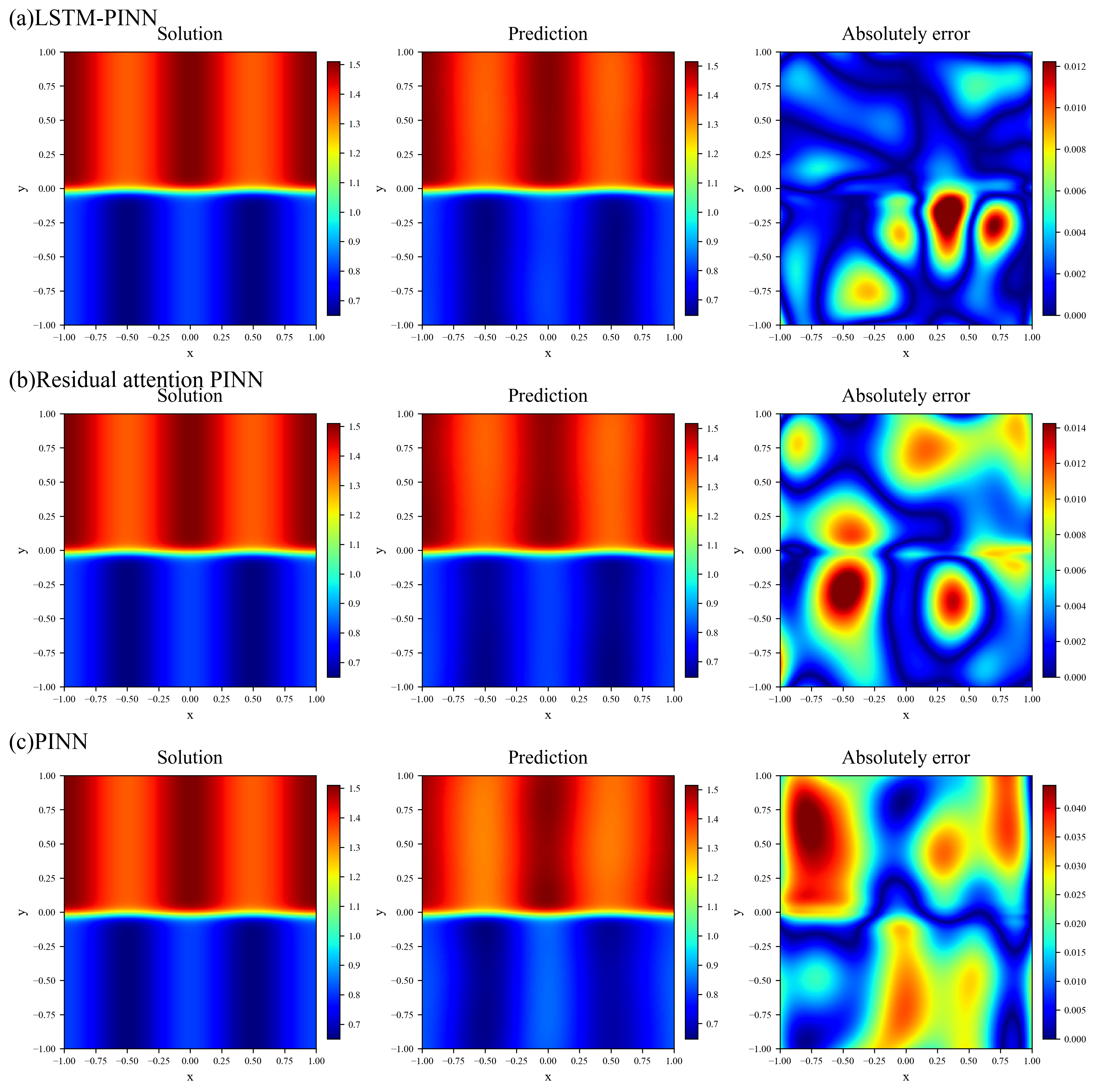}
    \caption{3$\times$3 reconstruction comparison for Case 06, focusing on the velocity-related panel to highlight the recovery of the shear layer coupled with the localized density pocket.}
    \label{fig:case06_ux_compare}
\end{figure}

\begin{figure}[H]
    \centering
    \includegraphics[width=0.95\textwidth]{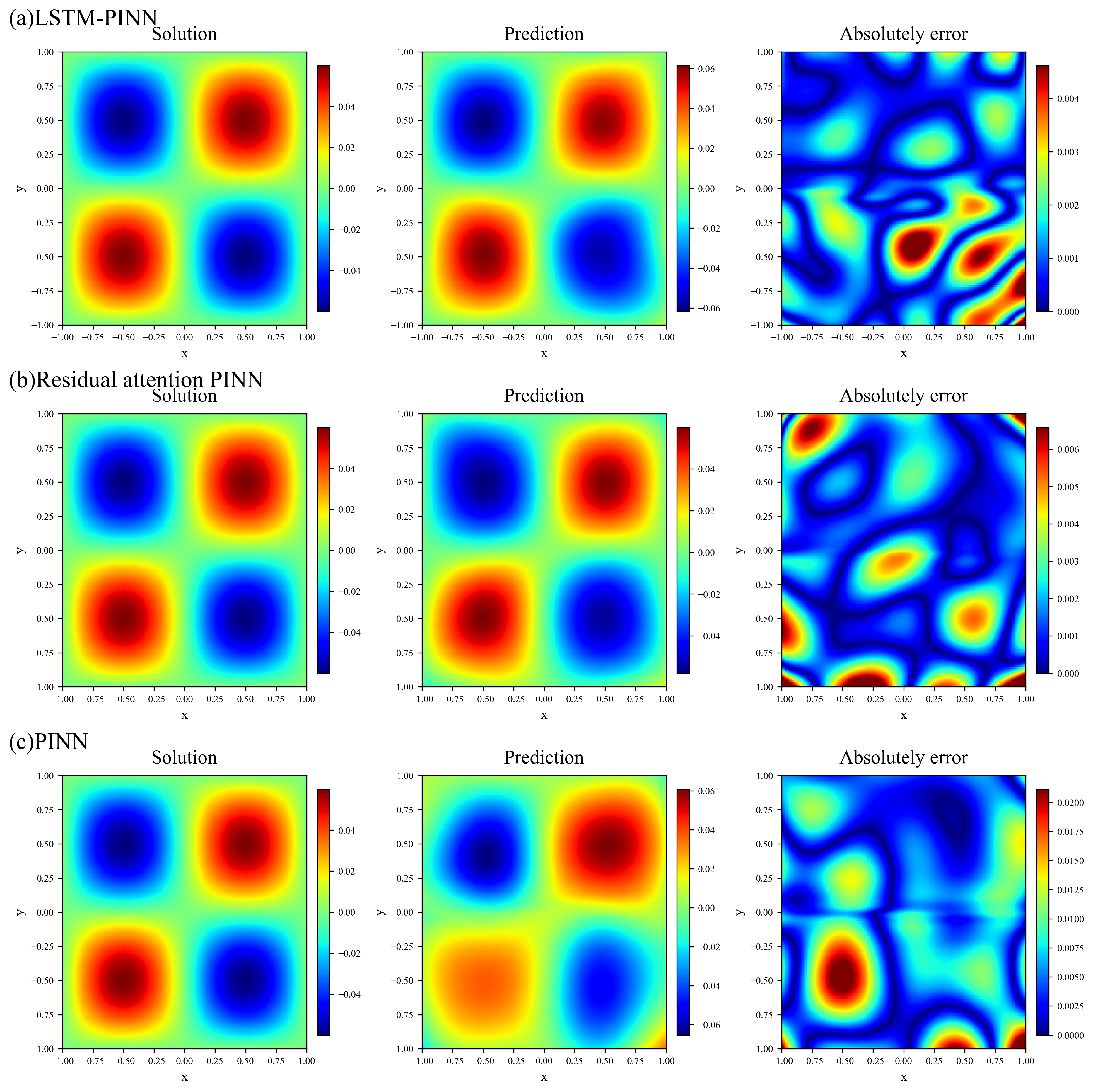}
    \caption{3$\times$3 reconstruction comparison of the \(u_y\)-field for Case 06.}
    \label{fig:case06_uy_panel_3x3}
\end{figure}

\begin{figure}[H]
    \centering
    \includegraphics[width=0.95\textwidth]{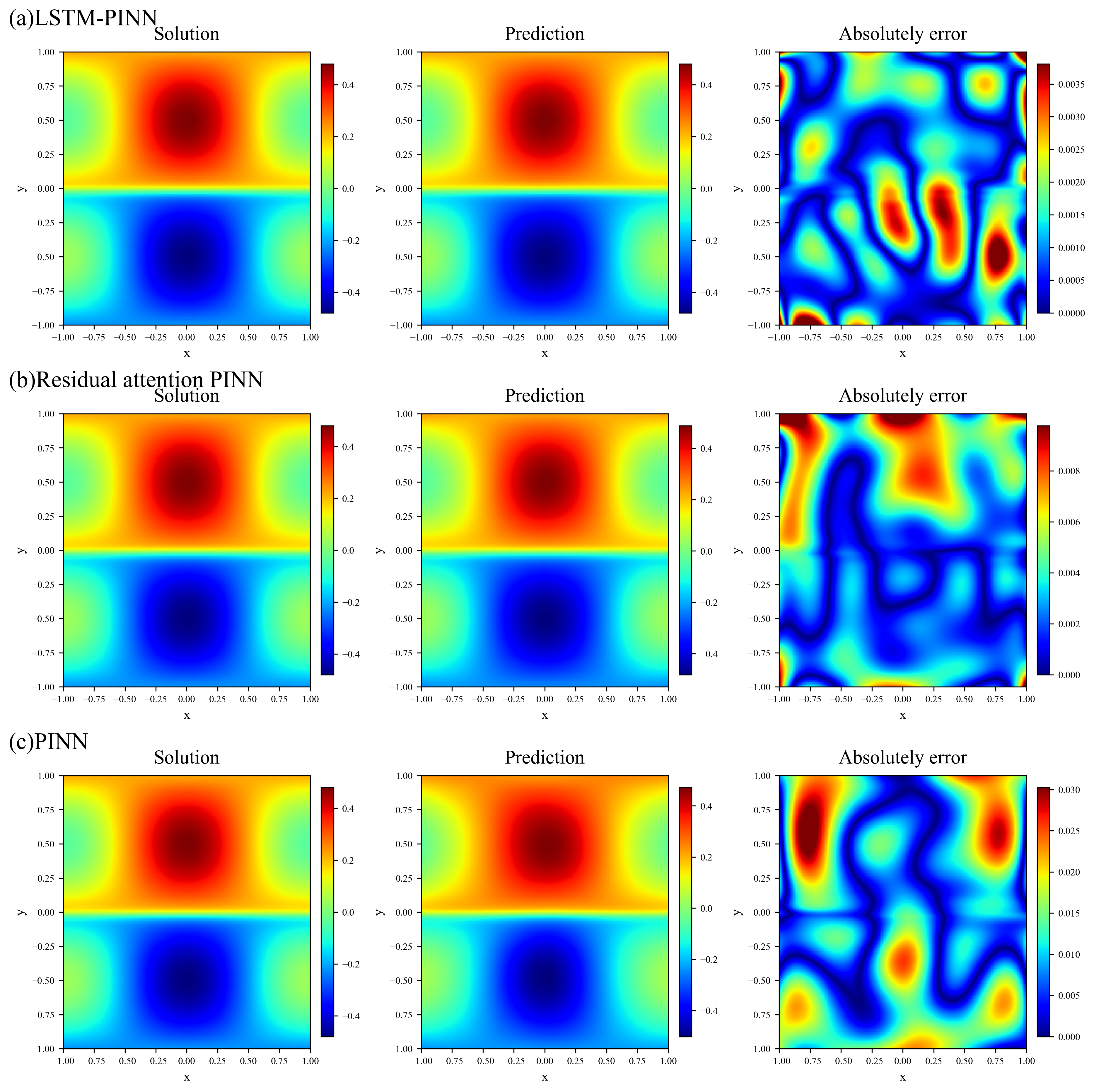}
    \caption{3$\times$3 reconstruction comparison of the \(\phi\)-field for Case 06.}
    \label{fig:case06_phi_panel_3x3}
\end{figure}

The reconstruction comparison in Fig.~\ref{fig:case06_ux_compare} focuses on the velocity-related panel because it more directly reflects the dominant role of the shear layer. LSTM-PINN maintains good continuity in both the layer region and the local abnormal region, whereas Standard PINN exhibits more scattered errors around the shear zone. The newly added \(n\), \(u_y\), and \(\phi\) panels in Figs.~\ref{fig:case06_n_panel_3x3}, \ref{fig:case06_uy_panel_3x3}, and \ref{fig:case06_phi_panel_3x3} show that this advantage is not confined to the dominant \(u_x\) component. In the density panel, LSTM-PINN keeps the small localized pocket tightly concentrated near the lower-right region while preserving a clean horizontal layer, whereas Standard PINN stretches both the pocket and the interfacial error over a much larger area. In the checkerboard-like \(u_y\) field, LSTM-PINN better maintains the background symmetry and confines the disturbance to a compact neighborhood of the pocket--layer interaction zone. The \(\phi\) panel shows the same pattern: the LSTM-PINN prediction retains the horizontal separation and the localized right-side anomalies more coherently, while Standard PINN produces a far noisier and more spatially scattered mismatch.

The quantitative results show that LSTM-PINN again provides the smallest overall RMSE, 0.0040, compared with 0.0055 for ResAtt-PINN and 0.0160 for Standard PINN. The same ranking is also observed for the overall MSE, MAE, and \(L_2\), suggesting that LSTM-PINN more effectively controls both the average error and the larger local deviations associated with the coupled shear-layer and pocket structure. The corresponding training history is shown in Fig.~\ref{fig:case06_loss_compare}.

\begin{center}
\captionsetup{hypcap=false}
\captionof{table}{Quantitative comparison for Case 06.}
\label{tab:case06_metrics}
\small
\setlength{\tabcolsep}{5pt}
\renewcommand{\arraystretch}{1.12}
\resizebox{\textwidth}{!}{%
\begin{tabular}{lcccccc}
\toprule
Model & \ohdr{overall RMSE}{RMSE} & \ohdr{overall MSE}{MSE} & \ohdr{overall MAE}{MAE} & \ohdr{overall \(L_2\)}{\(L_2\)} & Training time (s) & Peak GPU memory (GB) \\
\midrule
Standard PINN & 0.0160 & 2.56e-04 & 0.0123 & 0.0193 & \textbf{663.0} & 3.455 \\
ResAtt-PINN & 0.0055 & 3.02e-05 & 0.0037 & 0.0066 & 6032.1 & 10.190 \\
LSTM-PINN & \textbf{0.0040} & \textbf{1.59e-05} & \textbf{0.0021} & \textbf{0.0048} & 3575.3 & \textbf{1.699} \\
\bottomrule
\end{tabular}}
\end{center}

\subsection{Case 7: Double Front with Secondary Transverse Layer}

Case 07 combines a double front in the \(x\)-direction, a secondary transverse layer, and a Gaussian forcing pocket. It can be regarded as a composite EHD pattern under multi-segment driving conditions. This benchmark is more complicated than Cases~01--06 because it contains multiple interacting structures with different spatial scales.

\begin{figure}[H]
    \centering
    \includegraphics[width=0.95\textwidth]{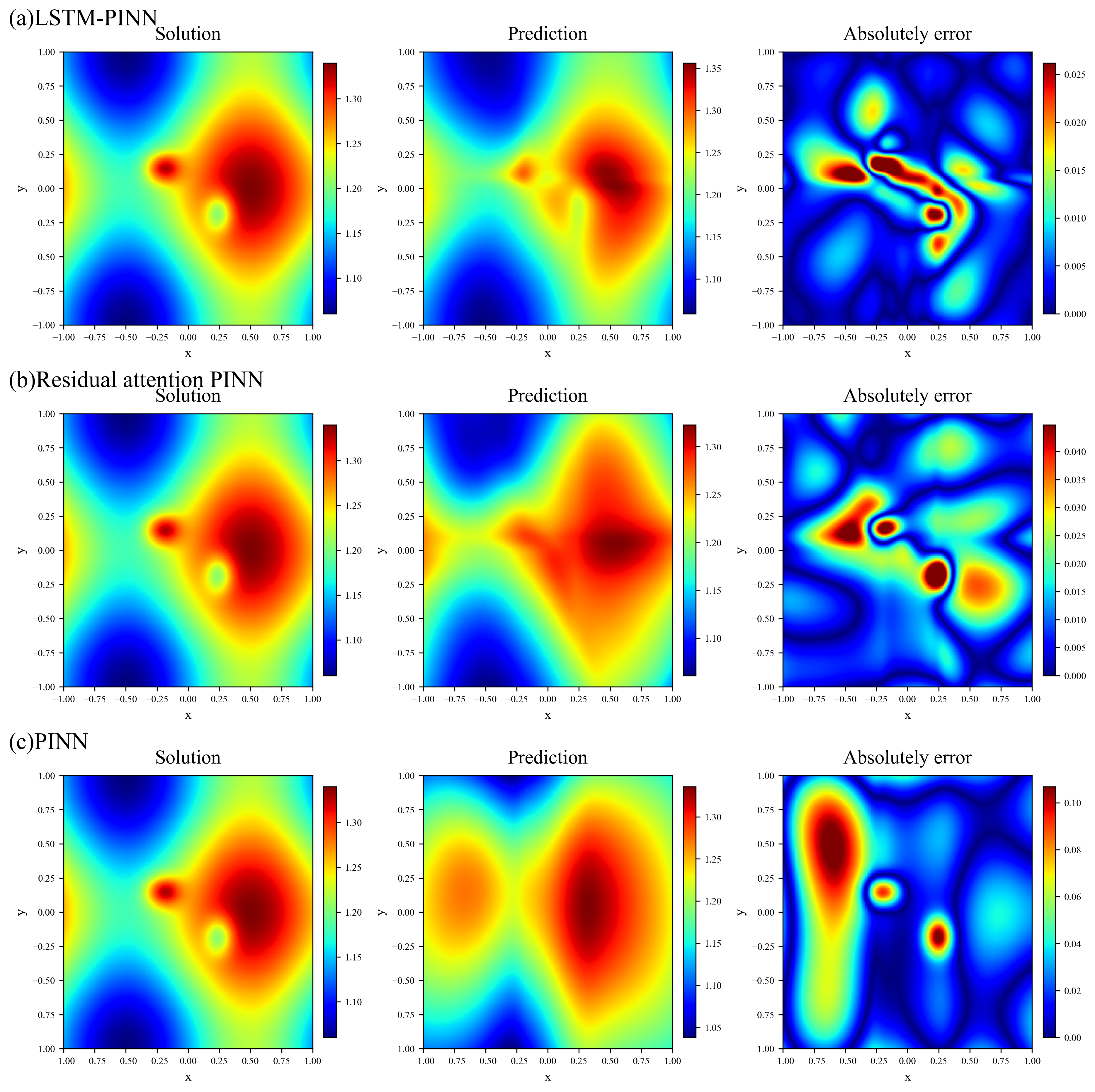}
    \caption{3$\times$3 reconstruction comparison of the \(n\)-field for Case 07, focusing on the preservation of the double front and the local Gaussian pocket.}
    \label{fig:case07_n_compare}
\end{figure}

\begin{figure}[H]
    \centering
    \includegraphics[width=0.95\textwidth]{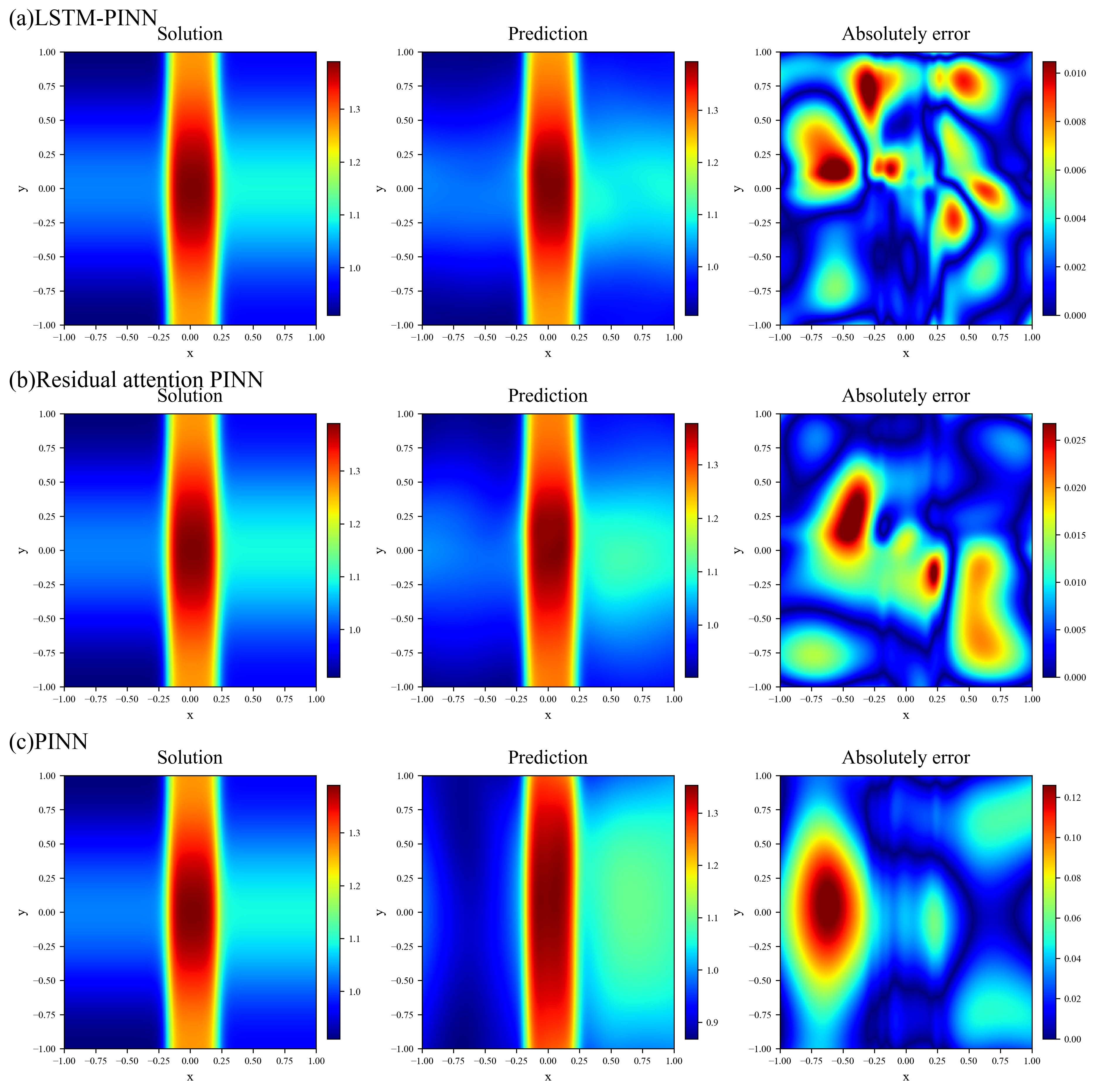}
    \caption{3$\times$3 reconstruction comparison of the \(u_x\)-field for Case 07.}
    \label{fig:case07_ux_panel_3x3}
\end{figure}

\begin{figure}[H]
    \centering
    \includegraphics[width=0.95\textwidth]{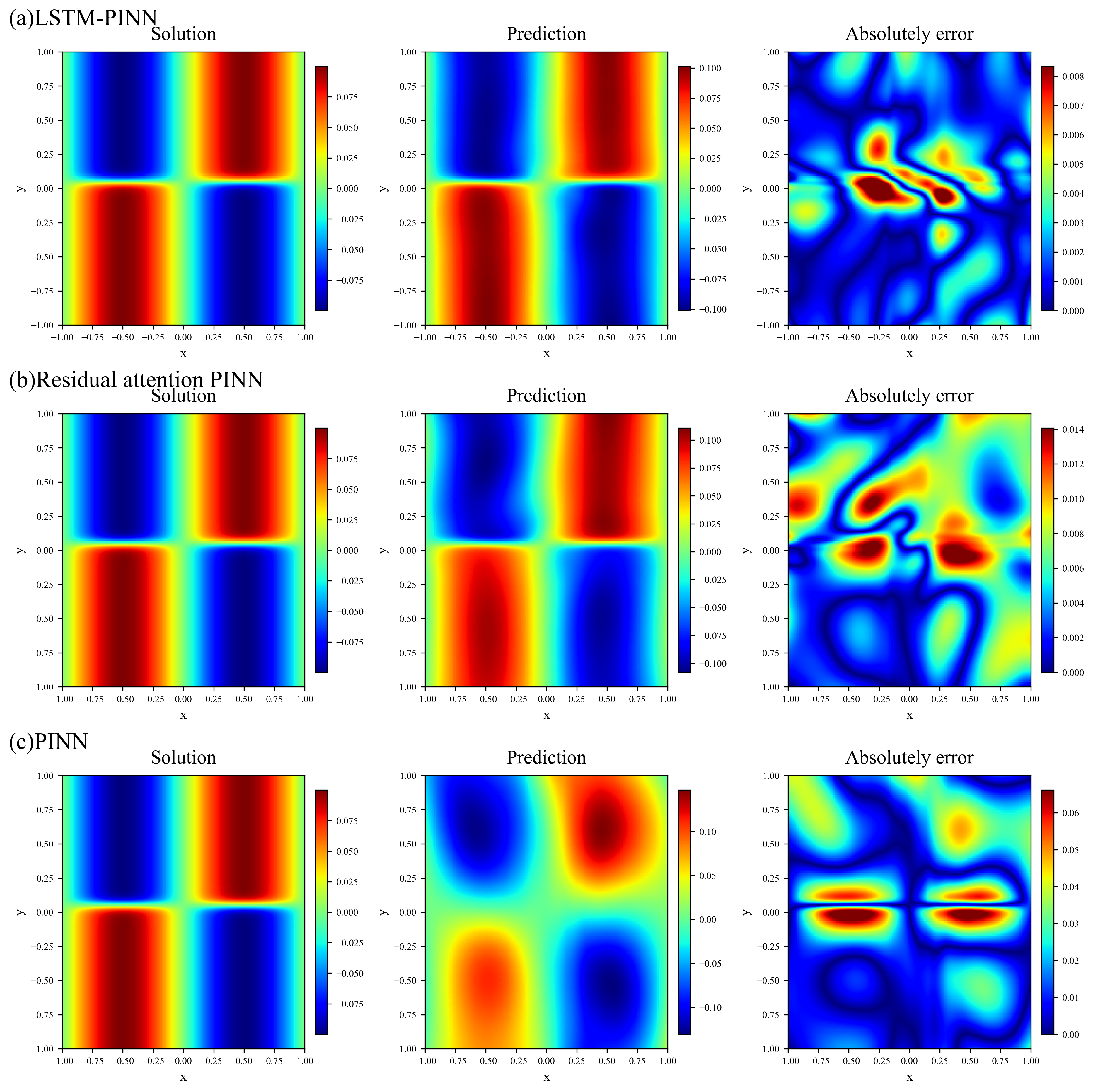}
    \caption{3$\times$3 reconstruction comparison of the \(u_y\)-field for Case 07.}
    \label{fig:case07_uy_panel_3x3}
\end{figure}

\begin{figure}[H]
    \centering
    \includegraphics[width=0.95\textwidth]{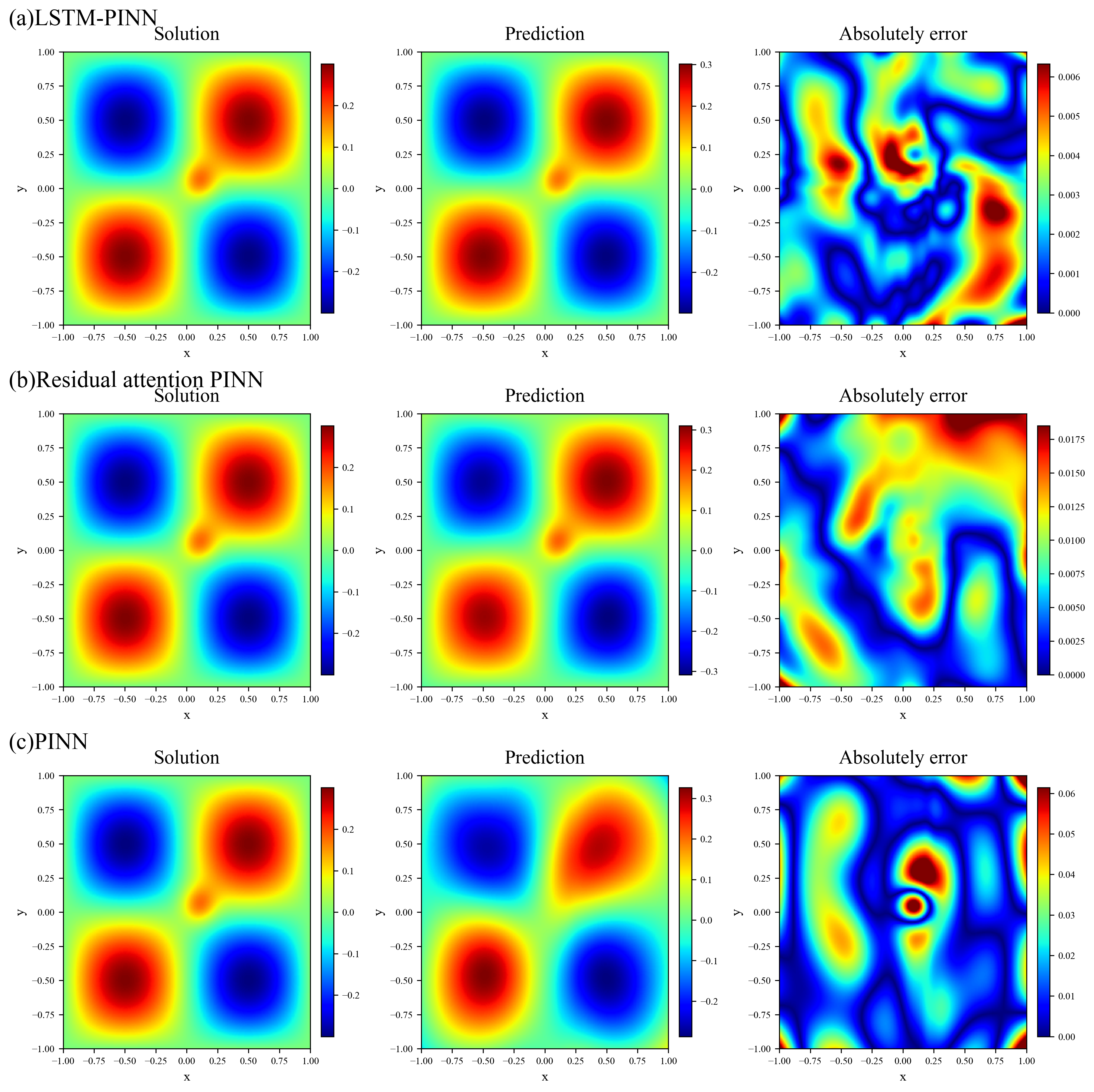}
    \caption{3$\times$3 reconstruction comparison of the \(\phi\)-field for Case 07.}
    \label{fig:case07_phi_panel_3x3}
\end{figure}

The visual comparison in Fig.~\ref{fig:case07_n_compare} shows that the simultaneous presence of the double front and the local Gaussian pocket gives this case a typical multi-peak, multi-level structure. LSTM-PINN preserves both front locations and the amplitude relationship in the region between them. ResAtt-PINN recovers the main structure correctly, but its reconstruction of the local pocket is weaker. Standard PINN is more likely to produce a blurred region between the two fronts. These observations indicate that LSTM-PINN provides better structural consistency for composite spatial patterns. The same conclusion is supported by the associated \(u_x\), \(u_y\), and \(\phi\) fields in Figs.~\ref{fig:case07_ux_panel_3x3}, \ref{fig:case07_uy_panel_3x3}, and \ref{fig:case07_phi_panel_3x3}. In the \(u_x\) panel, LSTM-PINN preserves the narrow central front and the nearby pocket-induced perturbations without visibly widening the main transition, whereas Standard PINN diffuses the front into a much broader high-error region on the left side. In the \(u_y\) and \(\phi\) fields, LSTM-PINN also keeps the symmetric four-cell background structure while localizing the pocket-related disturbance, whereas Standard PINN merges the local anomalies into a less organized error plume and ResAtt-PINN, although much better than Standard PINN, still loses more local detail than LSTM-PINN around the pocket-affected region.

\begin{landscape}
\begin{figure}[p]
    \centering
    \includegraphics[width=1\linewidth]{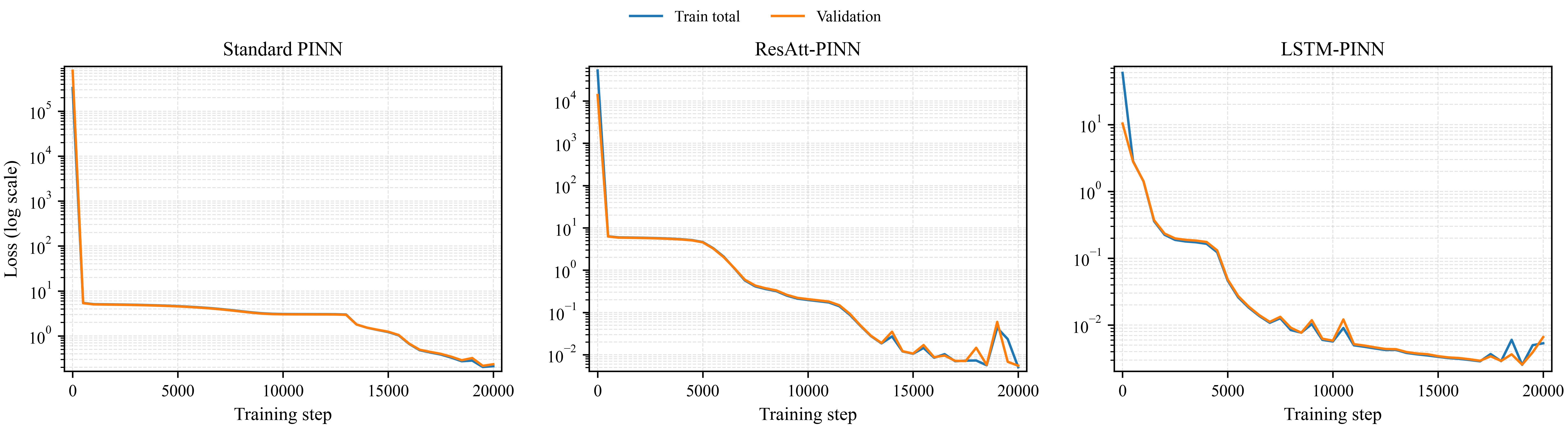}
    \caption{Training and validation loss curves for Case 07.}
    \label{fig:case07_loss_compare}

    \vspace{3em}

    \includegraphics[width=1\linewidth]{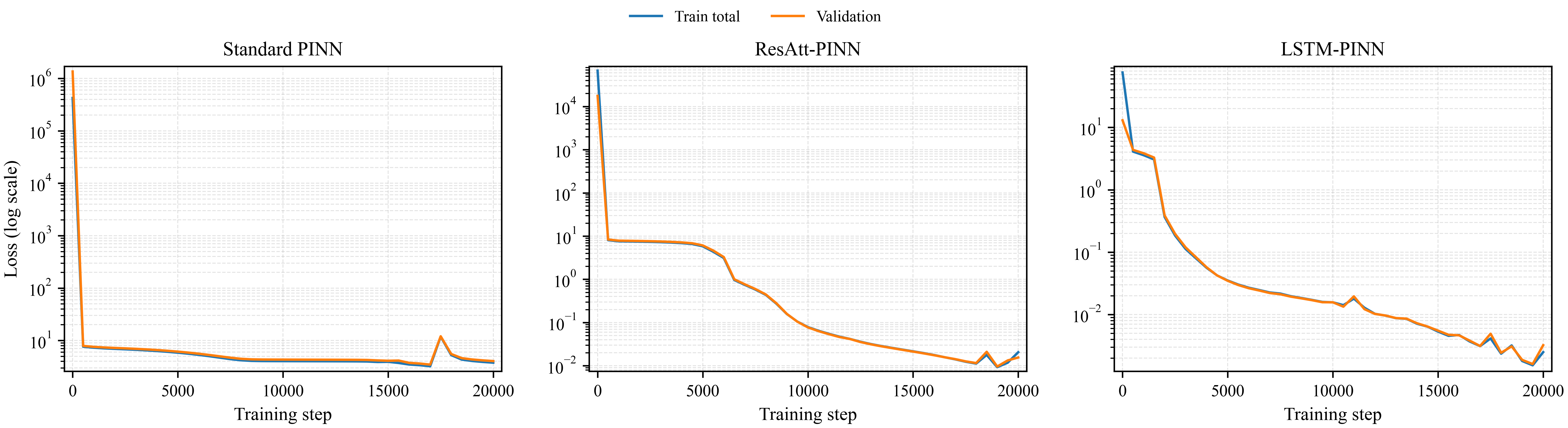}
    \caption{Training and validation loss curves for Case 08.}
    \label{fig:case08_loss_compare}
\end{figure}
\end{landscape}

For a quantitative comparison, Table~\ref{tab:case07_metrics} reports the overall RMSE, overall MSE, overall MAE, overall \(L_2\), training time, and peak GPU memory. LSTM-PINN achieves the smallest overall RMSE of 0.0042 and the smallest overall \(L_2\) error of 0.0052. It also attains the smallest overall MSE and MAE, indicating that its advantage extends consistently across multiple overall error measures. Although ResAtt-PINN is substantially more accurate than Standard PINN, it also requires much more training time and GPU memory. The quantitative results are consistent with the error distribution shown in Fig.~\ref{fig:case07_n_compare}, and the corresponding loss curves are shown in Fig.~\ref{fig:case07_loss_compare}, again demonstrating the better robustness and representation capability of the LSTM backbone for this kind of composite spatial structure.

\begin{center}
\captionsetup{hypcap=false}
\captionof{table}{Quantitative comparison for Case 07.}
\label{tab:case07_metrics}
\small
\setlength{\tabcolsep}{5pt}
\renewcommand{\arraystretch}{1.12}
\resizebox{\textwidth}{!}{%
\begin{tabular}{lcccccc}
\toprule
Model & \ohdr{overall RMSE}{RMSE} & \ohdr{overall MSE}{MSE} & \ohdr{overall MAE}{MAE} & \ohdr{overall \(L_2\)}{\(L_2\)} & Training time (s) & Peak GPU memory (GB) \\
\midrule
Standard PINN & 0.0323 & 1.04e-03 & 0.0236 & 0.0402 & \textbf{663.3} & 3.455 \\
ResAtt-PINN & 0.0100 & 1.00e-04 & 0.0075 & 0.0125 & 5147.3 & 10.190 \\
LSTM-PINN & \textbf{0.0042} & \textbf{1.73e-05} & \textbf{0.0027} & \textbf{0.0052} & 3839.5 & \textbf{1.699} \\
\bottomrule
\end{tabular}}
\end{center}

\subsection{Case 8: Hybrid Multiscale Benchmark}

Case 08 is the hardest multiscale benchmark in this study. It combines an oblique front, a local pocket, two-dimensional oscillations, and a transverse layer, and therefore serves as a comprehensive abstraction of a complex steady EHD structure. Because multiple structures overlap in space, this case is particularly suitable for testing the robustness of the model under highly coupled multiscale patterns.

\begin{figure}[H]
    \centering
    \includegraphics[width=0.95\textwidth]{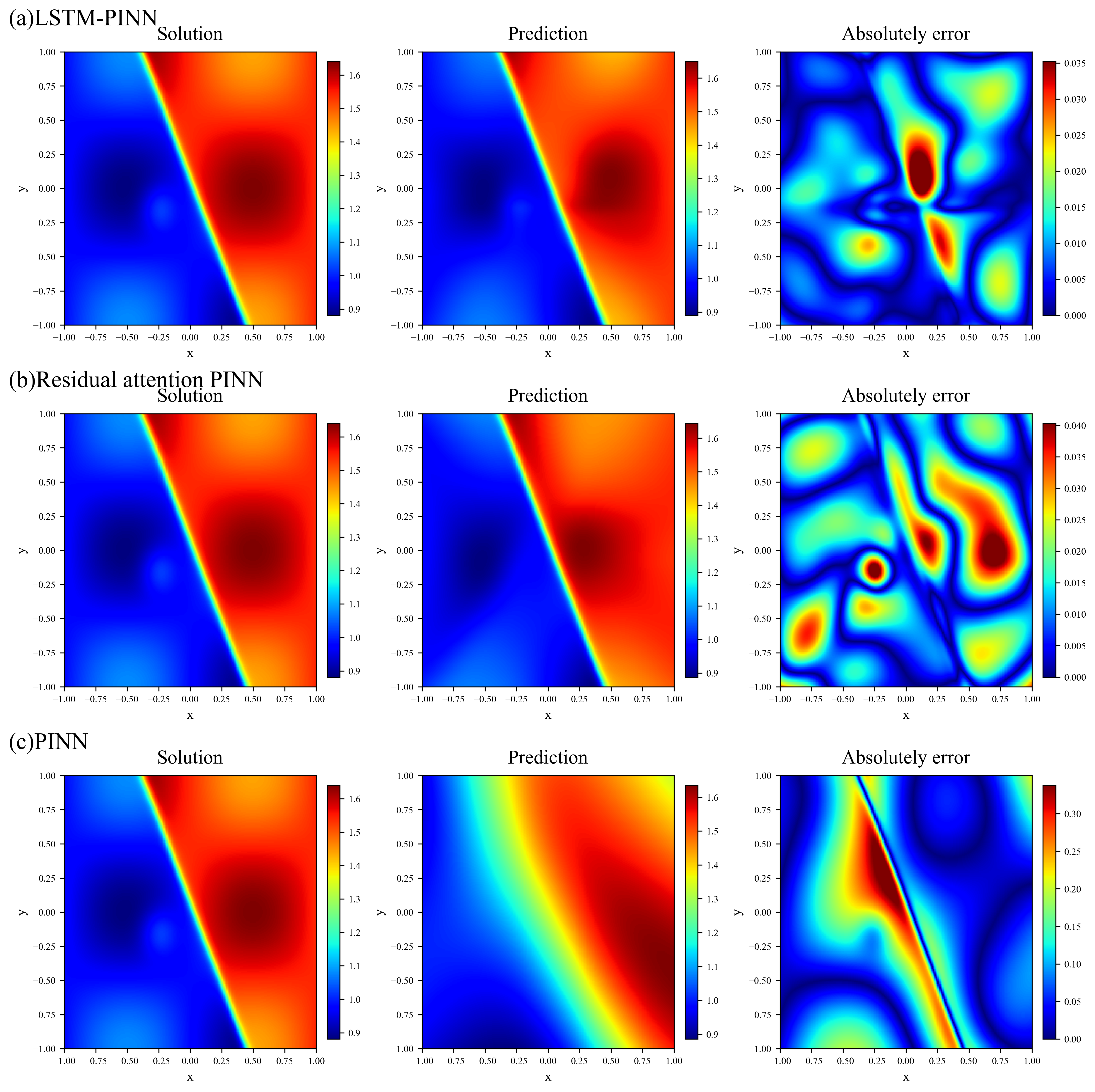}
    \caption{3$\times$3 reconstruction comparison of the \(n\)-field for Case 08, highlighting the reconstruction capability for the hardest multiscale benchmark with composite structures.}
    \label{fig:case08_n_compare}
\end{figure}

\begin{figure}[H]
    \centering
    \includegraphics[width=0.95\textwidth]{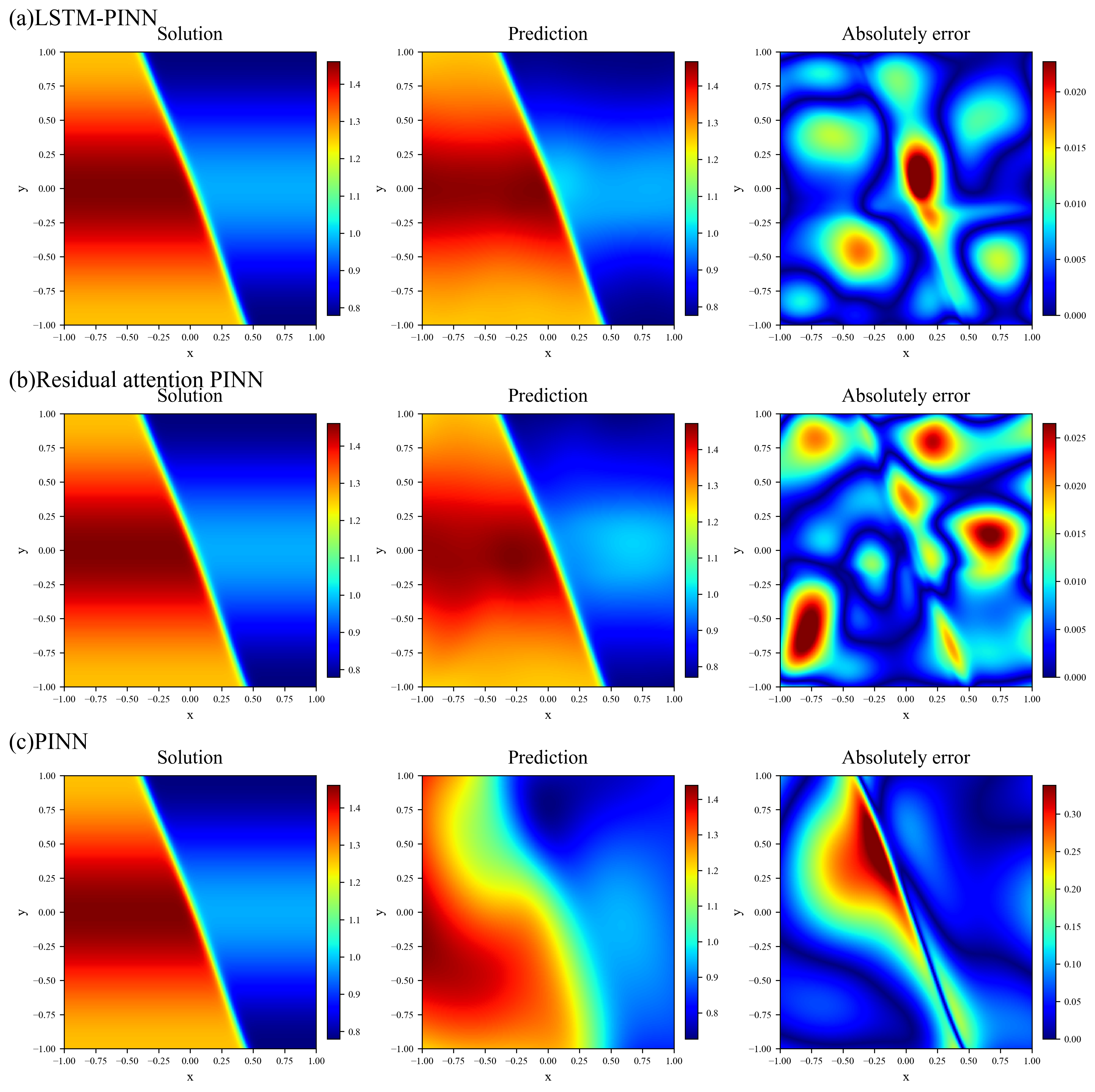}
    \caption{3$\times$3 reconstruction comparison of the \(u_x\)-field for Case 08.}
    \label{fig:case08_ux_panel_3x3}
\end{figure}

\begin{figure}[H]
    \centering
    \includegraphics[width=0.95\textwidth]{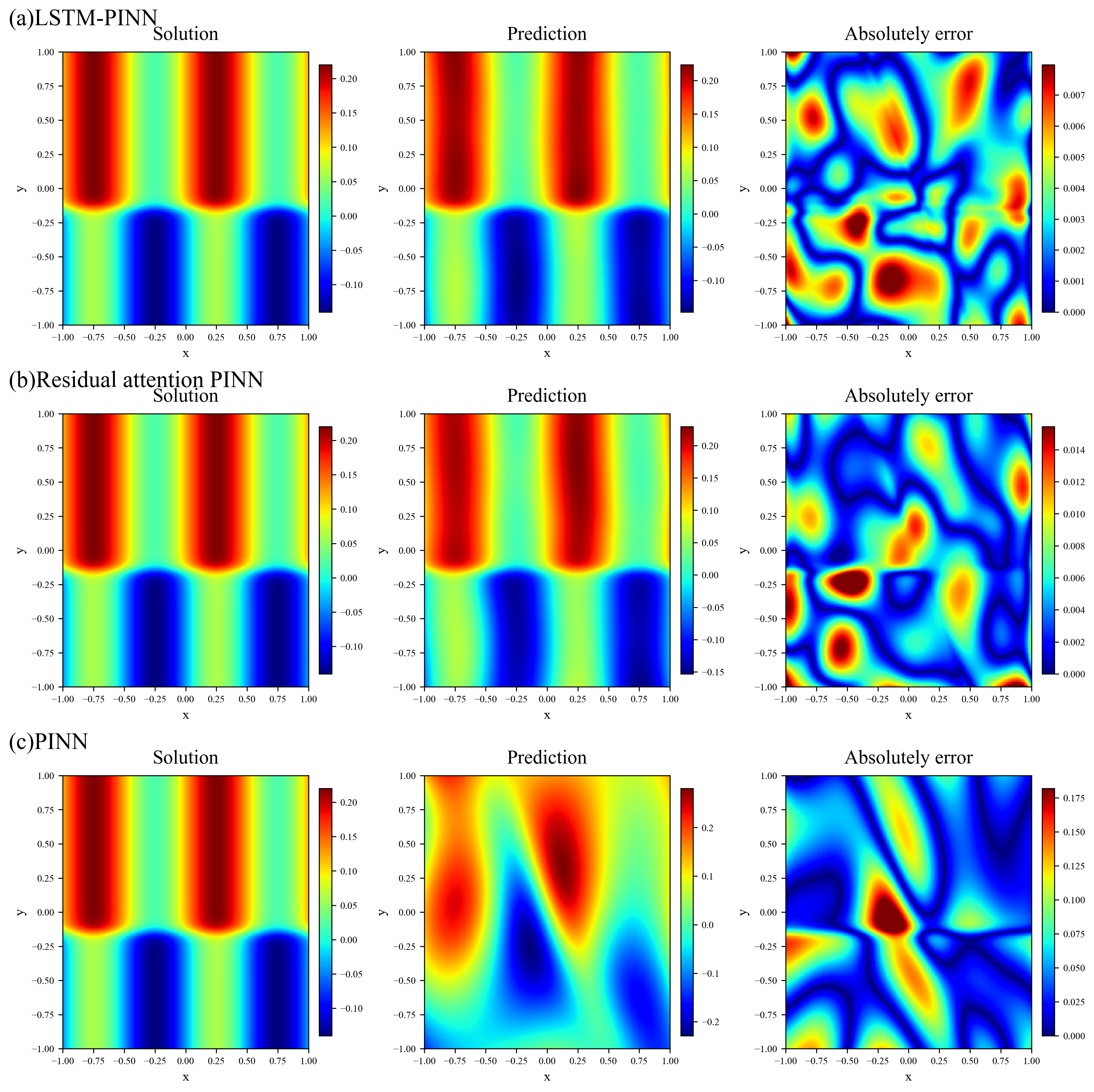}
    \caption{3$\times$3 reconstruction comparison of the \(u_y\)-field for Case 08.}
    \label{fig:case08_uy_panel_3x3}
\end{figure}

\begin{figure}[H]
    \centering
    \includegraphics[width=0.95\textwidth]{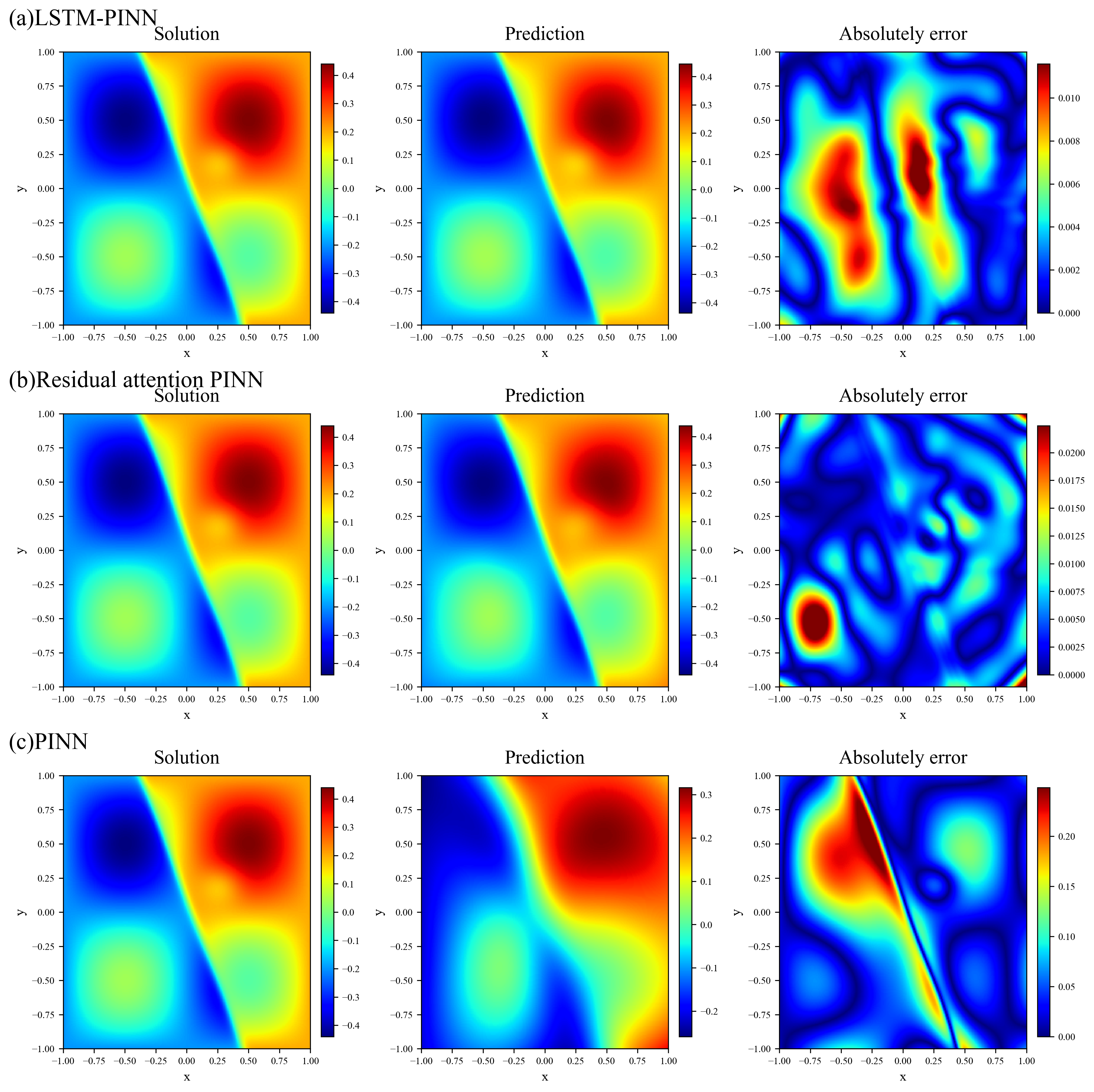}
    \caption{3$\times$3 reconstruction comparison of the \(\phi\)-field for Case 08.}
    \label{fig:case08_phi_panel_3x3}
\end{figure}

The reconstruction results in Fig.~\ref{fig:case08_n_compare} show that LSTM-PINN still preserves the overall consistency of both the main front and the local structures. Standard PINN exhibits the most significant error, especially in the overlapping region of the composite structures. This case most clearly demonstrates the advantage of LSTM-PINN in complicated multiscale settings. The added \(u_x\), \(u_y\), and \(\phi\) comparisons in Figs.~\ref{fig:case08_ux_panel_3x3}, \ref{fig:case08_uy_panel_3x3}, and \ref{fig:case08_phi_panel_3x3} show that the same robustness extends to the full coupled-field reconstruction. In the \(u_x\) panel, LSTM-PINN keeps the oblique interface narrow and leaves the smooth left and right backgrounds largely intact, whereas Standard PINN bends the front and introduces a broad cross-region contamination pattern. The \(u_y\) field is even more revealing: LSTM-PINN preserves the striped background on both sides of the oblique front, while Standard PINN collapses the stripe pattern into a strongly distorted V-shaped structure near the interface. The \(\phi\) panel shows the same hierarchy, with LSTM-PINN best maintaining the separation between the oblique transition and the surrounding smooth field, ResAtt-PINN remaining intermediate, and Standard PINN showing the largest residual distortion in the overlapping multiscale zone.

The quantitative comparison further confirms this conclusion. LSTM-PINN obtains the smallest overall RMSE of 0.0066, while ResAtt-PINN reaches 0.0101 and Standard PINN yields 0.0945. The overall MSE, MAE, and \(L_2\) values show the same ordering, which further supports the conclusion that LSTM-PINN provides the most reliable reconstruction under the hardest multiscale setting. Although ResAtt-PINN is much more accurate than Standard PINN, its training time and memory cost are both much higher than those of LSTM-PINN. The corresponding loss evolution is shown in Fig.~\ref{fig:case08_loss_compare}.

\begin{center}
\captionsetup{hypcap=false}
\captionof{table}{Quantitative comparison for Case 08.}
\label{tab:case08_metrics}
\small
\setlength{\tabcolsep}{5pt}
\renewcommand{\arraystretch}{1.12}
\resizebox{\textwidth}{!}{%
\begin{tabular}{lcccccc}
\toprule
Model & \ohdr{overall RMSE}{RMSE} & \ohdr{overall MSE}{MSE} & \ohdr{overall MAE}{MAE} & \ohdr{overall \(L_2\)}{\(L_2\)} & Training time (s) & Peak GPU memory (GB) \\
\midrule
Standard PINN & 0.0945 & 8.93e-03 & 0.0676 & 0.1089 & \textbf{662.8} & 3.455 \\
ResAtt-PINN & 0.0101 & 1.01e-04 & 0.0074 & 0.0116 & 5174.4 & 10.190 \\
LSTM-PINN & \textbf{0.0066} & \textbf{4.35e-05} & \textbf{0.0046} & \textbf{0.0076} & 4646.4 & \textbf{1.699} \\
\bottomrule
\end{tabular}}
\end{center}

\section{Comparative Discussion}

\subsection{Accuracy Comparison}

Table~\ref{tab:avg_metrics_acc} summarizes the average quantitative performance over all eight benchmark cases in terms of accuracy. LSTM-PINN achieves the best average overall RMSE of 0.0043, the best average overall MSE of \(1.98\times10^{-5}\), the best average overall MAE of 0.0029, and the best average overall \(L_2\) error of 0.0051. ResAtt-PINN ranks second, while Standard PINN performs the worst on all four overall error metrics. Moreover, the case-wise RMSE summary shows that LSTM-PINN attains the smallest overall RMSE on every one of the eight cases. This indicates that its advantage is not restricted to a specific front geometry, but extends consistently to vertical layers, horizontal layers, oblique fronts, interacting layers, radial fronts, and the hardest multiscale benchmark.

\begin{center}
\captionsetup{hypcap=false}
\captionof{table}{Average quantitative performance over all eight benchmark cases (accuracy).}
\label{tab:avg_metrics_acc}
\small
\setlength{\tabcolsep}{6pt}
\renewcommand{\arraystretch}{1.15}
\begin{tabular*}{0.92\textwidth}{@{\extracolsep{\fill}}lcccc}
\toprule
Model & Avg. RMSE & Avg. MSE & Avg. MAE & Avg. \(L_2\) \\
\midrule
LSTM-PINN & \textbf{0.0043} & \textbf{1.98e-05} & \textbf{0.0029} & \textbf{0.0051} \\
ResAtt-PINN & 0.0070 & 5.39e-05 & 0.0051 & 0.0083 \\
Standard PINN & 0.0408 & 2.19e-03 & 0.0290 & 0.0475 \\
\bottomrule
\end{tabular*}
\end{center}

\subsection{Efficiency Comparison}

From the perspective of efficiency, Standard PINN has the shortest average training time, 691.6~s, but this speed advantage comes at the cost of much larger errors. ResAtt-PINN improves the accuracy significantly relative to Standard PINN, yet it also incurs the highest average training time, 5342.5~s, and the largest average peak GPU memory, 9.555~GB. In contrast, LSTM-PINN provides the best balance between accuracy and computational cost: it achieves the best average accuracy across RMSE, MSE, MAE, and \(L_2\), while using only 1.699~GB of peak GPU memory on average, which is even lower than Standard PINN. Therefore, LSTM-PINN offers a much more favorable accuracy--efficiency trade-off than the other two baselines.

\begin{center}
\captionsetup{hypcap=false}
\captionof{table}{Average quantitative performance over all eight benchmark cases (efficiency).}
\label{tab:avg_metrics_eff}
\small
\setlength{\tabcolsep}{6pt}
\renewcommand{\arraystretch}{1.15}
\begin{tabular*}{0.92\textwidth}{@{\extracolsep{\fill}}lcc}
\toprule
Model & Avg. training time (s) & Avg. peak GPU memory (GB) \\
\midrule
LSTM-PINN & 3459.5 & \textbf{1.699} \\
ResAtt-PINN & 5342.5 & 9.555 \\
Standard PINN & \textbf{691.6} & 3.455 \\
\bottomrule
\end{tabular*}
\end{center}

\subsection{Robustness Analysis}

The case-by-case visual comparisons suggest that the main advantage of LSTM-PINN is not limited to reducing isolated local peak errors. More importantly, it preserves the global consistency of complex spatial structures more effectively. This feature is especially evident in cases with oblique fronts, crossing layers, curved radial interfaces, and strongly coupled multiscale structures. In other words, the LSTM backbone appears to be more robust in representing long-range spatial correlations after pseudo-sequential encoding, which explains its consistently superior performance across all eight benchmark cases. This robustness is also supported by the quantitative results, since LSTM-PINN remains the best-performing model not only in RMSE and \(L_2\), but also in MSE and MAE under the unified benchmark setting.

\begin{center}
\captionsetup{hypcap=false}
\captionof{table}{Case-wise overall RMSE comparison across the eight benchmark cases.}
\label{tab:casewise_rmse}
\small
\setlength{\tabcolsep}{6pt}
\renewcommand{\arraystretch}{1.15}
\begin{tabular*}{0.92\textwidth}{@{\extracolsep{\fill}}lccc}
\toprule
Case & Standard PINN & ResAtt-PINN & LSTM-PINN \\
\midrule
Case 01 & 0.0296 & 0.0066 & \textbf{0.0046} \\
Case 02 & 0.0231 & 0.0050 & \textbf{0.0047} \\
Case 03 & 0.0361 & 0.0085 & \textbf{0.0029} \\
Case 04 & 0.0413 & 0.0044 & \textbf{0.0041} \\
Case 05 & 0.0534 & 0.0062 & \textbf{0.0037} \\
Case 06 & 0.0160 & 0.0055 & \textbf{0.0040} \\
Case 07 & 0.0323 & 0.0100 & \textbf{0.0042} \\
Case 08 & 0.0945 & 0.0101 & \textbf{0.0066} \\
\bottomrule
\end{tabular*}
\end{center}

\section{Conclusion}

In this study, we have developed a comprehensive benchmark framework to systematically evaluate the capability of various physics-informed neural network architectures in resolving two-dimensional steady electrohydrodynamic shock-like problems. To address the profound computational challenges posed by strongly coupled fields with sharp gradients and multiscale spatial structures, we formulated a unified four-variable system governing charged-particle density, velocity components, and electric potential. Grounded in the fundamental principles of steady electrohydrodynamic transport, this formulation provides a physically meaningful and mathematically consistent foundation for eight rigorously designed benchmark cases that encompass diverse front geometries and intricate multiscale patterns. By employing a unified residual construction alongside a strictly controlled training protocol, we ensured a fair and highly reproducible comparison among the Standard PINN, ResAtt-PINN, and LSTM-PINN models under identical equations, source terms, and sampling strategies. Our extensive evaluations consistently demonstrate that the LSTM-PINN architecture achieves the highest overall accuracy and the most robust structural reliability across all test scenarios. Furthermore, the LSTM-PINN explicitly offers the most favorable balance between predictive precision and computational efficiency among the tested architectures. Ultimately, this work not only successfully extends the application of recurrent network backbones to a highly demanding four-variable charged-fluid system, but it also establishes a standardized and reproducible testbed to accelerate the future development of advanced mesh-free solvers for complex spatial structures in computational physics.

\section*{Declaration of competing interest}
The authors declared that they have no conflicts of interest to this work. 
\section*{Acknowledgment}
This work is supported by the developing Project of Science and Technology of Jilin Province (20250102032JC).  
%\section*{Acknowledgment}
%This work is supported by the developing Project of Science and Technology of Jilin Province (20240402042GH). 

\section*{Data availability}
All the code for this article is available open access at a Github repository available at https://github.com/Uderwood-TZ/-Two-Dimensional-Steady-Electrohydrodynamic-Flow-Based-on-LSTM-PINN.git.
\clearpage
\bibliographystyle{model1-num-names}
\bibliography{cas-refs}

\appendix

\clearpage
\begin{landscape}

\refstepcounter{table}
\begin{center}
\begin{minipage}{0.82\linewidth}
\raggedright
\scriptsize
\setlength{\tabcolsep}{2.3pt}
\renewcommand{\arraystretch}{1.03}

\textbf{Table \thetable}\\
\textit{Parameter summary table for all benchmark cases.}
\label{tab:appendix_parameter_summary}

\vspace{0.25em}

\centering
\begin{tabular}{@{}>{\raggedright\arraybackslash}p{0.09\linewidth}
                >{\raggedright\arraybackslash}p{0.13\linewidth}
                >{\raggedright\arraybackslash}p{0.70\linewidth}@{}}
\toprule
Case & Category & Parameters \\
\midrule

Case 01 & Density &
$n_0=1.25$, $A_n=0.18$, $A_s=0.42$, $x_n=0.02$, $\delta_n=0.06$ \\
Case 01 & Velocity &
$U_0=1.15$, $A_u=0.22$, $A_{us}=0.35$, $x_u=-0.03$, $\delta_u=0.05$, $A_v=0.10$ \\
Case 01 & Potential &
$A_{\phi o}=0.45$, $A_{\phi s}=0.24$, $x_\phi=-0.01$, $\delta_\phi=0.07$ \\
\midrule

Case 02 & Density &
$n_0=1.18$, $A_n=0.14$, $A_s=0.38$, $y_n=-0.05$, $\delta_n=0.05$ \\
Case 02 & Velocity &
$U_0=0.92$, $A_u=0.20$, $V_0=-0.15$, $A_v=0.10$, $A_{ys}=0.30$, $y_u=0.01$, $\delta_u=0.045$ \\
Case 02 & Potential &
$A_{\phi o}=0.32$, $A_{\phi s}=0.20$, $y_\phi=0.02$, $\delta_\phi=0.06$ \\
\midrule

Case 03 & Density &
$\theta=35^\circ$, $n_0=1.22$, $A_{n1}=0.10$, $A_{n2}=0.08$, $A_{ns}=0.40$, $s_n=0$, $\delta_n=0.055$ \\
Case 03 & Velocity &
$U_0=1.05$, $A_u=0.18$, $A_{us}=0.28$, $s_u=0.03$, $\delta_u=0.05$, $A_{v1}=0.08$, $A_{v2}=0.10$ \\
Case 03 & Potential &
$A_{\phi o}=0.28$, $A_{\phi s}=0.22$, $s_\phi=-0.015$, $\delta_\phi=0.07$ \\
\midrule

Case 04 & Density &
\parbox[t]{\linewidth}{$\theta_1=30^\circ$, $\theta_2=-40^\circ$, $n_0=1.30$, $A_n=0.08$, $A_{s1}=0.30$, $s_{n1}=-0.10$, $\delta_{n1}=0.06$,\\
$A_{s2}=0.24$, $s_{n2}=0.12$, $\delta_{n2}=0.05$} \\
Case 04 & Velocity &
\parbox[t]{\linewidth}{$U_0=1.02$, $A_u=0.12$, $A_{u1}=0.22$, $s_u=-0.08$, $\delta_u=0.06$, $V_0=0.05$, $A_v=0.07$,\\
$A_{v1}=0.18$, $s_v=0.10$, $\delta_v=0.05$} \\
Case 04 & Potential &
\parbox[t]{\linewidth}{$A_{\phi o}=0.22$, $A_{\phi 1}=0.16$, $s_{\phi 1}=-0.05$, $\delta_{\phi 1}=0.06$, $A_{\phi 2}=0.14$,\\
$s_{\phi 2}=0.03$, $\delta_{\phi 2}=0.05$} \\
\midrule

Case 05 & Density &
$x_c=-0.08$, $y_c=0.05$, $n_0=1.24$, $A_n=0.12$, $A_{rs}=0.32$, $r_n=0.42$, $\delta_n=0.045$ \\
Case 05 & Velocity &
$U_0=0.95$, $A_u=0.10$, $A_\omega=0.18$, $\beta=3.0$, $V_0=0.05$ \\
Case 05 & Potential &
$A_{\phi o}=0.20$, $A_{\phi s}=0.18$, $r_\phi=0.40$, $\delta_\phi=0.05$ \\
\midrule

Case 06 & Density &
\parbox[t]{\linewidth}{$y_s=-0.02$, $\delta_s=0.04$, $n_0=1.16$, $A_n=0.10$, $A_s=0.12$, $A_g=0.10$, $x_g=0.35$,\\
$y_g=-0.30$, $\sigma_x=0.12$, $\sigma_y=0.10$} \\
Case 06 & Velocity &
$U_0=1.08$, $A_u=0.35$, $A_{uo}=0.08$, $A_v=0.06$ \\
Case 06 & Potential &
$A_{\phi o}=0.26$, $A_{\phi s}=0.22$ \\
\midrule

Case 07 & Density &
\parbox[t]{\linewidth}{$n_0=1.20$, $A_{n1}=0.08$, $A_{n2}=0.06$, $A_{g1}=0.12$, $x_{g1}=-0.20$, $y_{g1}=0.15$, $\sigma_{x1}=0.10$,\\
$\sigma_{y1}=0.08$, $A_{g2}=-0.10$, $x_{g2}=0.25$, $y_{g2}=-0.18$, $\sigma_{x2}=0.08$, $\sigma_{y2}=0.10$} \\
Case 07 & Velocity &
\parbox[t]{\linewidth}{$U_0=1.00$, $A_{f1}=0.18$, $x_{u1}=-0.18$, $\delta_{u1}=0.05$, $A_{f2}=0.15$, $x_{u2}=0.22$, $\delta_{u2}=0.05$,\\
$A_u=0.06$, $A_v=0.10$, $y_v=0.05$, $\delta_v=0.04$} \\
Case 07 & Potential &
\parbox[t]{\linewidth}{$A_{\phi o}=0.30$, $A_{\phi g}=0.16$, $x_{\phi g}=0.10$, $y_{\phi g}=0.05$, $\sigma_{\phi x}=0.10$,\\
$\sigma_{\phi y}=0.10$} \\
\midrule

Case 08 & Density &
\parbox[t]{\linewidth}{$\theta=22.5^\circ$, $s_n=0.03$, $\delta_n=0.035$, $n_0=1.26$, $A_n=0.10$, $A_s=0.28$, $A_g=0.10$,\\
$x_g=-0.25$, $y_g=-0.15$, $\sigma_x=0.10$, $\sigma_y=0.12$} \\
Case 08 & Velocity &
\parbox[t]{\linewidth}{$U_0=1.12$, $A_u=0.24$, $A_{uo}=0.10$, $V_0=0.04$, $A_v=0.10$, $A_{vy}=0.08$,\\
$y_v=-0.15$, $\delta_v=0.05$} \\
Case 08 & Potential &
\parbox[t]{\linewidth}{$A_{\phi o}=0.24$, $A_{\phi s}=0.20$, $A_{\phi g}=-0.12$, $x_{\phi g}=0.28$, $y_{\phi g}=0.20$,\\
$\sigma_{\phi x}=0.12$, $\sigma_{\phi y}=0.10$} \\
\bottomrule
\end{tabular}
\end{minipage}
\end{center}

\vspace{0.3em}

\subsection*{A. Parameter Summary Table}

{\normalsize
Table~\ref{tab:appendix_parameter_summary} summarizes the density, velocity, and electric-potential parameters of all eight benchmark cases. For better readability, the parameter summary is organized in a three-row format for each case, rather than forcing all parameter groups into a single row.
}

\end{landscape}

\clearpage
\begin{landscape}
\section{Detailed Error Metrics by Variable}

Table~\ref{tab:detailed_error_metrics} reports the detailed variable-wise error metrics for all eight benchmark cases and all three models. In addition to the overall RMSE, overall MSE, overall MAE, and overall $L_2$ error reported in the main text, this appendix further lists the RMSE of the four physical variables, namely $n$, $u_x$, $u_y$, and $\phi$. These results provide a finer-grained view of how the prediction error is distributed across different field components.

\vspace{0.4em}

\begin{table}[H]
\centering
\small
\setlength{\tabcolsep}{4pt}
\renewcommand{\arraystretch}{1.05}
\caption{Variable-wise error metrics for all benchmark cases and all models.}
\label{tab:detailed_error_metrics}
\begin{tabular}{llcccccccc}
\toprule
Case & Model & $n$ RMSE & $u_x$ RMSE & $u_y$ RMSE & $\phi$ RMSE & \ohdr{overall RMSE}{RMSE} & \ohdr{overall MSE}{MSE} & \ohdr{overall MAE}{MAE} & \ohdr{overall $L_2$}{$L_2$} \\
\midrule
Case 01 & Standard PINN & 0.0377 & 0.0356 & 0.0166 & 0.0233 & 0.0296 & 8.76e-04 & 0.0226 & 0.0331 \\
Case 01 & ResAtt-PINN   & 0.0079 & 0.0087 & 0.0034 & 0.0049 & 0.0066 & 4.37e-05 & 0.0048 & 0.0074 \\
Case 01 & LSTM-PINN     & \textbf{0.0072} & \textbf{0.0045} & \textbf{0.0024} & \textbf{0.0025} & \textbf{0.0046} & \textbf{2.09e-05} & \textbf{0.0030} & \textbf{0.0051} \\

Case 02 & Standard PINN & 0.0360 & 0.0216 & 0.0161 & 0.0111 & 0.0231 & 5.36e-04 & 0.0165 & 0.0289 \\
Case 02 & ResAtt-PINN   & 0.0079 & 0.0045 & \textbf{0.0030} & 0.0031 & 0.0050 & 2.53e-05 & 0.0036 & 0.0063 \\
Case 02 & LSTM-PINN     & \textbf{0.0073} & \textbf{0.0031} & 0.0042 & \textbf{0.0024} & \textbf{0.0047} & \textbf{2.18e-05} & \textbf{0.0032} & \textbf{0.0058} \\

Case 03 & Standard PINN & 0.0599 & 0.0313 & 0.0210 & 0.0138 & 0.0361 & 1.30e-03 & 0.0257 & 0.0421 \\
Case 03 & ResAtt-PINN   & 0.0126 & 0.0100 & 0.0042 & 0.0039 & 0.0085 & 7.30e-05 & 0.0061 & 0.0100 \\
Case 03 & LSTM-PINN     & \textbf{0.0043} & \textbf{0.0025} & \textbf{0.0022} & \textbf{0.0019} & \textbf{0.0029} & \textbf{8.16e-06} & \textbf{0.0021} & \textbf{0.0033} \\

Case 04 & Standard PINN & 0.0558 & 0.0414 & 0.0381 & 0.0233 & 0.0413 & 1.71e-03 & 0.0317 & 0.0472 \\
Case 04 & ResAtt-PINN   & \textbf{0.0053} & 0.0054 & 0.0030 & 0.0033 & 0.0044 & 1.93e-05 & 0.0033 & 0.0050 \\
Case 04 & LSTM-PINN     & 0.0064 & \textbf{0.0041} & \textbf{0.0021} & \textbf{0.0024} & \textbf{0.0041} & \textbf{1.70e-05} & \textbf{0.0028} & \textbf{0.0047} \\

Case 05 & Standard PINN & 0.0813 & 0.0593 & 0.0195 & 0.0299 & 0.0534 & 2.85e-03 & 0.0316 & 0.0603 \\
Case 05 & ResAtt-PINN   & 0.0098 & 0.0064 & 0.0025 & 0.0030 & 0.0062 & 3.81e-05 & 0.0040 & 0.0070 \\
Case 05 & LSTM-PINN     & \textbf{0.0060} & \textbf{0.0036} & \textbf{0.0017} & \textbf{0.0021} & \textbf{0.0037} & \textbf{1.40e-05} & \textbf{0.0024} & \textbf{0.0042} \\

Case 06 & Standard PINN & 0.0171 & 0.0221 & 0.0083 & 0.0133 & 0.0160 & 2.56e-04 & 0.0123 & 0.0193 \\
Case 06 & ResAtt-PINN   & 0.0076 & 0.0058 & 0.0030 & 0.0045 & 0.0055 & 3.02e-05 & 0.0037 & 0.0066 \\
Case 06 & LSTM-PINN     & \textbf{0.0069} & \textbf{0.0033} & \textbf{0.0017} & \textbf{0.0017} & \textbf{0.0040} & \textbf{1.59e-05} & \textbf{0.0021} & \textbf{0.0048} \\

Case 07 & Standard PINN & 0.0352 & 0.0428 & 0.0223 & 0.0244 & 0.0323 & 1.04e-03 & 0.0236 & 0.0402 \\
Case 07 & ResAtt-PINN   & 0.0139 & 0.0089 & 0.0061 & 0.0096 & 0.0100 & 1.00e-04 & 0.0075 & 0.0125 \\
Case 07 & LSTM-PINN     & \textbf{0.0066} & \textbf{0.0035} & \textbf{0.0024} & \textbf{0.0028} & \textbf{0.0042} & \textbf{1.73e-05} & \textbf{0.0027} & \textbf{0.0052} \\

Case 08 & Standard PINN & 0.1163 & 0.1060 & 0.0640 & 0.0830 & 0.0945 & 8.93e-03 & 0.0676 & 0.1089 \\
Case 08 & ResAtt-PINN   & 0.0153 & 0.0095 & 0.0065 & 0.0064 & 0.0101 & 1.01e-04 & 0.0074 & 0.0116 \\
Case 08 & LSTM-PINN     & \textbf{0.0099} & \textbf{0.0065} & \textbf{0.0036} & \textbf{0.0044} & \textbf{0.0066} & \textbf{4.35e-05} & \textbf{0.0046} & \textbf{0.0076} \\
\bottomrule
\end{tabular}
\end{table}
\end{landscape}

\clearpage

\clearpage

\clearpage

\clearpage

\clearpage

\clearpage

\clearpage

\clearpage

\clearpage

\clearpage

\clearpage

\clearpage

\clearpage

\clearpage

\clearpage

\clearpage

\clearpage

\clearpage

\clearpage

\clearpage

\clearpage

\clearpage

\clearpage

\clearpage

\clearpage

\clearpage

\clearpage

\clearpage

\clearpage

\clearpage

\clearpage

\clearpage

\captionsetup{width=1\linewidth, skip=8pt}

\end{document}